\pdfoutput=1
\documentclass[10pt,twocolumn]{article} 
\usepackage{simpleConference}
\usepackage{times}
\usepackage{graphicx}
\usepackage{amssymb}
\usepackage[table,xcdraw]{xcolor} 
\usepackage{colortbl} 
\usepackage{lscape}
\usepackage{cite}
\usepackage{pdflscape}
\usepackage{amsmath,amssymb,amsfonts}
\usepackage{algorithmic}
\usepackage{graphicx}
\usepackage{makecell}
\usepackage{multirow}
\usepackage{subfig} 
\usepackage{textcomp}
\usepackage{xcolor}
\usepackage{float}

\begin{document}

\title{Visualization of missing data: a state-of-the-art survey\\}

\author{
{\makecell{{Sarah Alsufyani}\\ 
\textit{School of Computing} \\
Newcastle upon Tyne, UK\\
s.h.h.alsufyani2@ncl.ac.uk
}
} \and {\makecell{{Matthew Forshaw}
\\ 
\textit{School of Computing} \\
Newcastle upon Tyne, UK\\
matthew.forshaw@ncl.ac.uk
}
}
\and{\makecell{{Sara Johansson Fernstad}\\ 
\textit{School of Computing} \\
Newcastle upon Tyne, UK\\
sara.fernstad@ncl.ac.uk}}
}


\maketitle
\thispagestyle{empty}

\begin{abstract}
Missing data, the data value that is not recorded for a variable, occurs in almost all statistical analyses and may be caused by many reasons, such as lack of collection or a lack of documentation. Researchers need to adequately deal with this issue to provide a valid analysis. The visualization of missing values plays an important role in supporting the investigation and understanding of the missing data patterns. While some techniques and tools for visualization of missing values are available, it is still a challenge to select the right visualization that will fulfil the user requirements for visualizing missing data. This paper provides an overview and state-of-the-art report (STAR) of research literature focusing on missing values visualization. To the best of our knowledge, this is the first survey paper with a focus on missing data visualization. The goal of this paper is to encourage visualization researchers to increase their involvement with Missing data visualization.
\end{abstract}

\section{\textbf {Introduction}}
Missing values or incomplete data refers to the absence of data values in parts of a dataset. It is a common challenge in the data collection process across a variety of disciplines, causing many issues such as uncertain, biased, inaccurate, and unreliable results. Data values could be missing for many reasons, such as participants dropping out of studies, equipment malfunctions and entry errors, amongst many others.

The two main methods to deal with missing values are elimination and imputation. The elimination method basically removes the missing values prior to analysis, while the imputation method recreates missing values by replacing them based on a reasonable estimate, such as the mean or median value for that variable. Both methods may bias the analysis results, and their appropriateness may depend on the distribution and dependencies of the missing values. Furthermore, missing values could be informative and draw attention to potential problems with data collection, pre-processing, and analysis. Visualization of missing values plays an important role in exploring and understanding missingness in data.

Recent years have seen an increase in the design and evaluation of visualization methods for missing data, but there is a lack of overview and guidance to support designers and future research. To address this, we present an overview of missing data visualizations and a state-of-the-art report (STAR) of research literature focusing on missing data visualization. The contributions of this work include: 
\begin{itemize} 
\itemsep0em 
\item The first survey paper on the topic of missing data visualization, providing an extensive and up-to-date overview of recent publications. 
\item A novel classification of missing data visualization methods and literature.
\item An overview of future challenges and open research directions in the field.
\end{itemize}

This paper is organised as follows:  Section \ref{sec:Methodology} provides an overview of the literature search methodology, search queries, the review scope, and the terminology used. Section \ref{sec:Taxonomy} describes our literature classification methods. Section \ref{sec:RW} introduces related work. Section \ref{sec:VIsTech} describes 11 focus papers on missing data visualization techniques. We further categorize these papers into: novel visualization techniques and modified visualization techniques. Section \ref{sec:VisAppTool} describes 8 focus papers on missing data visualization applications and tools found from our literature search. Section \ref{sec:VisEval} describes 6 focus papers on evaluation of missing data visualization. Finally, section \ref{sec:Conclusion}  concludes the paper with a summary of the main contributions and future challenges.

\section {\textbf{{Methodology}}} \label{sec:Methodology}

\subsection{ \textbf{Literature Search Methodology }}\label{AA}
We started our literature search primarily on publications found through the following resources: 
\begin{itemize}
\item	IEEE VIS: Visualization \& Visual Analytics conferences 
\item	EuroVis: Eurographics Conference on Visualization conferences 
\item {IEEE TVCG:} IEEE Transactions on Visualization and Computer Graphics journal 
\item	Google Scholar
\item	IEEE Xplore
\end{itemize}



In addition to searches using the above sources, the related work sections and citations of reviewed papers were used to extend the search. We also used connected papers, a visual tool, to help finding papers relevant to visualizing missing data \cite{Behera2023}. A total of 42 papers were selected for initial review of which 25 were selected as focus papers for the survey, as described in more detail below.

\subsection{\textbf{Search queries}}
Our survey used the method of systematic literature review to search and gather literature, with focus on these keywords: missing data visualization, missing value visualization, incomplete data visualization, and imputed values visualization. 

\subsection{\textbf{Scope}}
In this report, we provide an overview of papers that focus on the visualization of missing values. All papers and research topics mentioned here present methods, techniques, applications, or tools that include visualization of missing values. The following criteria were used to define if a paper was in scope of the review or not.
\begin{itemize}
\item \textbf{In Scope:} In this work, we focus on missing data visualization. We include peer-reviewed literature, novel techniques, tools, and evaluation papers. No time restrictions were applied.
\item \textbf{Out of Scope:} Non-peer-reviewed publications, visualization of data quality aspects other than missing values, missing data analysis not focusing on visualization and other off-topic papers. 
\item \textbf{Off-topic:} We exclude papers that focus solely on the visualization of statistics related to missing values or on imputation methods.
\end{itemize}

\subsection{\textbf{Background and terminology }}
The following terminology and definitions are used throughout the report.
\begin{itemize}
\item \textbf{Data visualization}: Data visualization is an interdisciplinary field that represents information graphically to give a better understanding of the data.
\item \textbf{Missing data - Incomplete data}: Missing data, also known as missing value or incomplete data, occurs when no data value is stored for the variable in an observation.
\item \textbf{Imputation method}: The method of replacing the missing value with a substitute value, imputed value, to create complete data.
\item \textbf{Imputed value}: A calculated estimate of a value derived from other available information when an explicit value is missing. 
\item \textbf{Missingness mechanism}: It describes the process or reason behind why data is missing in a dataset.
\item \textbf{Missingness pattern}: It refers to the structure and distribution of missing data within a dataset, describing how the missing values are spread across different variables and observations.
\end{itemize}

\section{\textbf{Taxonomy and Classification of Missing Data Visualizations}} \label{sec:Taxonomy}
To classify the papers and projects, we developed a novel classification: 
\begin{itemize}
\item \textbf{Data type:} numerical, categorical, temporal, and other (network, flow or images).
\item \textbf{Imputation method}: if the paper addresses visualization for imputed data or not.
\item \textbf{Task:} visual representation of missing data, exploration of missing values, representation of the structure of missing values, identification of missing patterns, identification of missing mechanisms, and guided approach for the imputation of missing values. 
\item \textbf{Interactivity:} interactive visualization, including brushing, highlighting, filtering, zooming, and linking, and Non-interactive visualization that does not include any interactive features.
\end{itemize}
Table \ref{table_Classification3} provides a detailed description of the reviewed papers in context of this classification.

To improve the readability of this paper, we further categorized the focus papers based on paper type, as follows:  
\begin{itemize}
\item \textbf{Missing data visualization techniques} (10 papers): describing novel visualization techniques or modifying an existing one that support the visualization of missing data.
\item \textbf{Missing data visualization applications and tools} (7 papers): including web-based systems, graphical user interfaces, packages, and tools that visualize missing data. 
\item \textbf{Missing data visualization evaluation} (6 papers): including papers encompassing various evaluation methodologies, such as pilot studies, user studies, case studies, heuristic evaluations, and interviews.
\end{itemize}
We employed this classification as the primary framework and structured the papers accordingly.

For the missing data visualization evaluation papers, we classify them based on the task of the evaluation: effect on decision-making, estimation of value/trend/pattern, effect on perceived confidence, effect on perceived data quality, interpretation of data with missing, and identification of missingness structures (see Table \ref{tab: Evaluation paper}).
Table \ref{table_Classification2} shows a classification of papers that represent most common visualization methods based on visualization method and paper type, and \ref{table_Classification} shows a classification of papers that represent specialised visualization methods based on visualization method and paper type.

\begin{table*}[]
\centering
\begin{tabular}{|c|c|ccccc|}
\hline
 &
  {\color[HTML]{000000} } &
  \multicolumn{5}{c|}{Task of the evaluation} \\ \cline{3-7} 
\multirow{-2}{*}{} &
  \multirow{-2}{*}{{\color[HTML]{000000} Reference}} &
  \multicolumn{1}{c|} {\makecell{Effect on \\ decision-making}} &
  \multicolumn{1}{c|} {\makecell{Estimation of \\ value/trend/pattern}} &
  \multicolumn{1}{c|}{\makecell{Effect on \\ perceived confidence}} &
  \multicolumn{1}{c|}{\makecell{Effect on \\ perceived data quality}} &
  {\makecell{Interpretation of \\data with missing}} \\ \hline
1 &
 \cite{Bauerle2022} &
  \multicolumn{1}{c|}{{\color[HTML]{3F3F3F} \textbf{}}} &
  \multicolumn{1}{c|}{\textbf{•}} &
  \multicolumn{1}{c|}{\textbf{•}} &
  \multicolumn{1}{c|}{\textbf{}} &
  \textbf{} \\ \hline
2 &
  \cite{Song2021} &
  \multicolumn{1}{c|}{{\color[HTML]{3F3F3F} \textbf{•}}} &
  \multicolumn{1}{c|}{\textbf{}} &
  \multicolumn{1}{c|}{\textbf{}} &
  \multicolumn{1}{c|}{\textbf{}} &
  \textbf{} \\ \hline
3 &
  \cite{Andreasson2014} &
  \multicolumn{1}{c|}{{\color[HTML]{3F3F3F} \textbf{•}}} &
  \multicolumn{1}{c|}{\textbf{}} &
  \multicolumn{1}{c|}{\textbf{•}} &
  \multicolumn{1}{c|}{\textbf{}} &
  \textbf{} \\ \hline
4 &
  \cite{Eaton2005} &
  \multicolumn{1}{c|}{} &
  \multicolumn{1}{c|}{} &
  \multicolumn{1}{c|}{} &
  \multicolumn{1}{c|}{} &
  • \\ \hline
5 &
  \cite{Fernstad} &
  \multicolumn{1}{c|}{} &
  \multicolumn{1}{c|}{•} &
  \multicolumn{1}{c|}{} &
  \multicolumn{1}{c|}{} &
   \\ \hline
6 &
  \cite{Song2019} &
  \multicolumn{1}{c|}{} &
  \multicolumn{1}{c|}{} &
  \multicolumn{1}{c|}{} &
  \multicolumn{1}{c|}{•} &
   \\ \hline
\end{tabular}

\caption{The evaluation papers}
\label{tab: Evaluation paper}
\end{table*}
\vspace{-3mm}

\begin{table}[]
\centering 

\scalebox{0.7}{
\begin{tabular}{|l|lll|l|}
\hline
\multicolumn{1}{|c|}{\multirow{2}{*}{\textbf{\makecell{Common \\visualization \\methods}}}} & \multicolumn{3}{c|}{\textbf{Paper type}}                                                                                 & \multicolumn{1}{c|}{\multirow{2}{*}{{\textbf{ \makecell{Total \\unique \\papers}}}}} \\ \cline{2-4}
\multicolumn{1}{|c|}{}                                      & \multicolumn{1}{l|}{ {\textbf{\makecell{Techniques \\papers}}} }& \multicolumn{1}{l|}{ {\textbf{ \makecell{Applications \\and tools \\papers}}}} & { \textbf{\makecell{ Evaluation \\papers }}}& \multicolumn{1}{c|}{}                                     \\ \hline
Bar chart                                                   & \multicolumn{1}{l|}{\cite{Tierney2023}\cite{Theus1997}\cite{Alsufyani2024}}                  & \multicolumn{1}{l|}{\cite{Templ2011} \cite{Cheng}\cite{Ruddle2022}}                              &   \cite{Song2019}                &  7                                                         \\ \hline
Line graph                                                  & \multicolumn{1}{l|}{\cite{Bogl2015}}                  & \multicolumn{1}{l|}{}                              &       \cite{Andreasson2014} \cite{Eaton2005} \cite{Song2019}          & 4                                                         \\ \hline
Histogram                                                   & \multicolumn{1}{l|}{\cite{Tierney2023} \cite{Templ2012}\cite{Theus1997}}                  & \multicolumn{1}{l|}{\cite{Templ2011} \cite{Cheng}\cite{Ruddle2022} }                             &                   &             6                                              \\ \hline
Box plot                                                    & \multicolumn{1}{l|}{\cite{Templ2012}\cite{Theus1997}\cite{Alsufyani2024}}                  & \multicolumn{1}{l|}{}                              &                   &            3                                               \\ \hline
Scatter plot                                                & \multicolumn{1}{l|}{\cite{Tierney2023}\cite{Templ2012}\cite{Theus1997}}                  & \multicolumn{1}{l|}{\cite{Templ2011} \cite{Cheng}}                             &            \cite{Song2021}       &  6                                                         \\ \hline
Heatmap                                                     & \multicolumn{1}{l|}{\cite{Tierney2023}\cite{Fernstad2022}\cite{Ruddle2022}}                  & \multicolumn{1}{l|}{\cite{Yeon2020}  }                            &              &     4                                                      \\ \hline
Parallel coordinates         & \multicolumn{1}{l|}{\cite{Tierney2023}\cite{Templ2012}\cite{Fernstad2022}}                  &  \multicolumn{1}{l|}{ \cite{Lu2012} \cite{Templ2011}}             &      \cite{Fernstad} \cite{Bauerle2022}            & 7                                                          \\ \hline
Maps                                                        & \multicolumn{1}{l|}{\cite{Templ2012}}                  & \multicolumn{1}{l|}{\cite{Templ2011}}                              &                   &  2                                                         \\ \hline
Matrix plot                                                      & \multicolumn{1}{l|}{\cite{Templ2012}}                  & \multicolumn{1}{l|}{\cite{Templ2011}}                              &  \cite{Fernstad}                 &       3                                                    \\ \hline
\textbf{Total unique papers }                                        & \multicolumn{1}{l|}{7}                  & \multicolumn{1}{l|}{5}                              &   6                &                                                     \\ \hline
\end{tabular}
}
\caption{Classification of papers representing most common visualization methods based on visualization and paper type}
\label{table_Classification2}
\end{table}

\begin{table}[]
\centering
 
\scalebox{0.8}{
\begin{tabular}{|l|lll|}

\hline
\multicolumn{1}{|c|}{} & \multicolumn{3}{c|}{\textbf{Paper type}} \\ \cline{2-4} 
\multicolumn{1}{|c|}{\multirow{-2}{*}{\textbf{\makecell{ Specialised \\ visualization method}}}} & \multicolumn{1}{l|}{ \textbf{\makecell{Techniques \\papers}}} & \multicolumn{1}{l|}{ \textbf{\makecell{Applications \\and tools papers}}} &  \textbf{\makecell{Evaluation \\papers}} \\ \hline
Sankey diagram & \multicolumn{1}{l|}{} & \multicolumn{1}{l|}{\cite{Yeon2020}} &  \\ \hline
Spine plot & \multicolumn{1}{l|}{} & \multicolumn{1}{l|}{\cite{Templ2011} \cite{Cheng}} &  \\ \hline
Aggregation plot & \multicolumn{1}{l|}{\cite{Templ2012}} & \multicolumn{1}{l|}{\cite{Templ2011}} &  \\ \hline
Upset plot & \multicolumn{1}{l|}{\cite{Tierney2023}} & \multicolumn{1}{l|}{} &  \\ \hline
Chord char & \multicolumn{1}{l|}{\cite{Alemzadeh2017}} & \multicolumn{1}{l|}{} &  \\ \hline
Lasagna plot & \multicolumn{1}{l|}{\cite{Jimenez2022}} & \multicolumn{1}{l|}{} &  \\ \hline
Bean plot & \multicolumn{1}{l|}{\cite{Alemzadeh2017}} & \multicolumn{1}{l|}{} &  \\ \hline
Missingess map & \multicolumn{1}{l|}{\cite{Honaker2011} \cite{Alsufyani2024}} & \multicolumn{1}{l|}{\cite{Alemzadeh2020} \cite{Cheng}} &  \\ \hline
Spinogram & \multicolumn{1}{l|}{\cite{Templ2012}} & \multicolumn{1}{l|}{\cite{Cheng}} &  \\ \hline
Contour plots & \multicolumn{1}{l|}{} & \multicolumn{1}{l|}{\cite{Cheng}} &  \\ \hline
Pattern plot & \multicolumn{1}{l|}{\cite{Valero-Mora2019}} & \multicolumn{1}{l|}{} &  \\ \hline
Mosaic plot & \multicolumn{1}{l|}{\cite{Templ2012} \cite{Theus1997}} & \multicolumn{1}{l|}{} &  \\ \hline
Marginplot & \multicolumn{1}{l|}{\cite{Templ2012}} & \multicolumn{1}{l|}{} &  \cite{Fernstad}\\ \hline
Strip plot & \multicolumn{1}{l|}{\cite{Alemzadeh2017}} & \multicolumn{1}{l|}{} &  \\ \hline
\end{tabular}
}
\caption{Classification of papers representing specialised visualization methods based on visualization and paper type.}
\label{table_Classification}
\end{table}


\section{\textbf{Related Work}} \label{sec:RW} 
In this section, we provide overview of related STARs that review data visualization literature. The existing works generally focus on visualization for specific visualization or data types such as maps and graphs, or specific domains such as healthcare and environment. Our STAR differs from previous ones by focusing only on the visualization of missing data. Isaacs et al.\ \cite{Isaacs2014} focus solely on existing approaches and techniques in performance visualization. They created context-based classification to organize the survey (hardware, software, tasks, and applications). The scale and goal of the visualization were used in the classification as well. Vehlow et al.\ \cite{Vehlow2015} reports research in visualizing group structures as part of graph diagrams to consider only techniques that support the visualization of both the group structure and the graph topology, and use an explicit encoding of the group structure. They also included evaluations and applications of the presented techniques. Their work identified major research challenges that could guide future research. Shen et al.\ \cite{Information} reviews state-of-the-art information visualization methods and techniques in the context of farmed data in defence simulations comprehensively. It presents common visualization methods, specialized visualization techniques, and distributed visualisation. It provides guidelines for future research in the information visualization field. Additionally, it presents a concept demonstrator designed to visualize joint force simulations. One of their limitations is the reliance on farmed data within simulations for the concept demonstrator, with insufficient consideration given to distinctions between farmed and real data. Also, the proposed conceptual framework for future research directions is based on current trends and gaps in the field, indicating that it might not comprehensively cover all potential areas for development. Hogr\"{a}fer et al.\ \cite{Hografer2020} survey the literature on map-like visualization from two perspectives: imitation and schematization of cartographic maps. These perspectives are divided into four principal categories: point, line, area, and field, based on the visual element that is affected. They furthermore present literature based on tasks for map-like visualization: identify locations, retrieve values, assess distances, and trace paths. Fischer and Keim \cite{Power} provide a comprehensive overview of the latest methods developed for the interactive visualization of network data on a geographical map for power grids, along with a qualitative comparison of these methods based on various criteria, such as complexity, visualization techniques, interaction, and evaluation.
Wang and Laramee \cite{EHR} reports the state-of-the-art in interactive visualizations of electronic health records (EHR) and population health records (PopHR) in open-access healthcare data sources. The papers were classified based on six multidisciplinary research themes derived from the investigation of interactive EHR visualization. Several major research themes in processing and visualizing EHRs were used to classify the papers (Machine Learning, Natural Language Processing, Event Sequence Simplification, Geospatial Visualization, Visual Analytics with Clustering, and Visual Analytics with Comparison). Also, they adopted a medical terminology standard, the Unified Medical Language System (UMLS), to classify each paper. Firat et al.\ \cite{Firat} focus on interactive visualization literacy with a special focus on evaluation. They identified five categories of evaluation methods: in the wild, controlled user study, classroom-based evaluation, crowd-sourced evaluation, and a meta-review of related literature. Shakeel et al.\ \cite{Shakeel} reviews theoretical, analytical, and statistical models and techniques for improving the performance of visualization. They also investigated the powerful applications of data visualization in various domains. They reviewed several challenges in data visualization and provided future research directions and opportunities. Kale et al.\ \cite{Kale2023} presents the state of the art in visualizing dynamic graphs, focusing only on animated diagrams and static charts based on a timeline. They applied tagging to structure the publications and developed a novel hierarchical taxonomy of dynamic graph visualization techniques. They also illustrated the evaluation methods and the most mature application in dynamic visualization. Patel \cite{Patel2023} focused on the techniques and tools used for the visualization of environmental data only. They surveyed the tools’ features, applicability, strengths and limitations. They also examined several directions of future research in data visualization. 

\section{\textbf{Missing data visualization techniques}}
\label{sec:VIsTech} This section describes 10 focus papers on Missing data visualization techniques found in our literature search. The first part focuses on publications presenting novel visualization techniques for the visualization of missing data. The second part focuses on modifications of existing visualization techniques to support the representation of missing data.

 \subsection{\textbf{Novel visualization techniques:} }
\begin{figure}[h]
       \centering
       	\includegraphics[width=5 cm]{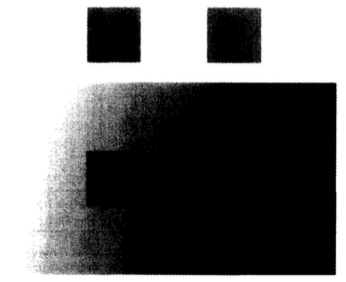}
       \caption{Two grey squares of equal luminance are placed on a background of varying luminance, the square on the lighter background appears darker and the square on the darker background appears lighter (Figure from \cite{Twiddy1994}).}
       	\label{Fig:Restorer}
     
 \end{figure}

 Twiddy et al. \cite{Twiddy1994} were one of the first to address missing values visualization issues. Their approach was inspired by a technique used in the restoration of paintings, where missing parts are replaced and missing areas are filled in. They used colour techniques to visually distinguish between recorded values and missing values.  It combines pseudocolouring and gray scale in the same colour map to fill in the gaps caused by missing data. Recorded data is mapped to colours in the colour map, while missing data in this technique were mapped to the grey scale portion of the colour map. The approach uses the luminance values of the colours surrounding the missing data regions to interpolate over these regions (Figure \ref{Fig:Restorer}). It visually blends the restored areas with the original data while still remaining distinguishable with closer inspection. It eliminates distracting gaps caused by missing data and shows a clear distinction between recorded data and missing data. This technique has been applied on different data sets, including three-dimensional data and it was completely successful in filling areas where black was surrounded by a single hue, and less successful in areas where black overlapped two or four hues of contrasting luminance. In addition to Laplacian Fill and Rank Fill algorithms, a Modified Rank Fill algorithm has been developed to handle cases where the missing data regions are too large for the Rank Fill Algorithm. This algorithm combines the output from the Rank Fill Algorithm with the Laplacian Fill Algorithm to fill the remaining areas of the missing data regions.
 
\begin{figure}[h]%
       \centering
       \subfloat[]
       [Main variable dashboard in Amelia View.]{
       	\includegraphics[width=3.45 cm]{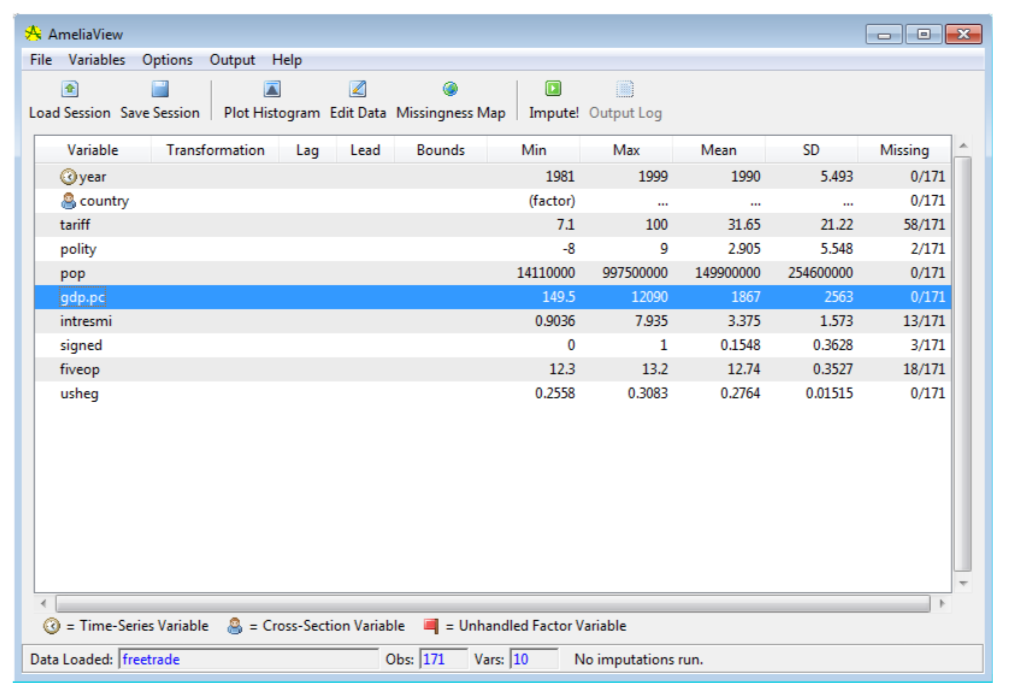}
       	\label{Fig:Amelia1}
       }%
       \qquad
       \subfloat[][Variable options via dashboard right-click menu.]{
       	\includegraphics[width=3.15 cm]{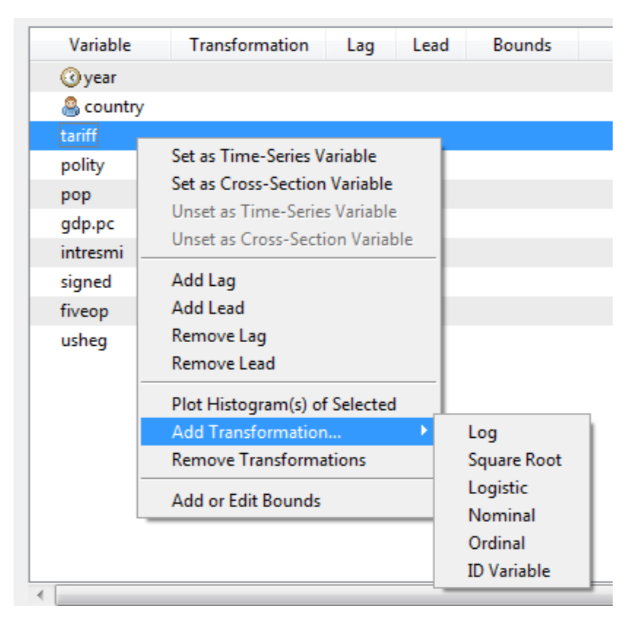}
       	\label{Fig:Amelia2}
       }%
       \qquad
       \subfloat[][Output log showing Amelia output for a successful imputation.]{
       \centering
       	\includegraphics[width=3.4 cm]{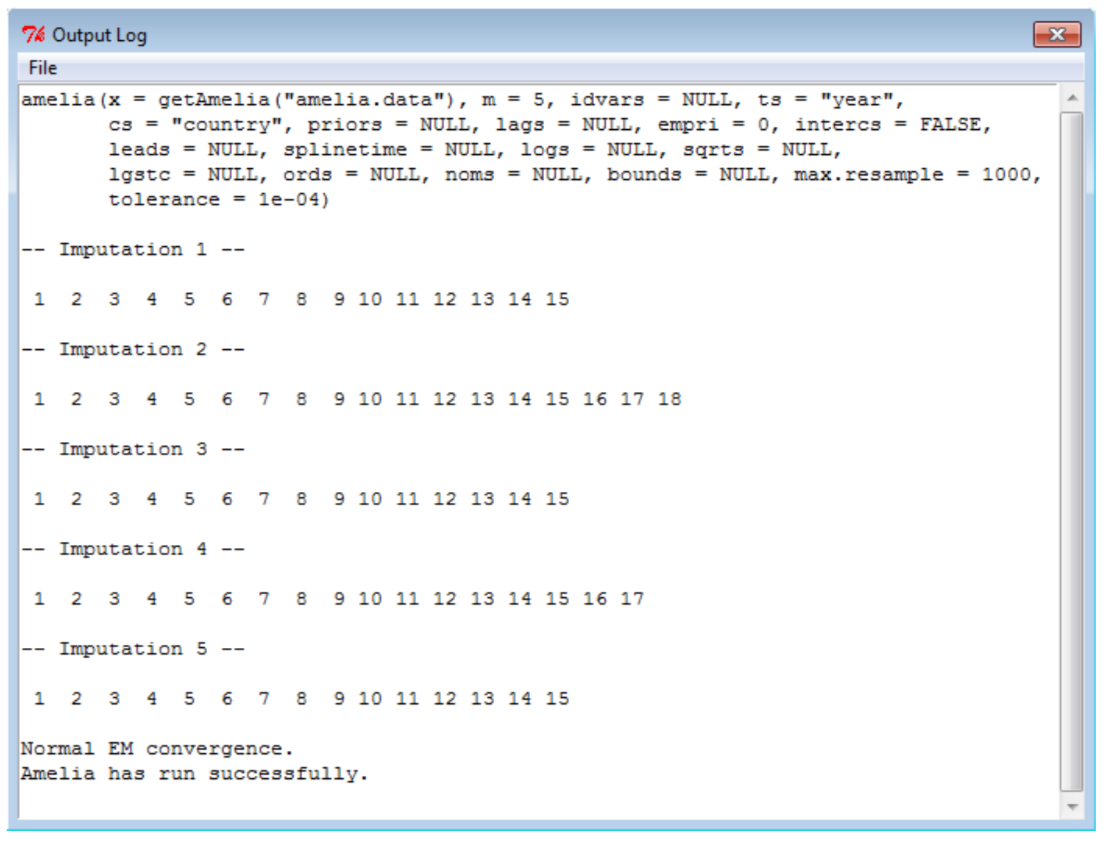}
      \label{Fig:Amelia3}
       	}%
       \qquad
       \subfloat[][Detail for Diagnostics dialog.]{
       	\includegraphics[width=3.7 cm]{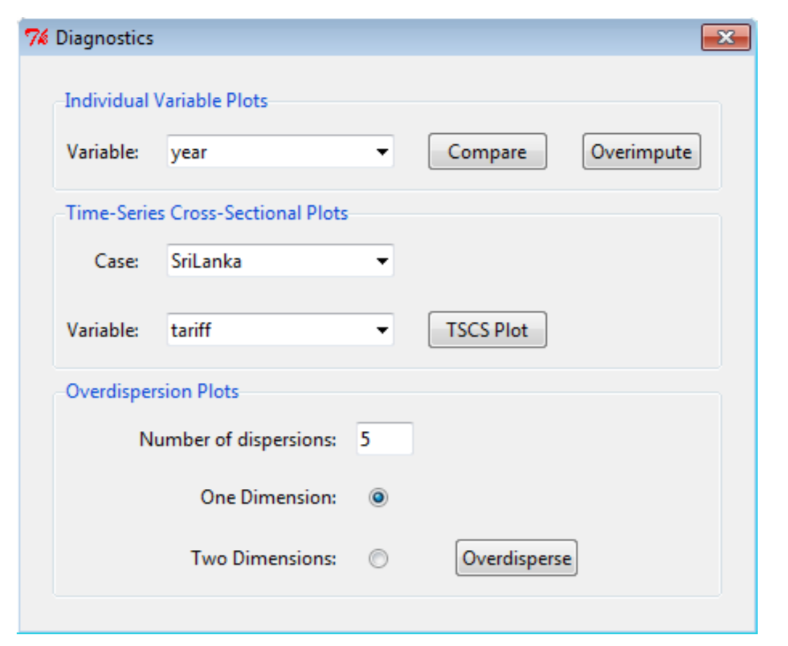}
       	\label{Fig:Amelia4}
       }
       \caption{The user interface of Amelia View \cite{Honaker2011}.}%
       \vspace{-2mm}
       \label{Fig:AmeliaUI}%
\end{figure}

\begin{figure}[h]
       
       \centering
       	\includegraphics[width=5 cm]{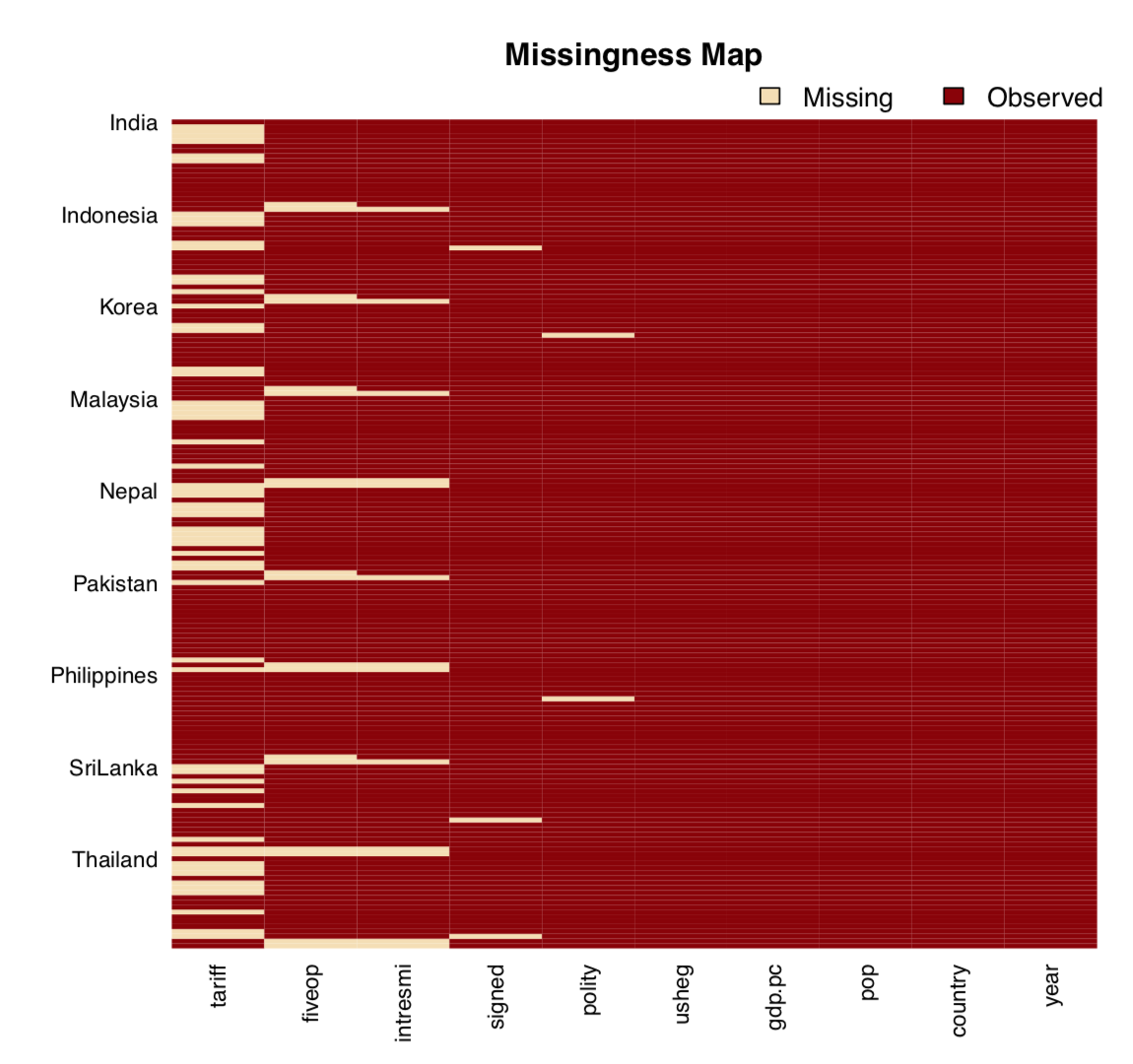}
       \caption{Missingness map in Amelia \cite{Honaker2011}.}
       	\label{Fig:MissingnessAmilia}

       
       
      
 \end{figure}

 Honaker et al.\ \cite{Honaker2011} provides a full R package (Amelia II) for multiple imputations of missing values with a powerful and user-friendly approach to handling missing data, with a focus on speed, accuracy, and ease of use. It provides useful user guidelines on how to use this package, as shown in Figures \ref{Fig:Amelia1}, \ref{Fig:Amelia2}, and \ref{Fig:Amelia3}. The missingness map function, Figure \ref{Fig:MissingnessAmilia}, visualizes missing data patterns easily and clearly which could help to improve the imputation model and the collection process. The paper uses a novel bootstrapping algorithm which is faster, works with larger numbers of variables, and is easier to use compared to other approaches, providing essentially the same results. A full set of graphical diagnostics is provided to check the validity of the imputation model (Figure \ref{Fig:Amelia4}). This package creates multiple “filled in” versions of incomplete datasets, allowing for the appropriate use of all available information and reducing bias compared to ad-hoc methods of imputation.  

\begin{figure}[h]
       \centering
       	\includegraphics[width=8 cm]{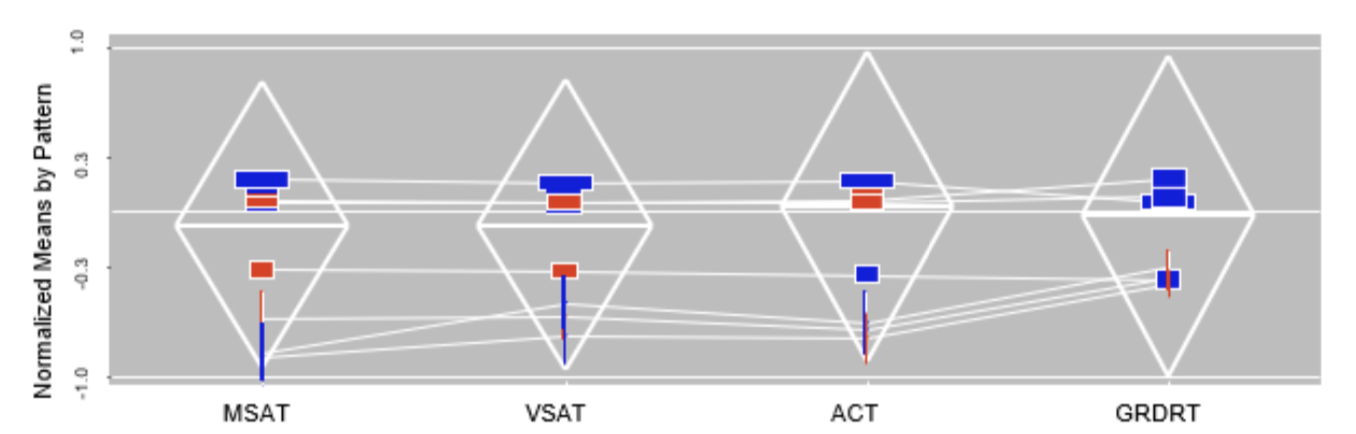}
       
       \caption{Plot of the missing-values patterns in college data \cite{Valero-Mora2019}.}
     	\label{Fig:PedroValeroMora}
 \end{figure}

Valero-Mora et al. \cite{Valero-Mora2019} emphasize the importance of understanding the quality of available data and the consequences of its deficiencies and highlight the lack of specialized tools for this purpose. They propose a plot that provides a visual representation of missing data patterns and allows for interactive exploration. The plot includes relevant information for visualizing these patterns and permits the highlighting of elements that are more relevant according to certain criteria which can aid in exploratory data analysis. They also provide an example using college data to illustrate the capabilities of the plot. Figure \ref{Fig:PedroValeroMora} shows their plot, where each pattern corresponds to one rectangle for each variable, blue colour for observed values and red colour for missing values. These rectangles are centered vertically on the average of the variable. The horizontal size of the rectangles represents the number of cases in the pattern. They state some limitations in this plot and suggest using dynamic interactive features to effectively declutter visualizations or highlight noteworthy patterns. 

\begin{figure}[h]
       \centering
       	\includegraphics[width=6 cm]{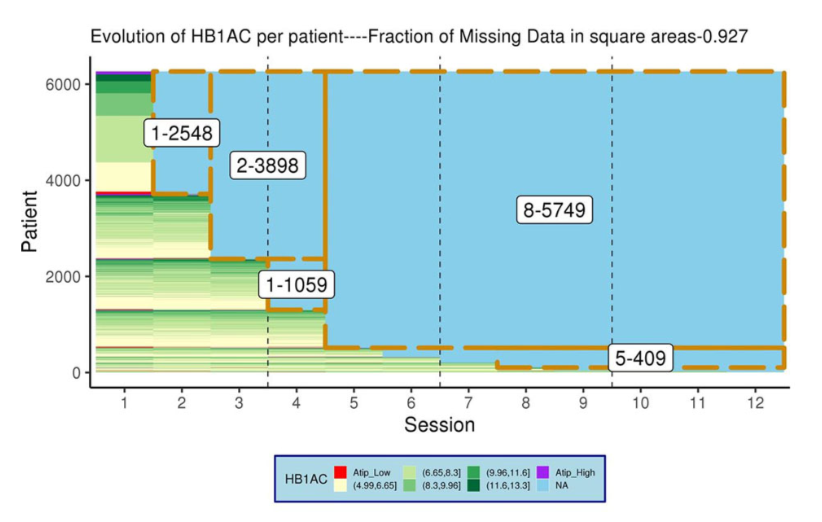}
       \caption{Diabetes data set with a lasagna plot \cite{Jimenez2022}.}
       	\label{Fig:Graphical}
 \end{figure}

 Jim\'{e}nez and Mac\'{\i}as \cite{Jimenez2022} propose algorithms for creating heat maps, specifically lasagna plots \ref{Fig:Graphical}, that integrate grouping, ordering, and sampling to visualize missing data in longitudinal studies. The ordering algorithm aids in the construction of a graphical overview of data using proportions, allowing for the visualization of monotone missingness patterns. The sampling technique involves using proportions through sampling to provide sufficient data for visualization in a matrix plot. Lastly, the grouping algorithm is based on descriptors of missing data, allowing for the identification of intermittent missingness patterns using known algorithms such as k-means.
 These algorithms focus on identifying the presence of monotone missingness patterns and localizing intermittent patterns in large data sets. The aim is to provide quick decision tools for ensuring data quality and identifying useful patterns for analysis in the exploratory stages. The techniques are applied to real-world data sets from different domains to demonstrate their effectiveness in identifying and visualizing missing data patterns.
 The proposed algorithms are evaluated using four real-world data sets. 

\begin{figure}[h]%
       \centering
       \subfloat[][The amount missing in each variable.]{
       	\includegraphics[width=3.2cm]{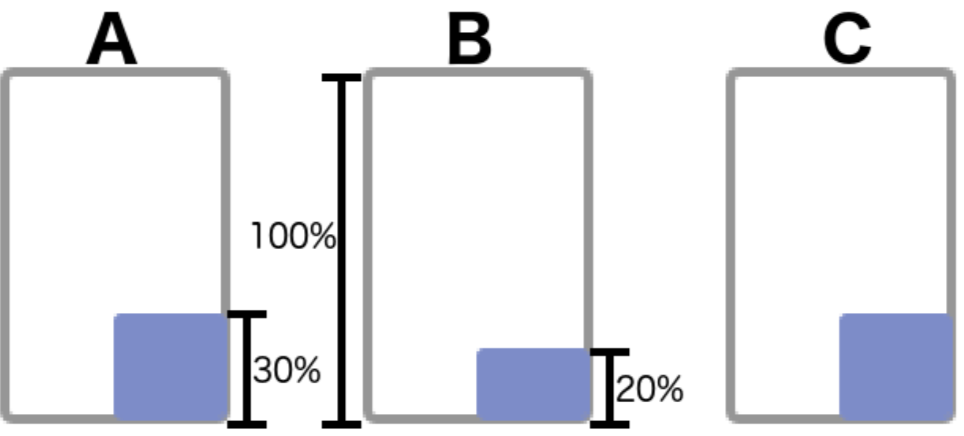}
       	\label{Fig:MissVisA}
       }%
       \qquad
       \subfloat[][Recorded items as histograms, with D being categorical.]{
       	\includegraphics[width=3.2cm]{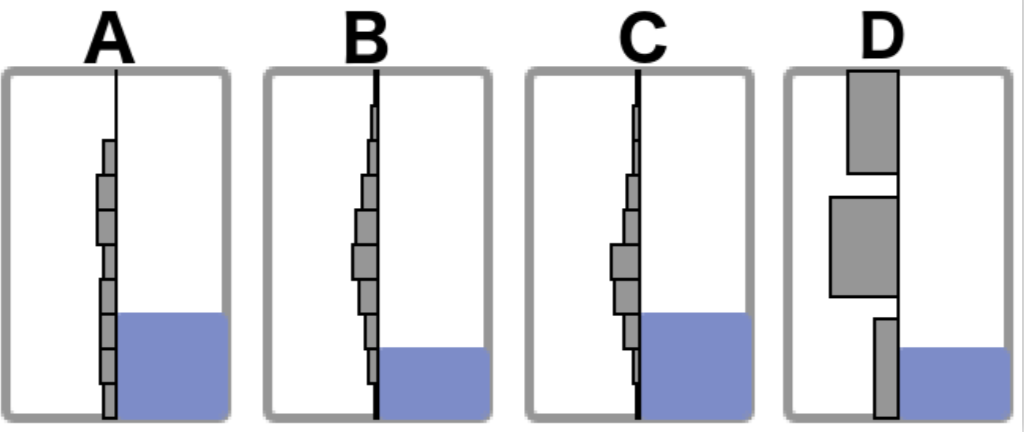}
       \label{Fig:MissVisB}
       }%
       \qquad
       \subfloat[][The joint missingness with a selected variable (C) is represented as a red block.]{
       	\includegraphics[width=3.2cm]{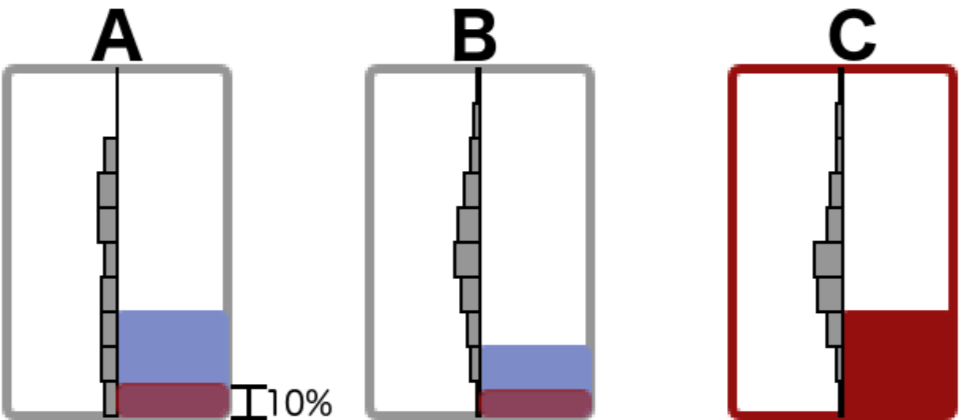}
       	\label{Fig:MissVisC}
       	}%
       \qquad
       \subfloat[][Red histograms represent distribution items that are missing in the selected variable (C).]{
       	\includegraphics[width=3.2cm]{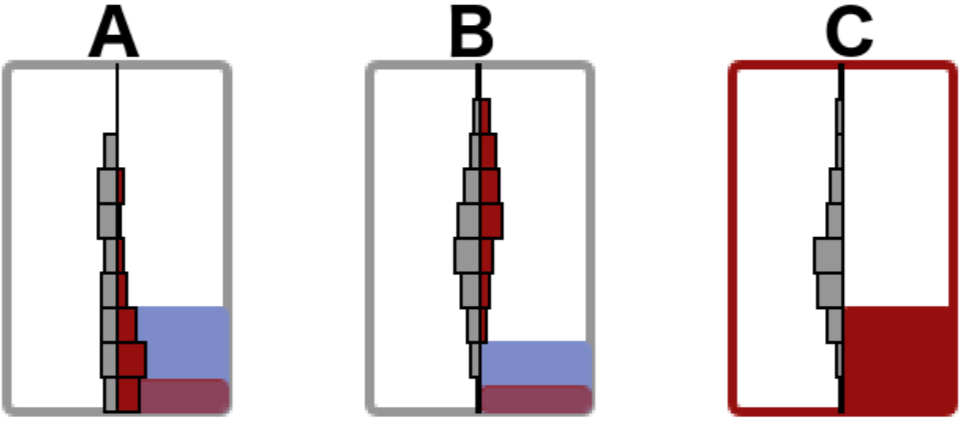}
       	\label{Fig:MissVisD}
       }
       \caption{The basic structure of MissiG \cite{Fernstad2022} for three or four variables. Variable C is selected in \ref{Fig:MissVisC} and \ref{Fig:MissVisD}}%
       \vspace{-2mm}
       \label{Fig:MissiG}%
\end{figure}

 Johansson Fernstad and Johansson Westberg \cite{Fernstad2022} presented MissiG, a glyph-based visualization technique designed to support exploration of missingness structures in N-dimensional data. MissiG (Figure \ref{Fig:MissiG}) utilise a combination of histograms and bar charts to represent the amount missing in variables, the joint missingness across variables, the distribution of recorded values and the relationship between missing and recorded across variables, using the missingness patterns defined by Johansson Fernstad \cite{Fernstad}. The glyph can be integrated with other visualization methods, and was designed to cater for both numerical and categorical data. The authors suggested the use of selection+highlighting to represent missingness structures across multiple variables.

       \begin{figure}[h]
    \centering
    \includegraphics[width=8 cm]{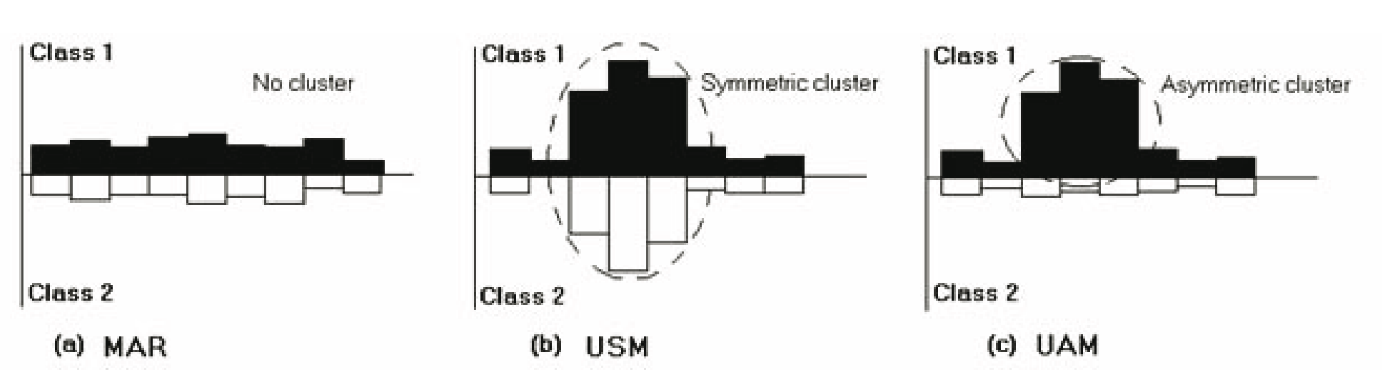}
    \caption{The Critical Patterns of Missing Values in Classification Data \cite{Wang2007}.}
    \label{fig:SOM}
\end{figure}
\begin{figure}[H]%
       \centering
       \subfloat[][Graphical summaries of missingness in the air-quality data.]{
       	\includegraphics[width=7 cm]{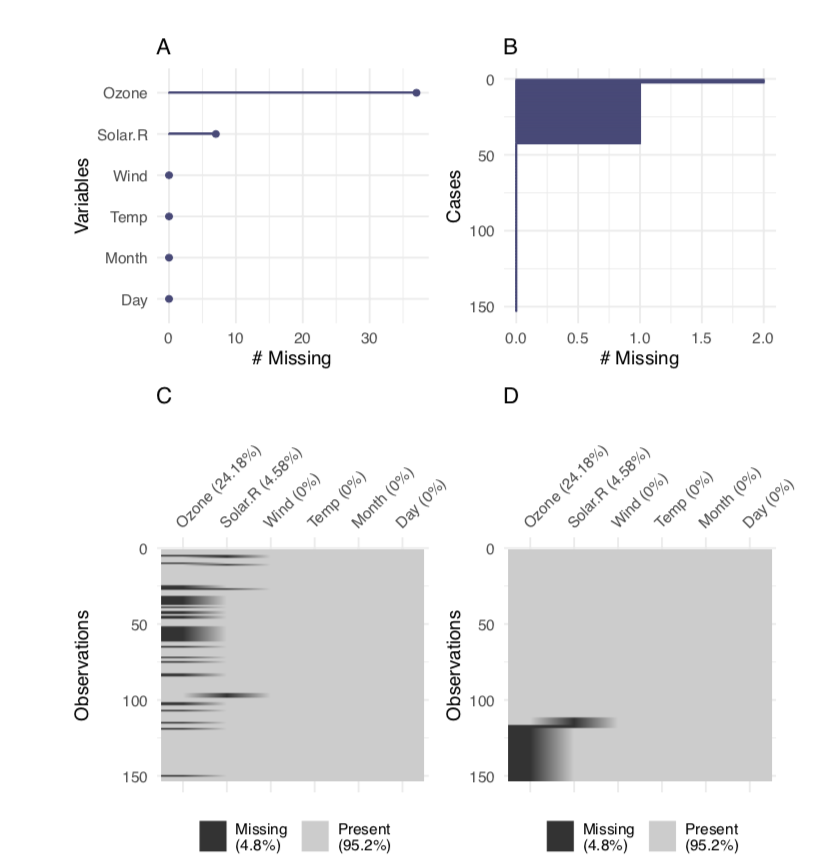}
       	\label{Fig:naniar1}
       }%
       \qquad
       \subfloat[][The pattern of missingness in the airquality dataset shown in an upset plot. There are 2 cases where both Solar.R and Ozone have missing values.]{
       	\includegraphics[width=7 cm]{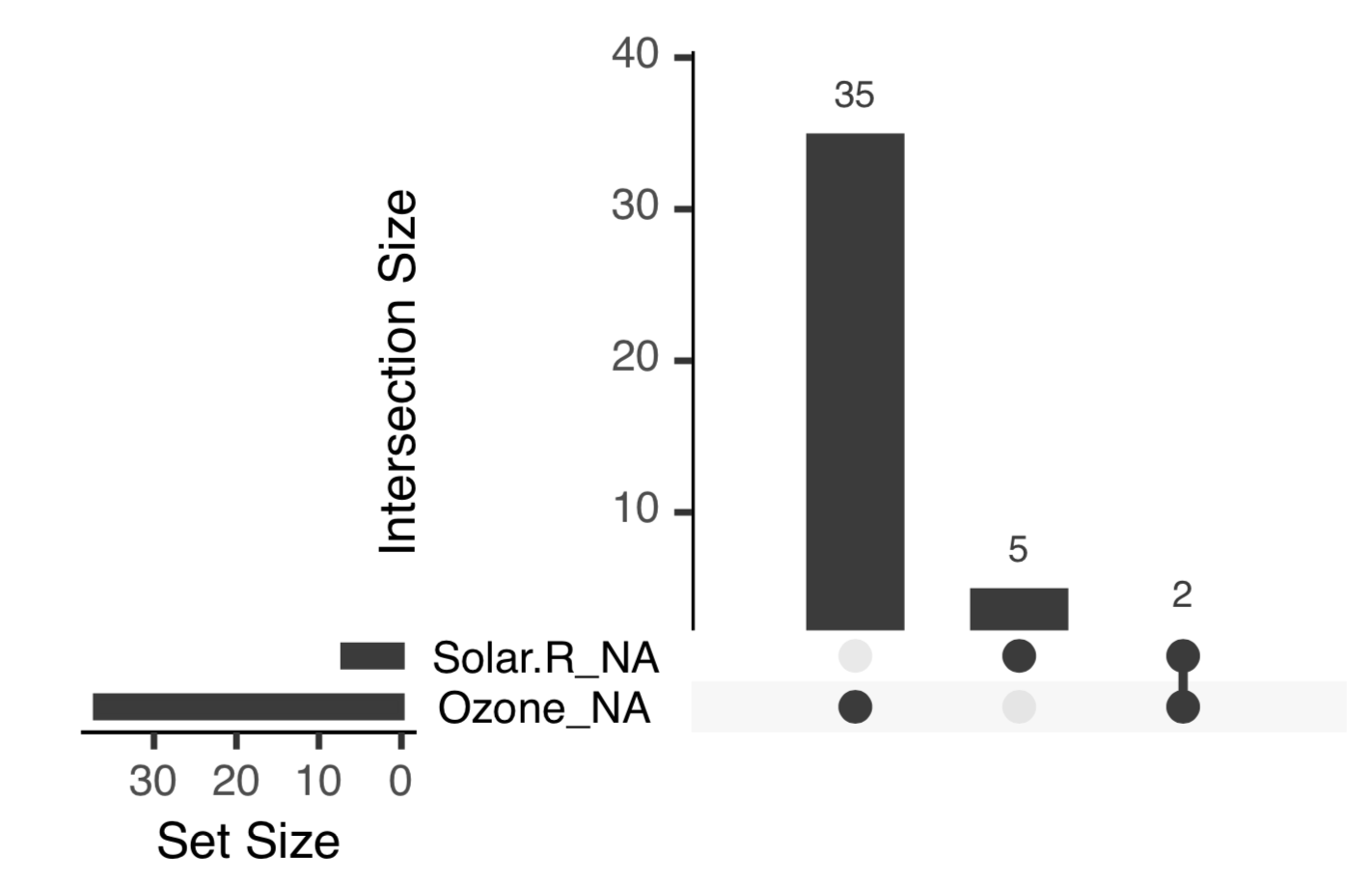}
       	\label{Fig:naniar2}
       }%
       \qquad
       \subfloat[][Parallel coordinate plot shows missing values imputed 10 \% below range for the oceanbuoys dataset.]{
       	\includegraphics[width=7 cm]{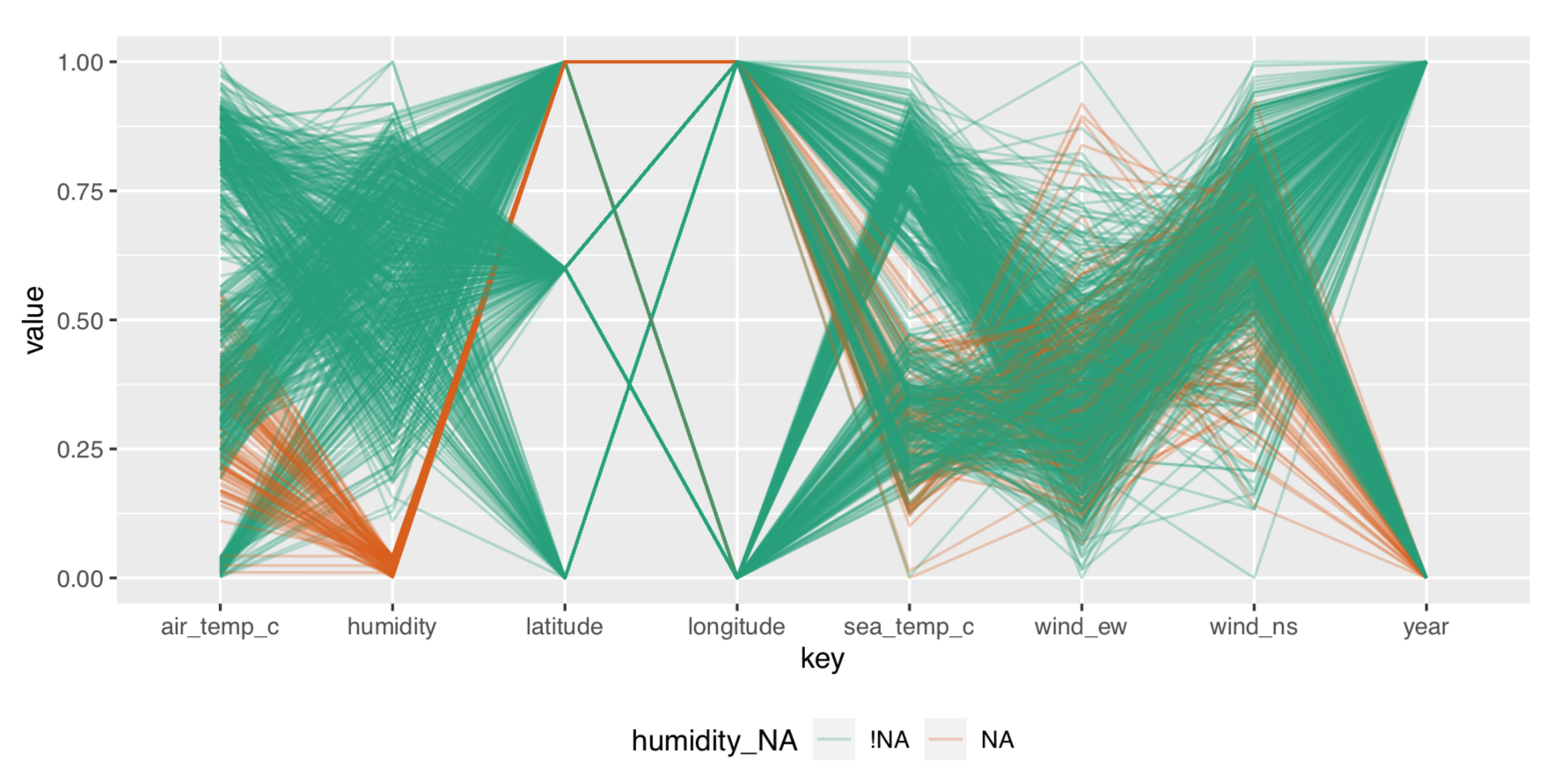}
       	\label{Fig:naniar3}
       }
       \caption{Examples of visualization available in naniar \cite{Tierney2023}.}%
       \vspace{-2mm}
       \label{Fig:naniar}%
\end{figure}

 Tierney and Cook \cite{Tierney2023} present a new methodology for handling missing values based on tidy data principles to consider handling missing data as a key part of the data analysis process. It builds upon tidy data principles and introduces a new data structure and a suite of new operations to facilitate the handling, exploration, and imputation of missing values.
 The methodology addresses the limitations of existing tools for handling missing data and aims to provide consistent interfaces and outputs for better integration with existing imputation methodology, visualization, and modelling. The new data structure proposed is called "nabular data" and is designed to facilitate the exploration of missing data based on the concept of a "shadow matrix" from past research, with four additional features to aid in the analysis of missing data. The methodology also emphasizes the use of tidy data tools for exploring missing data, translating existing methods from the missing data graphics literature into tidy data and tidy tools to enable more effective data visualizations. These methods are available in the R package naniar (Figure \ref{Fig:naniar}).

 \subsection{\textbf{Modified visualization techniques:}}
Wang and Wang \cite{Wang2007} focus on determining if the missing values were distributed randomly, unevenly, or biased towards a certain class. They emphasised the significance of the dataset's missing value distribution, concentrating on whether or not the values are distributed randomly or unevenly across variables and classes. They proposed three missingness patterns in a classification data set: 1) Missing At Random (MAR), when values are randomly distributed in the sample space (See (a) in Figure \ref{fig:SOM}) ; 2) Uneven Symmetric Missing (USM), when values are missing more often in some variables and missing values in these variables may be correlated (See (b) in Figure \ref{fig:SOM}); and 3) Uneven Asymmetric Missing (UAM), when values are missing unevenly in the data and may be biased towards a particular class (See (c) in Figure \ref{fig:SOM}). They suggested a self-organizing map (SOM) based cluster analysis approach to provide visual presentations for missing data patterns.

\begin{figure}[H]
       \centering
       	\includegraphics[width=3.5 cm]{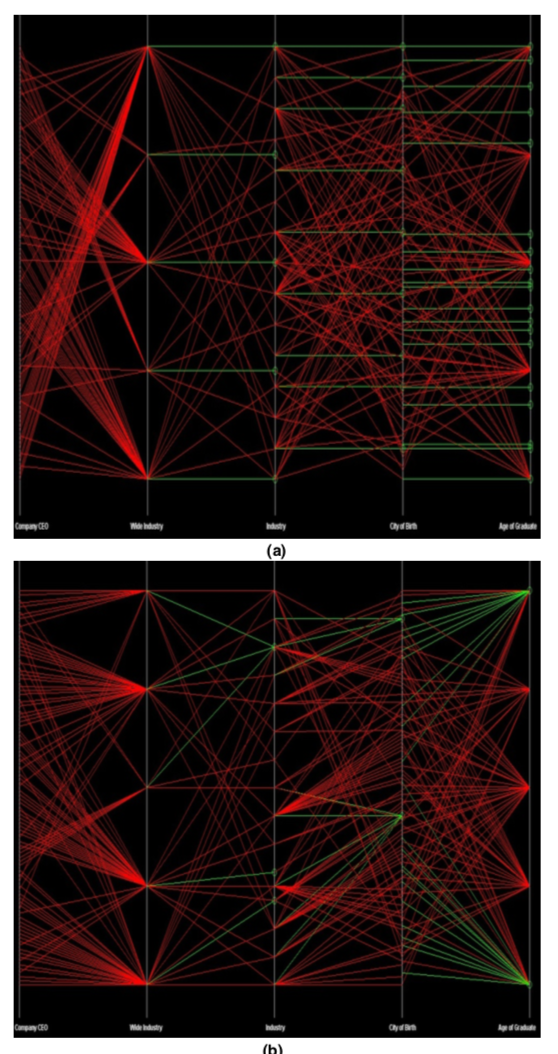}
       \caption{A dataset with 5 variables visualized in parallel coordinate visualization (The lines in red and green behave realistic and uncertain data respectively.): (a) Original plot; (b) after clutter reduction \cite{Lu2012}.}
       	\label{Fig:CltterReduction}
\end{figure}

Lu et al.\ \cite{Lu2012} leverage Sugiyama’s layered directed graph drawing method into parallel coordinates for uncertainty visualization to reduce clutter (edge crossings) and improve readability, particularly in the case of incomplete multi-dimensional data. Figure (a) in  \ref{Fig:CltterReduction} shows 5, 10, and 20 uncertain values in the last three dimensions, and Figure (b) 
shows that in total 35 dummy vertices on axes could be reduced to 7. This proposed method for clutter reduction through the reduction of edge (polyline) crossings has been demonstrated to improve the visual quality of both complete and incomplete data items in visualization. It involves the use of dummy vertices and clustered polylines to reduce the noise of incomplete data items with missing values, leading to a significant reduction in visual clutter. They also conclude that the number of missing values in the data has an impact on clutter reduction, with a higher number of missing values leading to decreased clutter reduction. Moreover, clutter reduction in multi-dimensional data visualizations is more effective for data with non-numerical values as they offer more opportunities for re-ordering and a higher potential for clutter reduction compared to data with numerical values.

\begin{figure}[H]
       \centering
       	\includegraphics[width=8 cm]{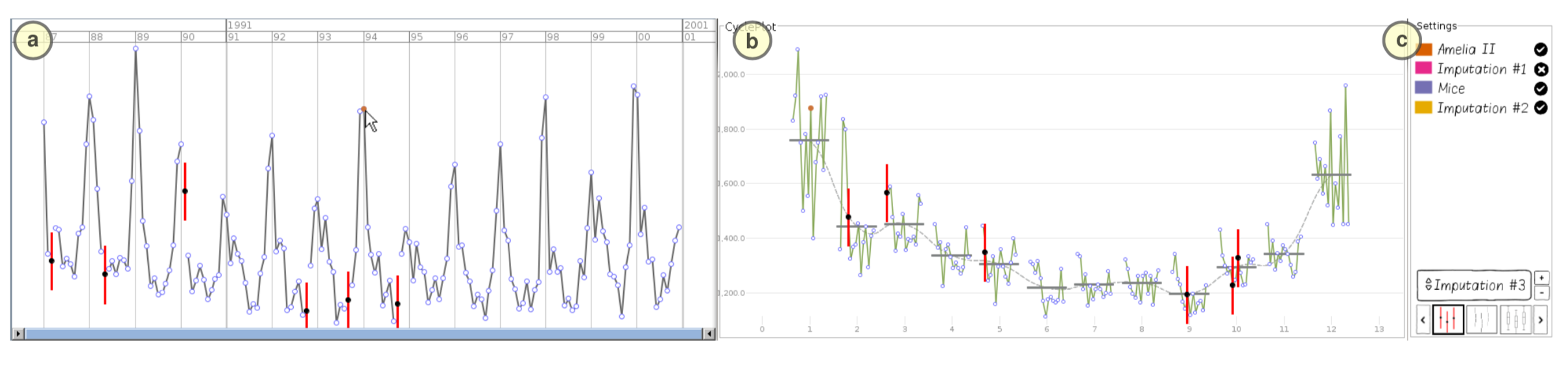}
       \caption{The estimated values (black dots) of missing values and boundaries (red bars) are displayed \cite{Bogl2015}.}
       	\label{Fig:TimeSeries}
\end{figure}

In Bogl et al.\ \cite{Bogl2015}, missing values are estimated using initial methods implemented in the statistical environment R. The estimated values from different imputation methods are combined and represented as black dots, with error boundaries or confidence intervals represented by red vertical bars (Figure \ref{Fig:TimeSeries}. This allows for the quantification and communication of the uncertainty of the imputation methods. 

\begin{figure}[H]
       \centering
     \includegraphics[width=6 cm]{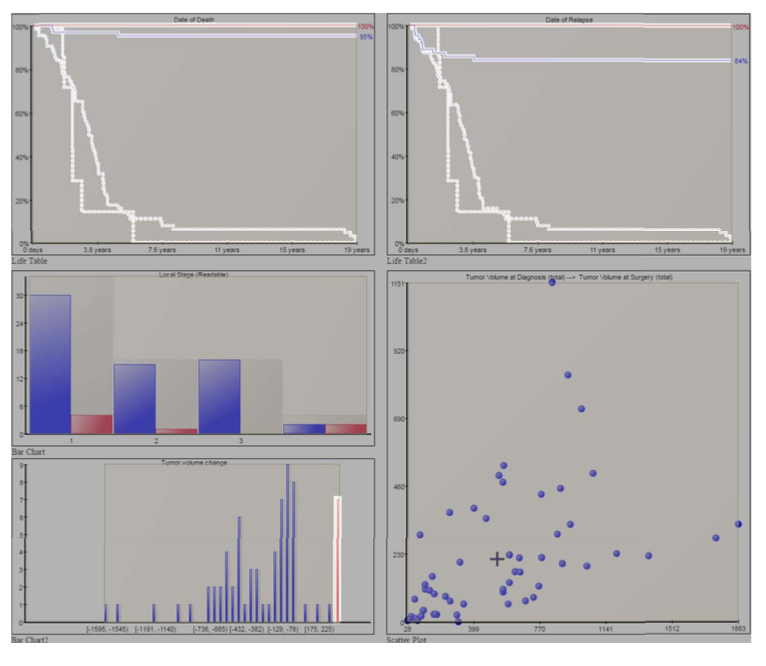}
       \caption{The bottom bar chart has a “No Value” category, the rightmost bar, coloured red for the missing values \cite{Sjobergh2017}.}
       	\label{Fig:vis1}
\end{figure}

\begin{figure}[]%
       \centering
       \subfloat[][Missing values in the heat map are coloured in black.]{
       	\includegraphics[width=6 cm]{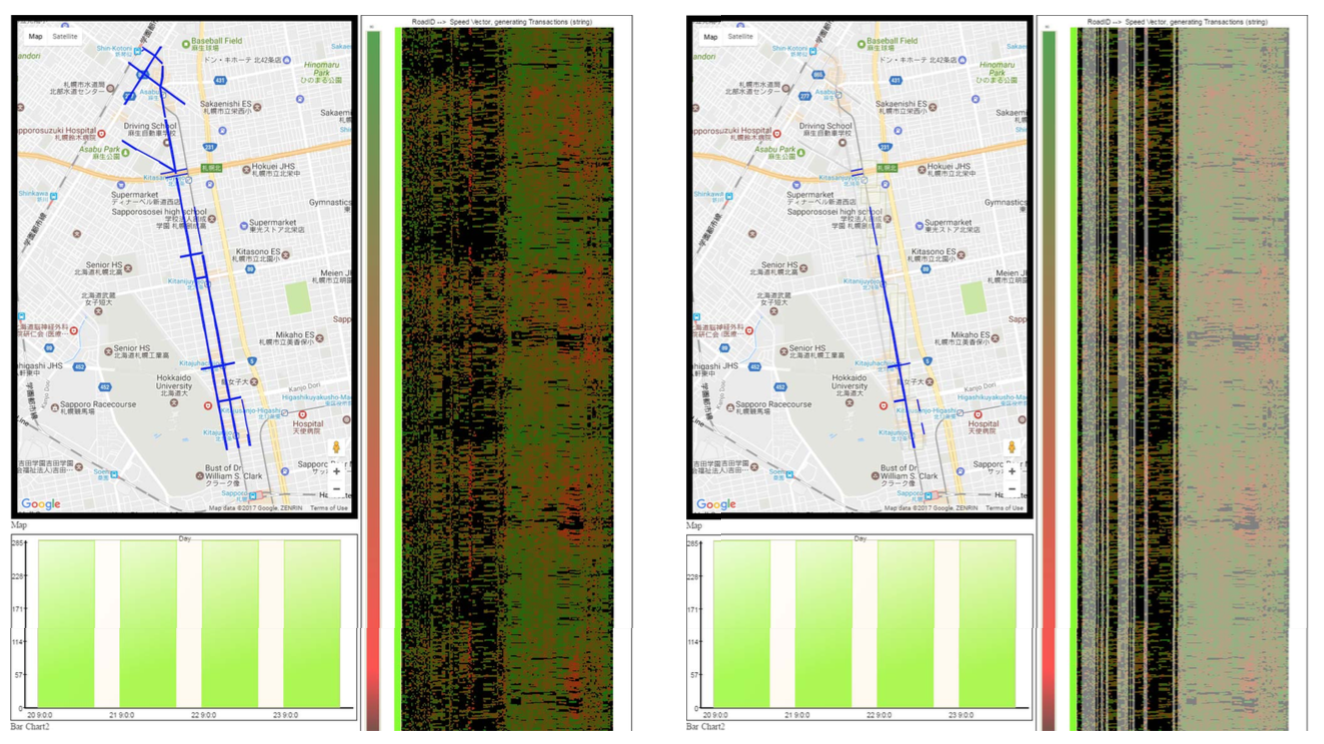}
       	\label{Fig:Vis2}
       }%
       \qquad
        \subfloat[][Imputed missing values for the scatter plot are colored in red, and recorded are blue. The heat maps have missing data and rows corresponding to missing data are left blank and grouped together.]{
       	\includegraphics[width=6 cm]{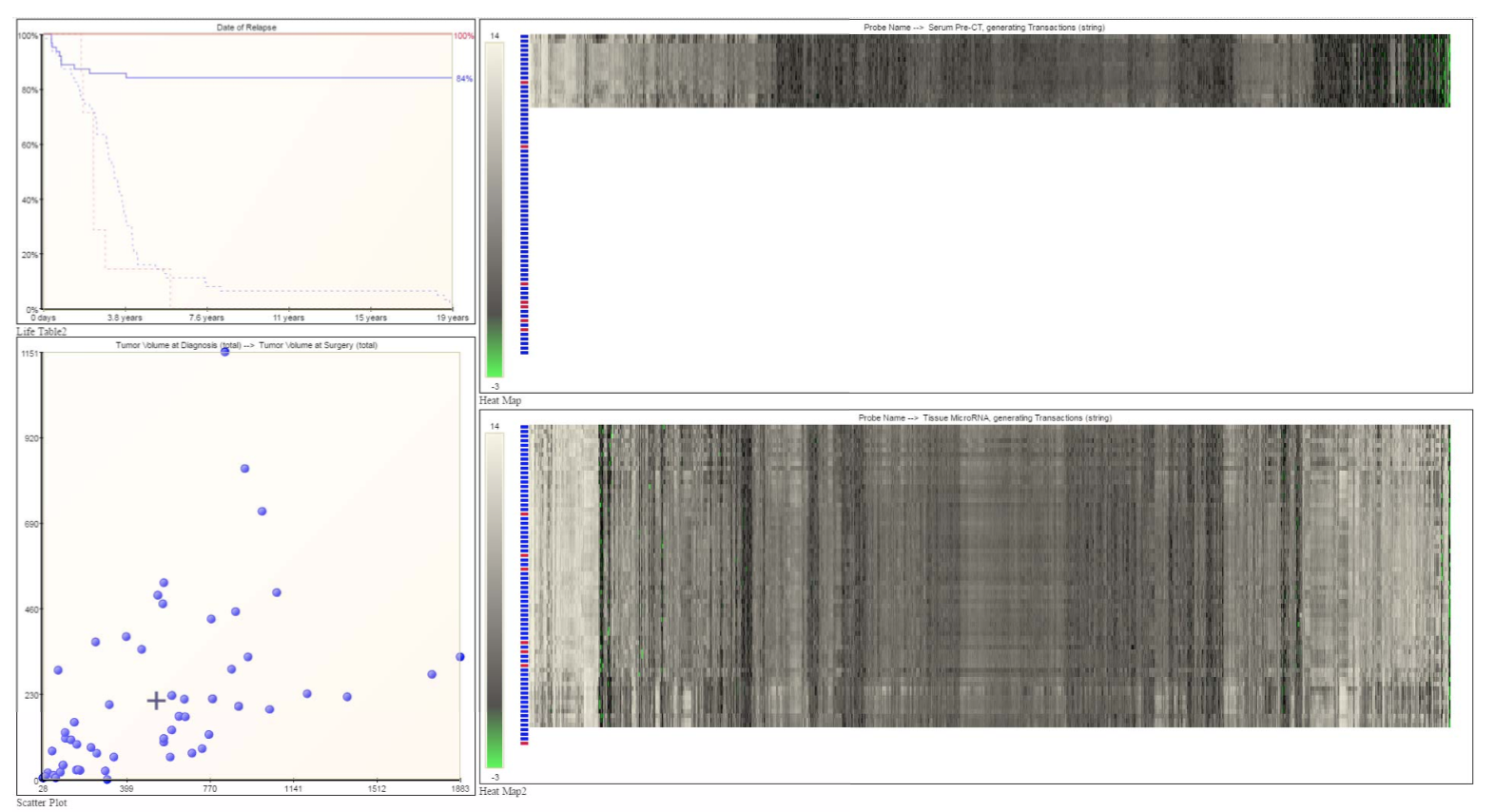}
       	\label{Fig:Vis3}
       }%
       \qquad
              \subfloat[][A chemical plant with various sensors. The time series plots in the middle. On the right the actual readings (in red) and predicted values (in blue) from a sensor location with a lower sample frequency, i.e. many times with missing values, are shown.]{
       	\includegraphics[width=6 cm]{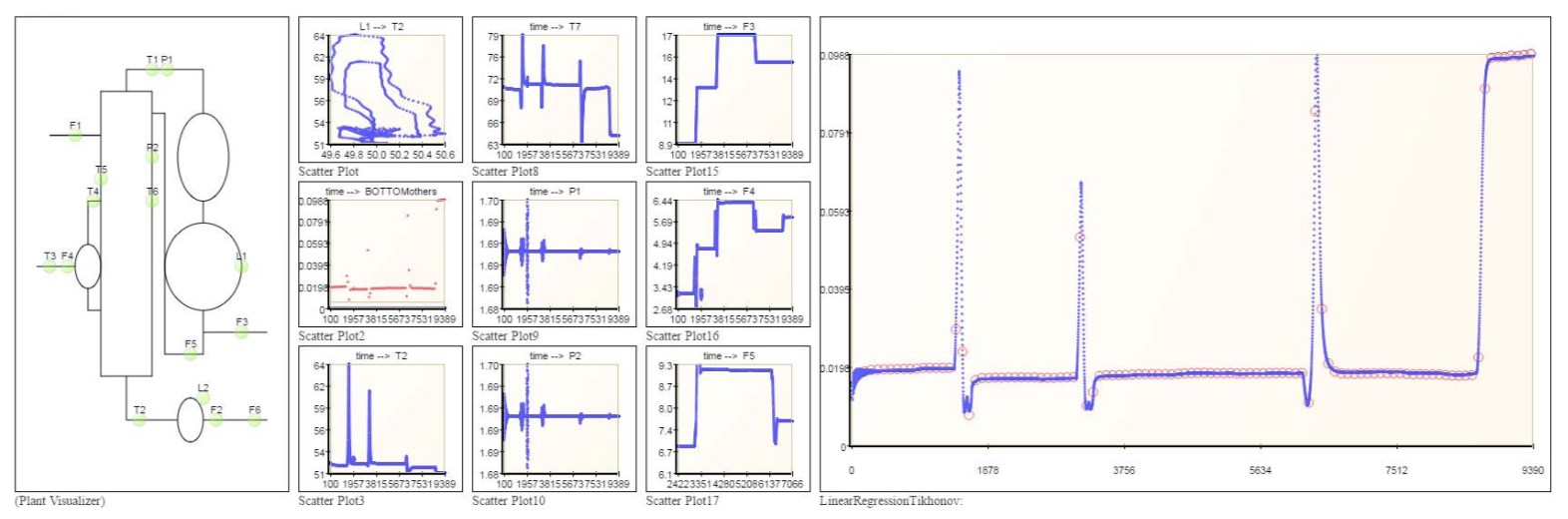}
       	\label{Fig:Vis4}
       }%
       \qquad
       \subfloat[][Missing values in  parallel coordinate visualization are drawn outside the coordinate axes.]{
       	\includegraphics[width=6 cm]{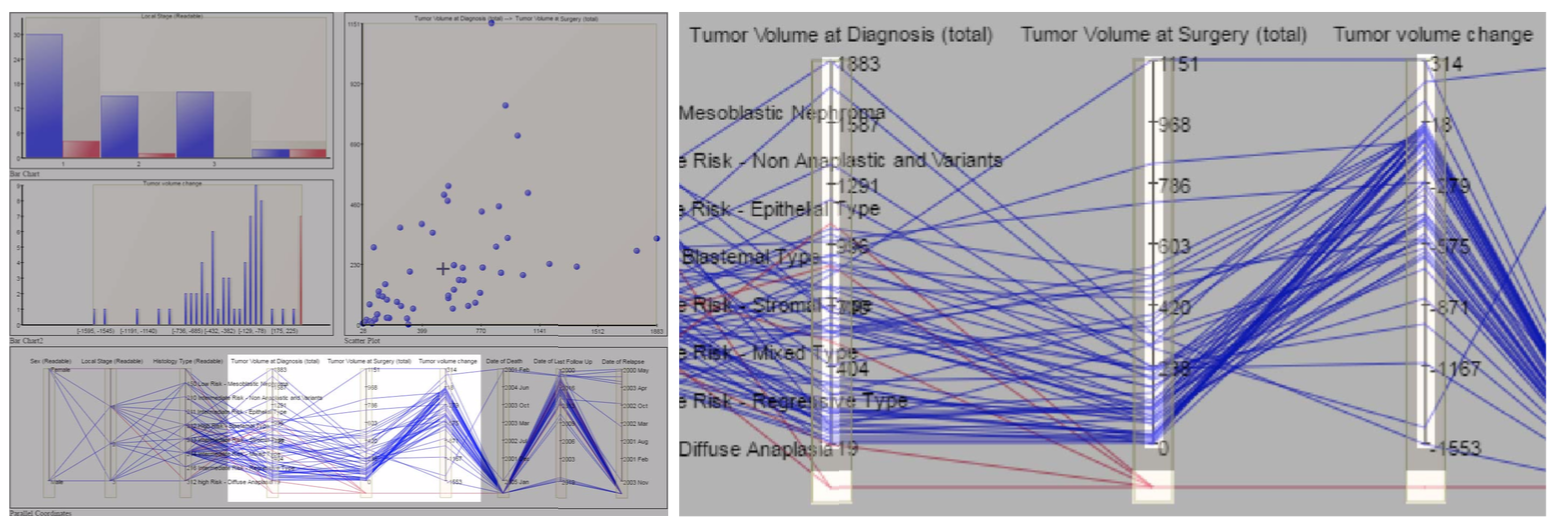}
       	\label{Fig:vis5}
       }
       \caption{Missing data visualization suggested by Sj\"{o}bergh and Tanaka \cite{Sjobergh2017}}%
       \vspace{-2mm}
       \label{Fig:vis}%
\end{figure}

Sj\"{o}bergh and Tanaka \cite{Sjobergh2017} discuss the issue of missing values in data sets, particularly in the context of sensor data in various applications such as chemical plants and medical research. It highlights the various reasons for missing data, including sampling rate differences, data transmission failures, and incomplete sensor coverage. They also mention the challenges of visualizing data with missing values, suggest methods for handling them, and illustrate some examples of different ways to visualize missing values, such as the missing value category in the bar chart (Figure \ref{Fig:vis1}), special value in the heat map and special location in the parallel coordinates (Figure \ref{Fig:vis}), and imputed value in red colour based on linear regression model). Their provided system is based on the coordinated multiple views that help handle the misleading issues of missing values.

\begin{figure}
     \centering
     \subfloat[][The MissVisG]{
     \includegraphics[width=0.5\linewidth]{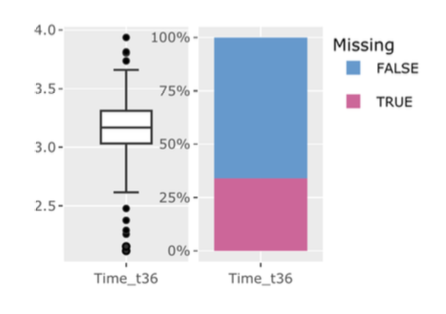}
     \label{Fig:MissVisG}
       }%
       \qquad
              \subfloat[][MissVis plot: A) The heat map. B) The box plot. C) The stacked bar chart.]{
              \includegraphics[width=0.8\linewidth]{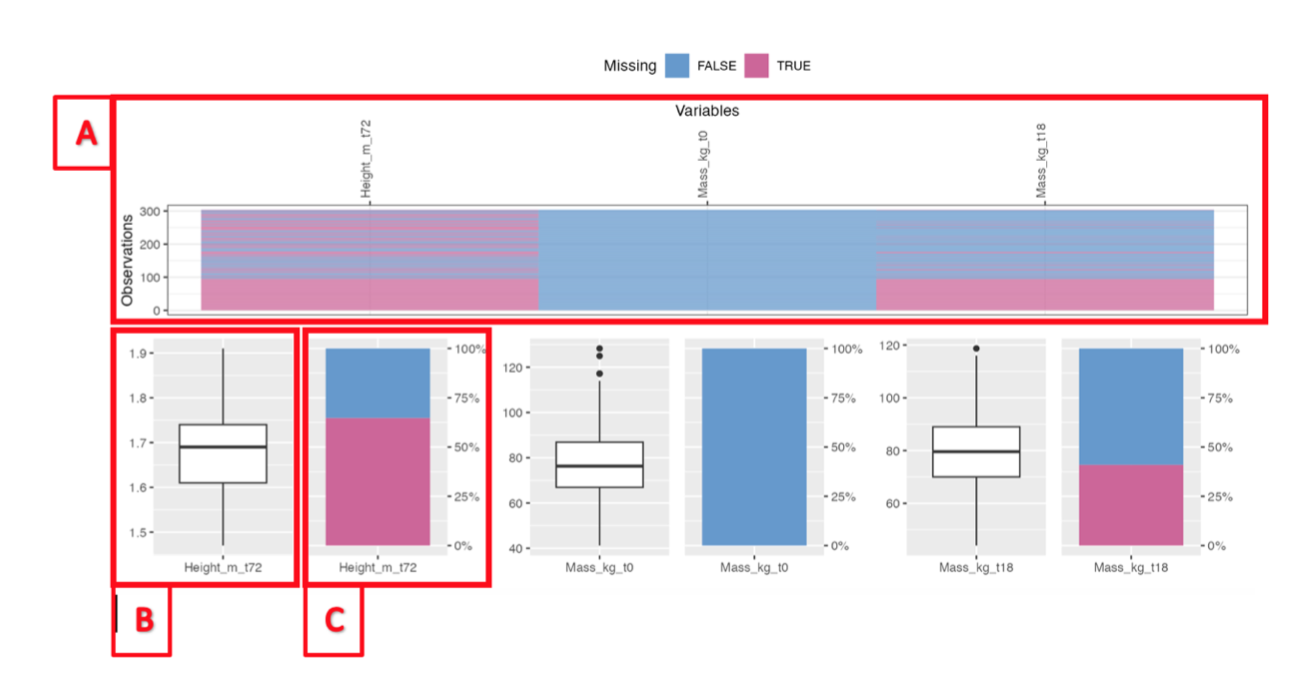}
     \label{Fig:MissVisPlot}
       }
     \caption{The MissVisG and MissVis plot.}
     \label{fig:MissVis}
 \end{figure}

 Alsufyani et al. \ \cite{Alsufyani2024} introduced two visualization techniques: the MissVisG glyph style \ref{Fig:MissVisG} and the MissVis plot \ref{Fig:MissVisPlot}. The visualization combines features such as stacked bar charts, box plots, and heatmaps, allowing for detailed examination of missingness patterns, relationships between missing and recorded values, and identification of outliers. The MissVisG glyph style represents a variable in the dataset using a rectangular shape that integrates stacked bar chart and box plot to display the percentage of missing values (in red), recorded values (in blue), the distribution of the variable which is important for choosing an appropriate imputation method, and outliers. The MissVisG glyph can function independently or be integrated with other visualization methods to enhance the understanding of the missingness. The MissVis plot is a visualization technique introduced in the paper to explore missing data patterns across multiple variables in a dataset. It builds on the MissVisG glyph to provide a more comprehensive view of missingness by combining several visual elements that support comparison and analysis. A heat map is used to display whether a data point is missing (in red) or recorded (in blue),which allows users to quickly identify missing values across multiple variables and examine patterns of missingness within and between variables. Below each variable in the heatmap, the MissVisG glyph is used to represent the percentage of missing and recorded values, along with the distribution and outliers of recorded values.
 
\section{\textbf{Missing data visualization applications and tools}} \label{sec:VisAppTool}
 This section describes the 7 focus papers on missing data visualization applications and tools, which were found in our literature search. 

\begin{figure}[H]%
       \centering
       \subfloat[][The Missing value chart draws a horizontal bar. The left part of the bar represents the proportion of recorded data, and the right part represents the proportion of missing values.]{
       	\includegraphics[width=3.5 cm]{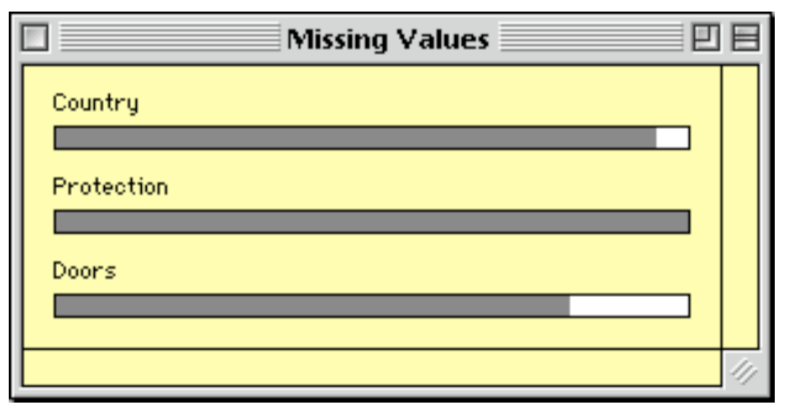}
       	\label{Fig:Manet1}
       }%
       \qquad
              \subfloat[][Box plots: a missing values plot is drawn for each box plot of a variable that has missing values.]{
       	\includegraphics[width=3.5 cm]{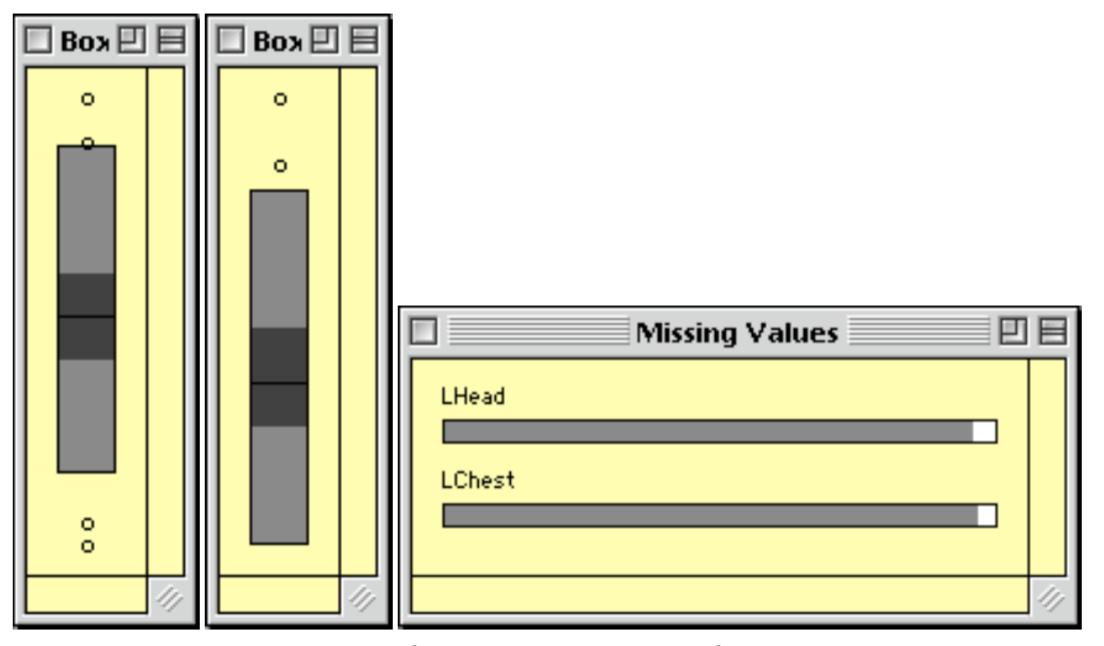}
       	\label{Fig:Manet3}
       }%
       \qquad
        \subfloat[][Weighted histograms: missing values in the weighting variable have the weight zero and an extra bin is plotted below the bar for each class that has missing values.]{
       	\includegraphics[width=3.5 cm]{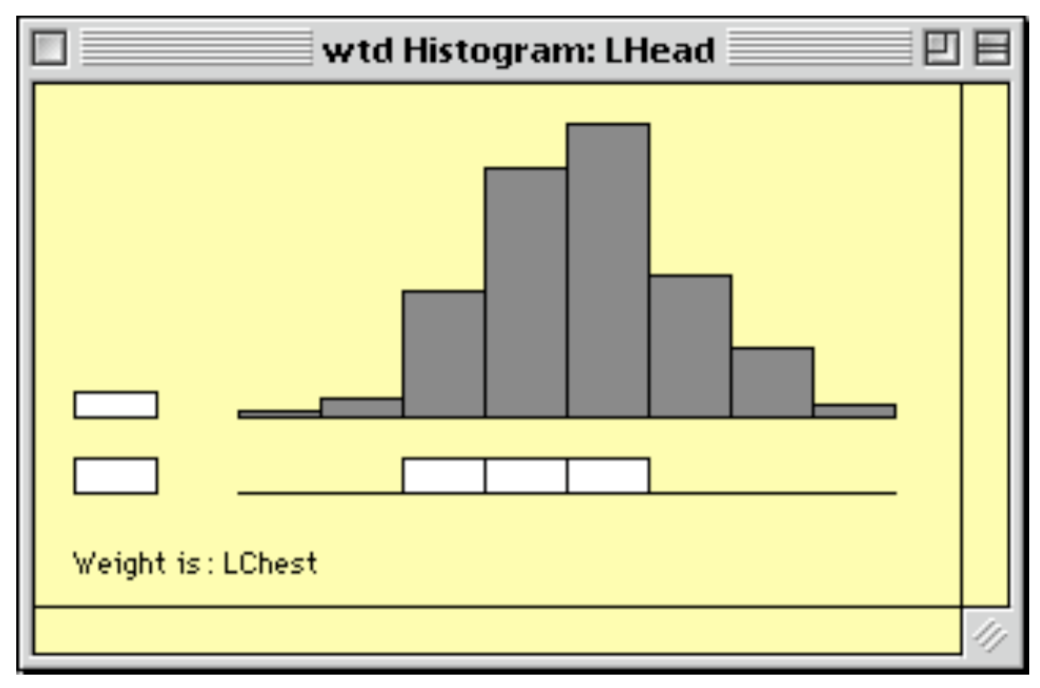}
       	\label{Fig:Manet2}
       }%
       \qquad
       \subfloat[][Scatter plots: additional boxes at the bottom of each scatter plot represent the proportion of missing x-values (left box), the proportion of missing x- and y-values (middle box), and the proportion of missing y-values (right box).]{
       	\includegraphics[width=3.5 cm]{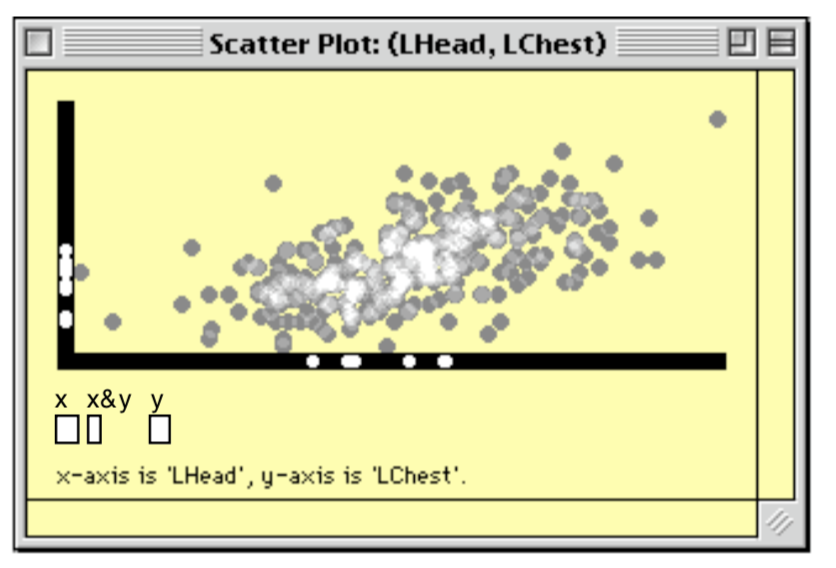}
       	\label{Fig:Manet4}
       }
       \caption{Missing data visualization in MANET \cite{Theus1997}.}%
       \label{Fig:Manet}%
\end{figure}

 Theus et al.\ \cite{Theus1997} developed object-oriented applications that handle missing values in an interactive graphical environment by integrating missing data into the standard statistical graphics in the MANET package. It offers interactive graphs like mosaic plots, weighted histograms, and weighted bar charts, along with generalised brushing and extensive interactive features (see Figure \ref{Fig:Manet} for examples). MANET allows users to weigh the values of one variable by another, visualizing the impact of missing values in the weighting variable on histograms and bar charts. In weighted histograms and bar charts, missing values are represented by extra bins or bars, enabling users to assess the influence of these missing weights. MANET also draws missing value plots alongside box plots, providing a visual representation of missing data. Multiple box plots can be combined with corresponding missing value charts. In scatter plots, MANET incorporates missing values by projecting them onto the axes, allowing users to identify observations where one or both values are missing. Additional boxes in the plot indicate the proportion of missing values for better analysis. Mosaic plots handle missing values by adding an extra category for all missing values of a variable. This interactive visualization method helps distinguish empty categories from those with few values.

\begin{figure}[H]%
       \centering
\subfloat[][the aggregation plot ]{        \includegraphics[width=3.7 cm]{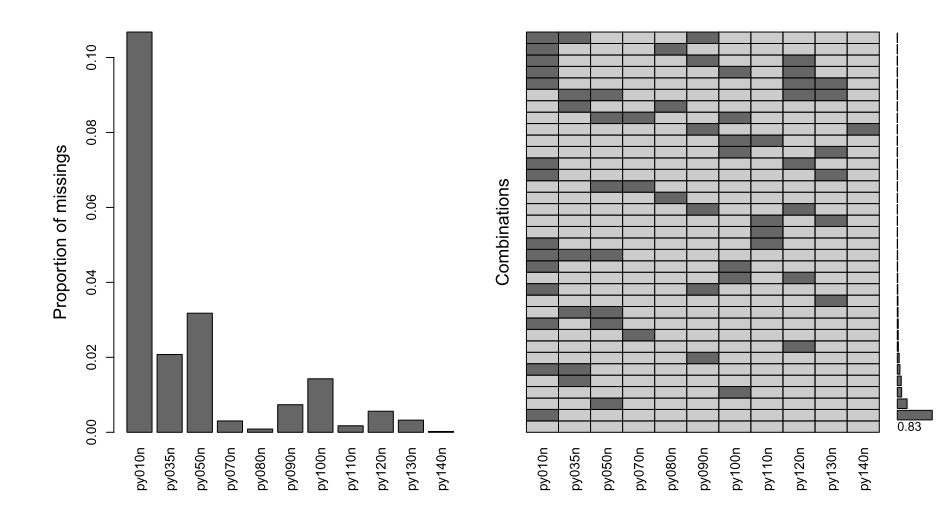}
        \label{Fig:VIM1}
       }%
        \qquad
       \subfloat[][ Histogram and spinogram]{        \includegraphics[width=3.5 cm]{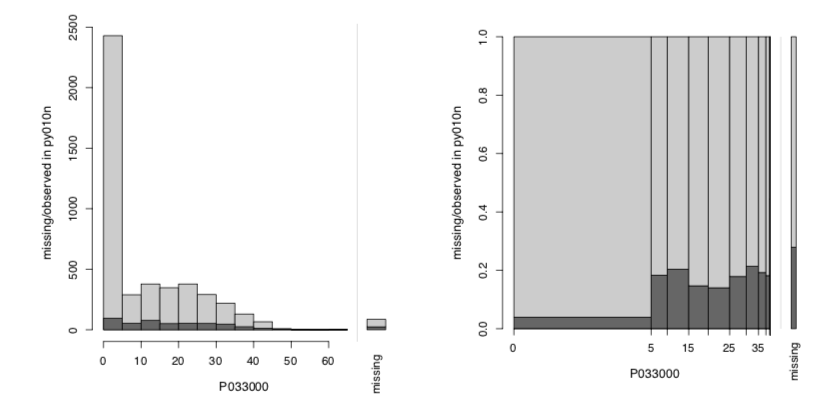}
        \label{Fig:VIM3}
       }%
       \qquad
        \subfloat[][Parallel box plots]{
       \includegraphics[width=3.5 cm]{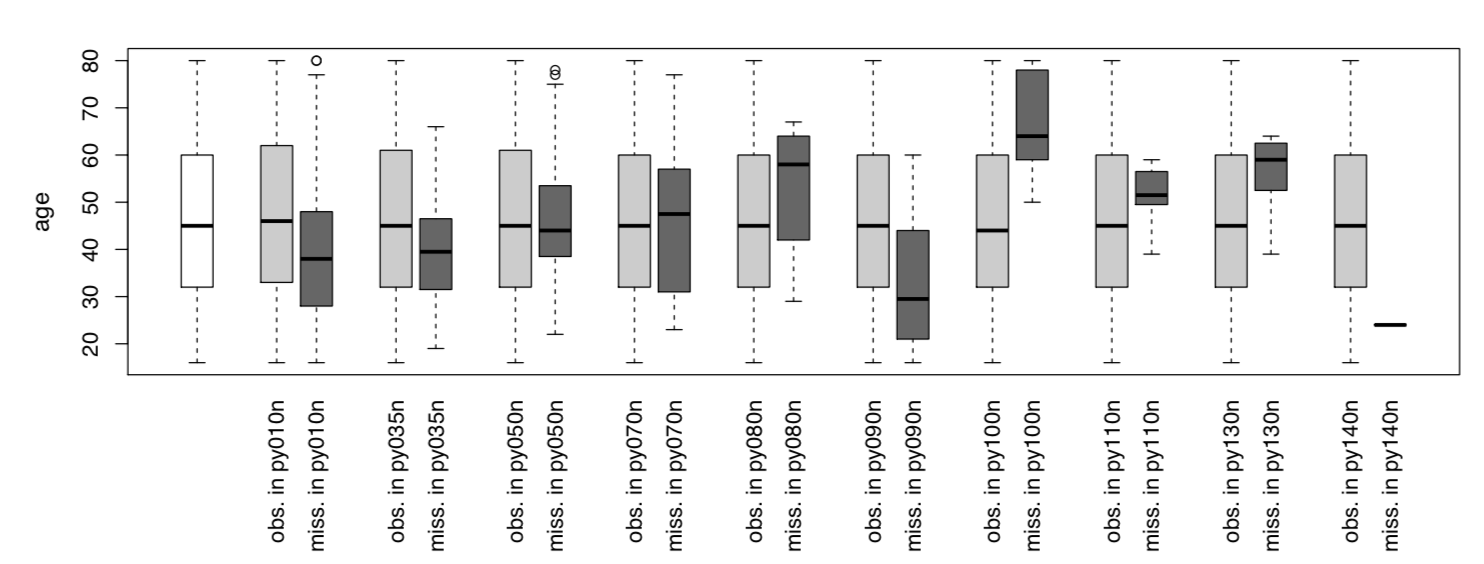}
        \label{Fig:VIM4}
       }%
       \qquad
        \subfloat[][Parallel coordinates]{
       	\includegraphics[width=3.5 cm]{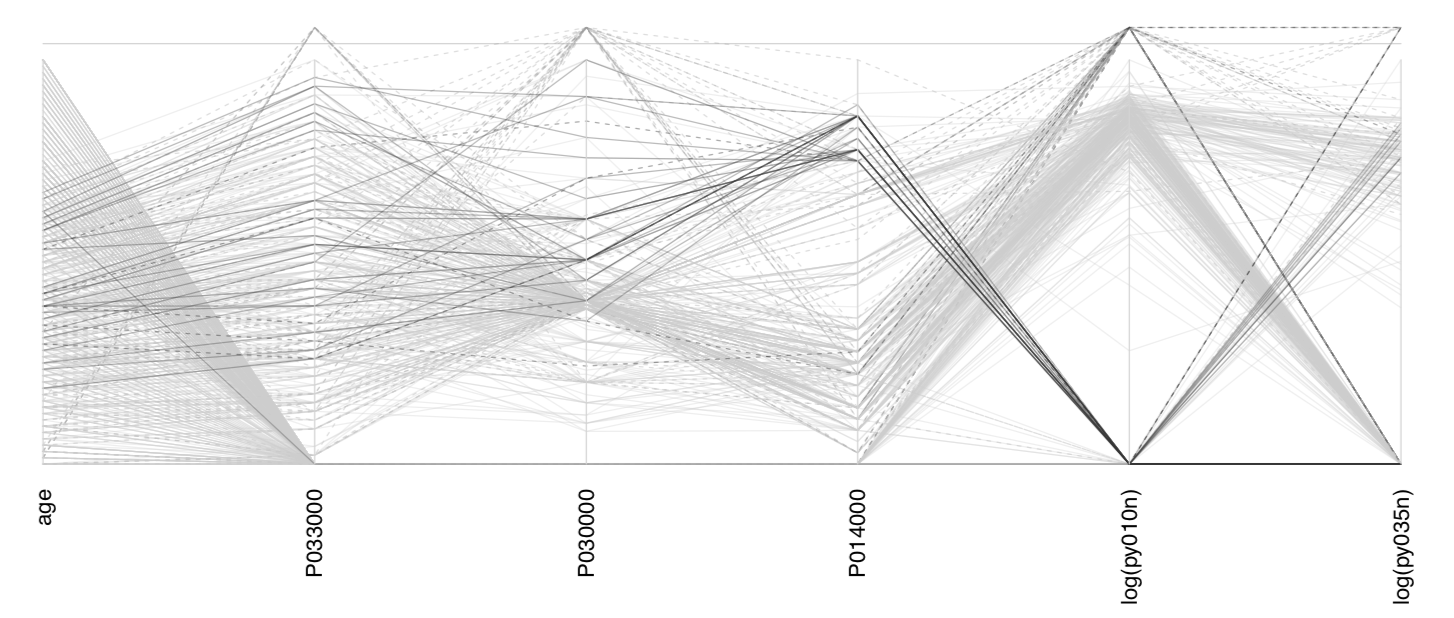}
        \label{Fig:VIM5}
       }%
       \qquad
        \subfloat[][Matrix plot]{
       \includegraphics[width=3.5 cm]{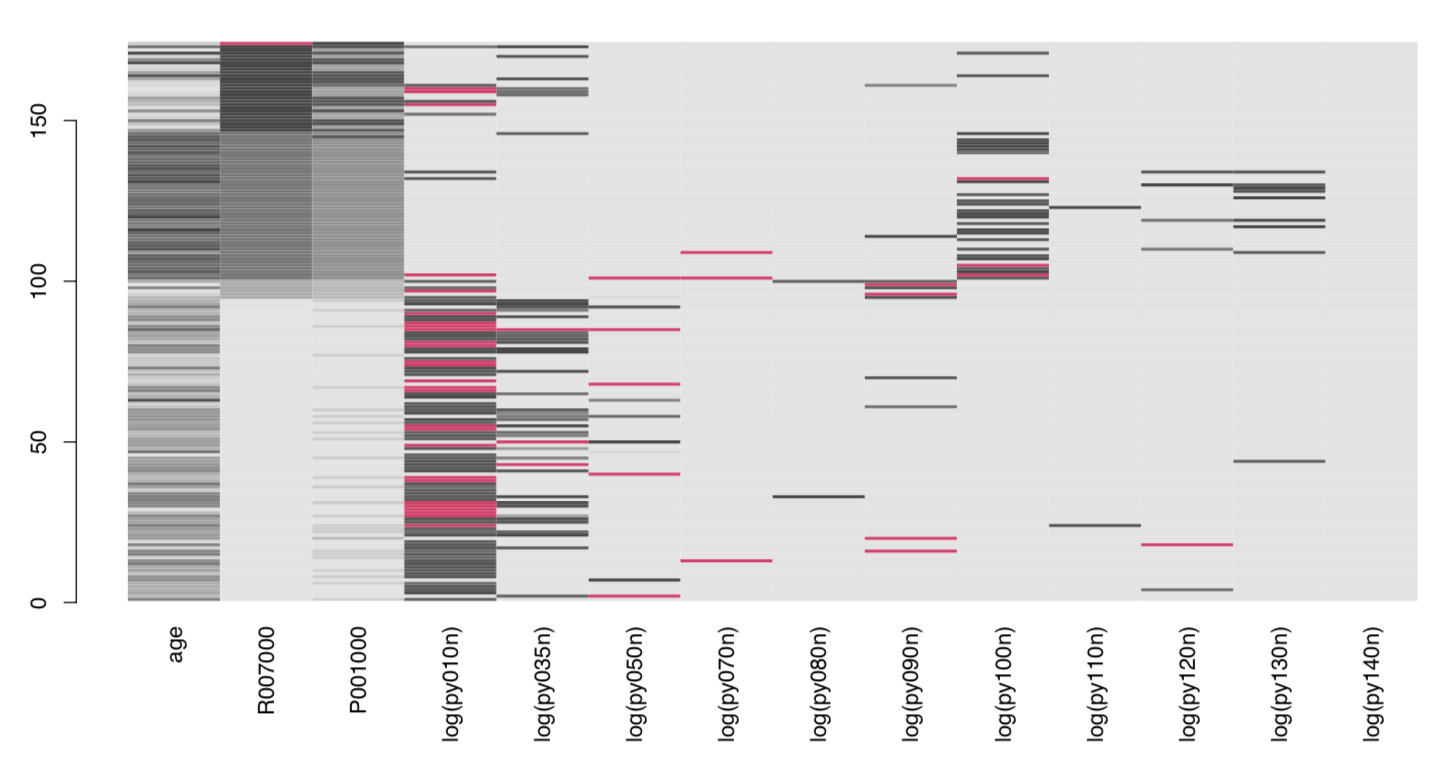}
        \label{Fig:VIM6}
       }%
        \qquad
       \subfloat[][ Left: growing dot map.  Right: the regional map]  {\includegraphics[width=3.7 cm]{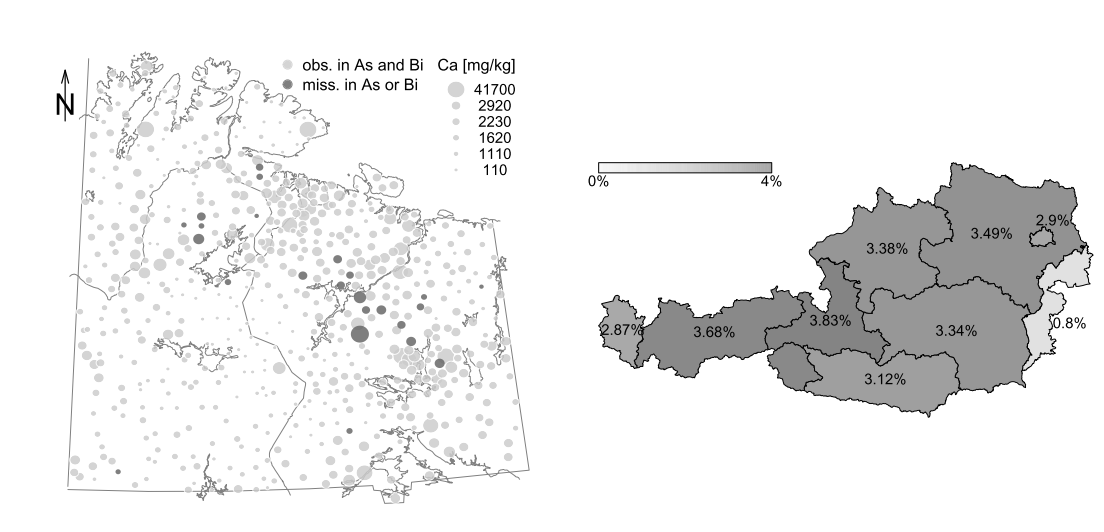}
        \label{Fig:VIM2}
       }%
       \qquad
       \subfloat[][Scatter/Margin plot]{
       	\includegraphics[width=3 cm]{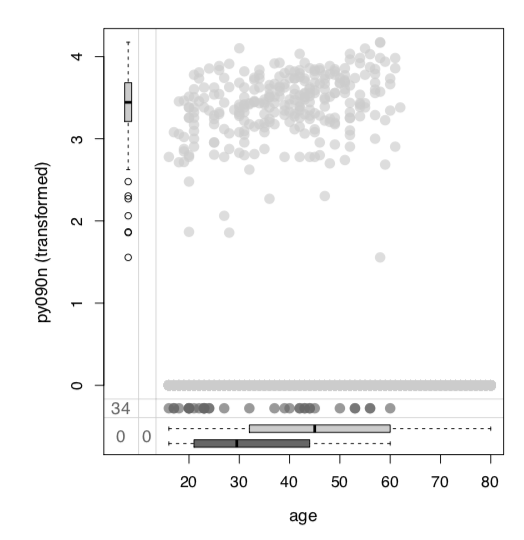}
        \label{Fig:VIM7}
       } %
         \qquad
       \subfloat[][Mosaic plot]{
       	\includegraphics[width=3.7 cm]{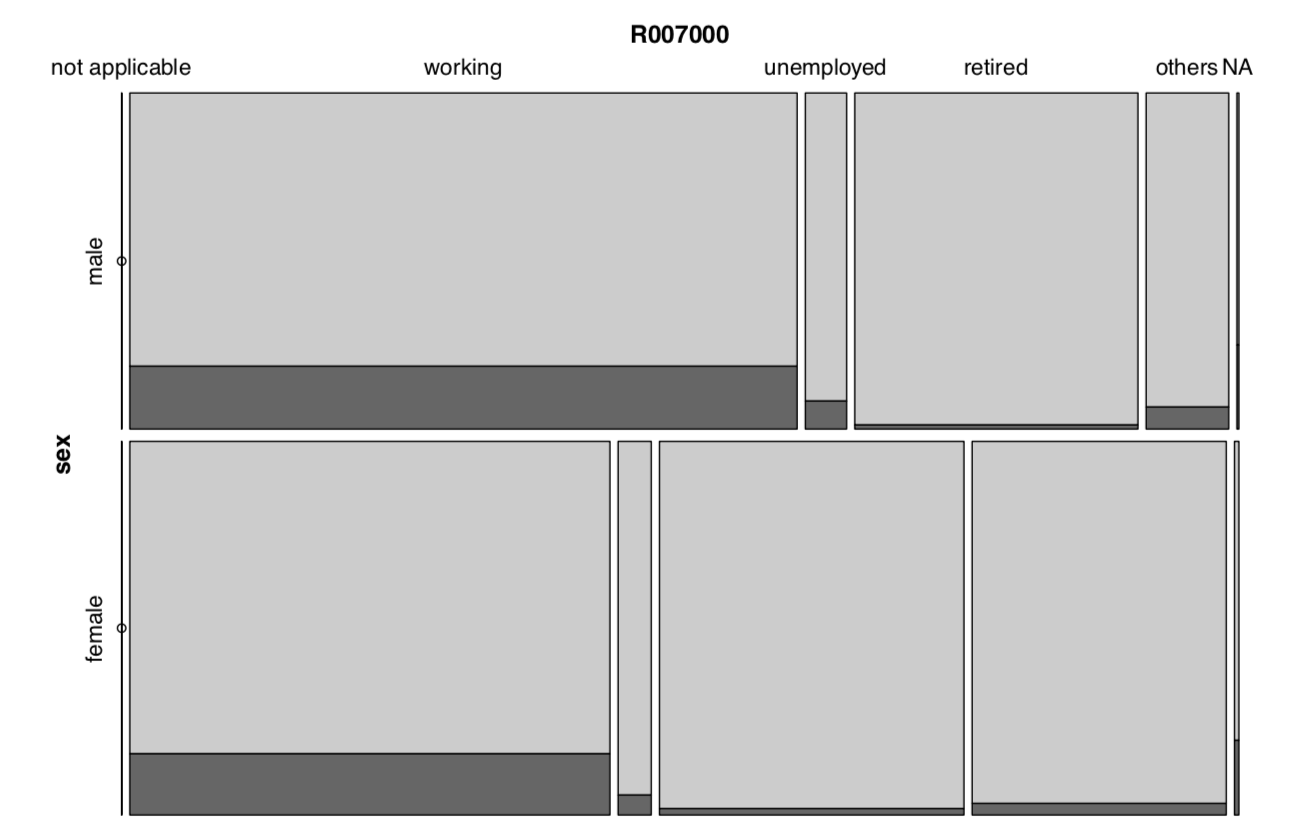}
        \label{Fig:VIM8}
         }
       \caption{Examples of missing data visualization in VIM \cite{Templ2011}. }%
       \label{Fig:VIMsub2}%
\end{figure}

 Templ et al.\ \cite{Templ2012, Templ2011} provided a mature tool and a graphical user interface for exploring the structure of missing values. They highlighted the importance of exploring missing values visually and confirmed that by visualizing incomplete data, researchers can identify patterns and relationships in missing values, which can aid in making informed decisions about data preparation, such as contacting respondents, calibrating data for missing values, or performing imputation. They utilize a collection of visualization techniques for incomplete data implemented in the R package VIM (Visualization and Imputation of Missing values), including histograms, scatter plots, parallel coordinate plots and heat maps. They used these visualizations to provide information about the missingness mechanism, the missing values characteristics, and the missing values structures. The aggregation plot (Figure \ref{Fig:VIM1}) is used to provide an overview of the amount of missing values and detect patterns. The proportion of missing values in each of the selected variables is visualized in the bar plot on the left side. All existing combinations of missing and non-missing values in the observations are represented on the right side and the frequencies of the combinations are visualized by small horizontal bars. In the histogram (Figure \ref{Fig:VIM3}), an extra bar is added to show the amount of missing values and recorded values. Parallel box plots (Figure \ref{Fig:VIM4}) are used to compare the conditional distributions according to a set of variables with observed values and  missing values. In a parallel coordinates plot (Figure \ref{Fig:VIM5}), missing values are represented by lines that are connected above the corresponding coordinate axes, separated from the observed data by a horizontal line. The matrix plot (Figure \ref{Fig:VIM6}) used a continuous colour scheme for observed values and a clearly distinguishable colour, such as red, for missing values. Scatter plots (Figure \ref{Fig:VIM7}) show information about missing values, such as the frequencies of missing values. Mosaic plot (Figure \ref{Fig:VIM8}) shows the distribution of the missing values in the dataset. Missing values are here highlighted in dark grey to help explore their structure. VIM also provides interactive maps to visualize missing data. In Figure \ref{Fig:VIM2}, in the original map, the amount of missing values in a variable are visualized for each region using continuous colour scheme. If an observation has missing values in another variable, the corresponding growing dots can be colour-coded with darker shade in the growing dots map.

\begin{figure}[H]
            \centering
       \subfloat[][The four types of graphs available:  (left to right) barchart, histogram, spineplot, and spinogram.]{
       	\includegraphics[width=8 cm]{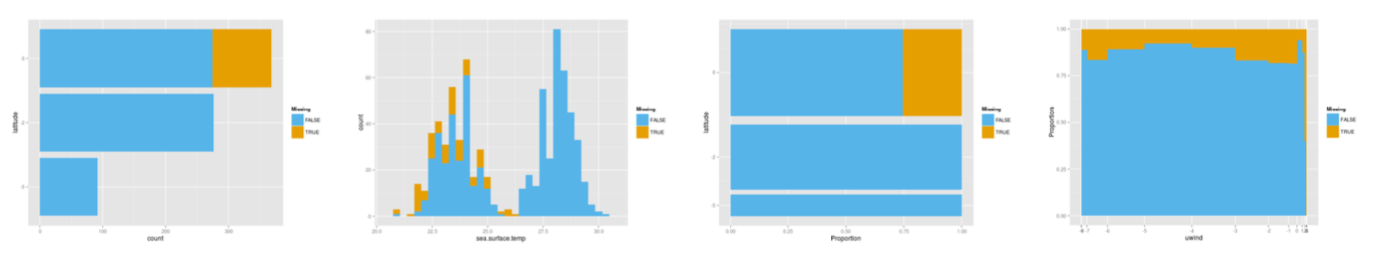}
        \label{Fig:cheng}
        }%
       \qquad
       \subfloat[][Examples of missingness maps.]{
       	\includegraphics[width=8 cm]{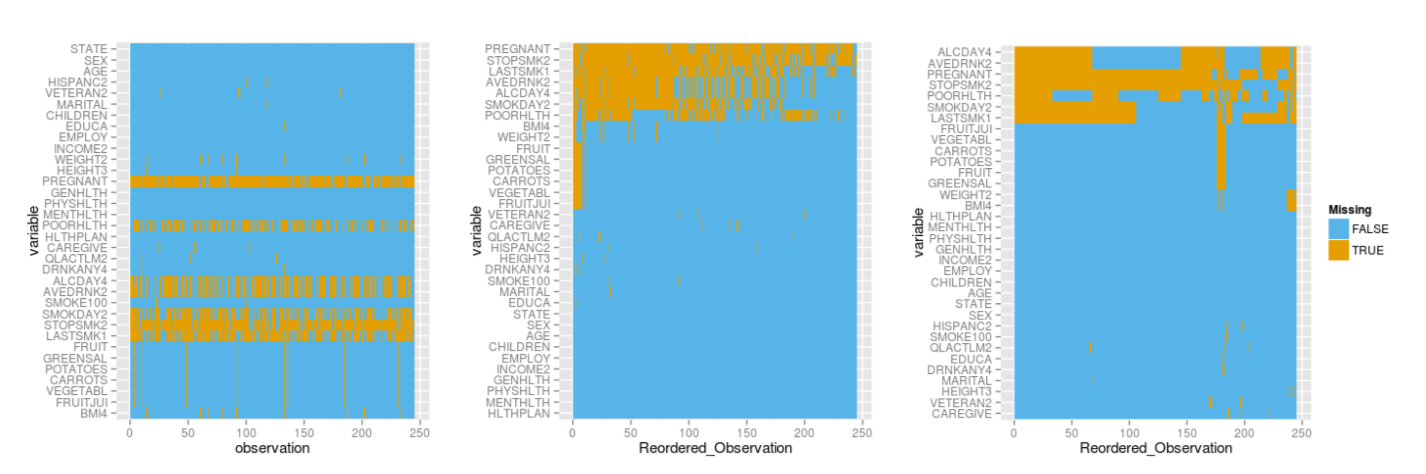}
        \label{Fig:Ch1}
        }%
       \qquad
        
         \subfloat[][Pairwise plots and parallel coordinates plots.]{
       	\includegraphics[width=8 cm]{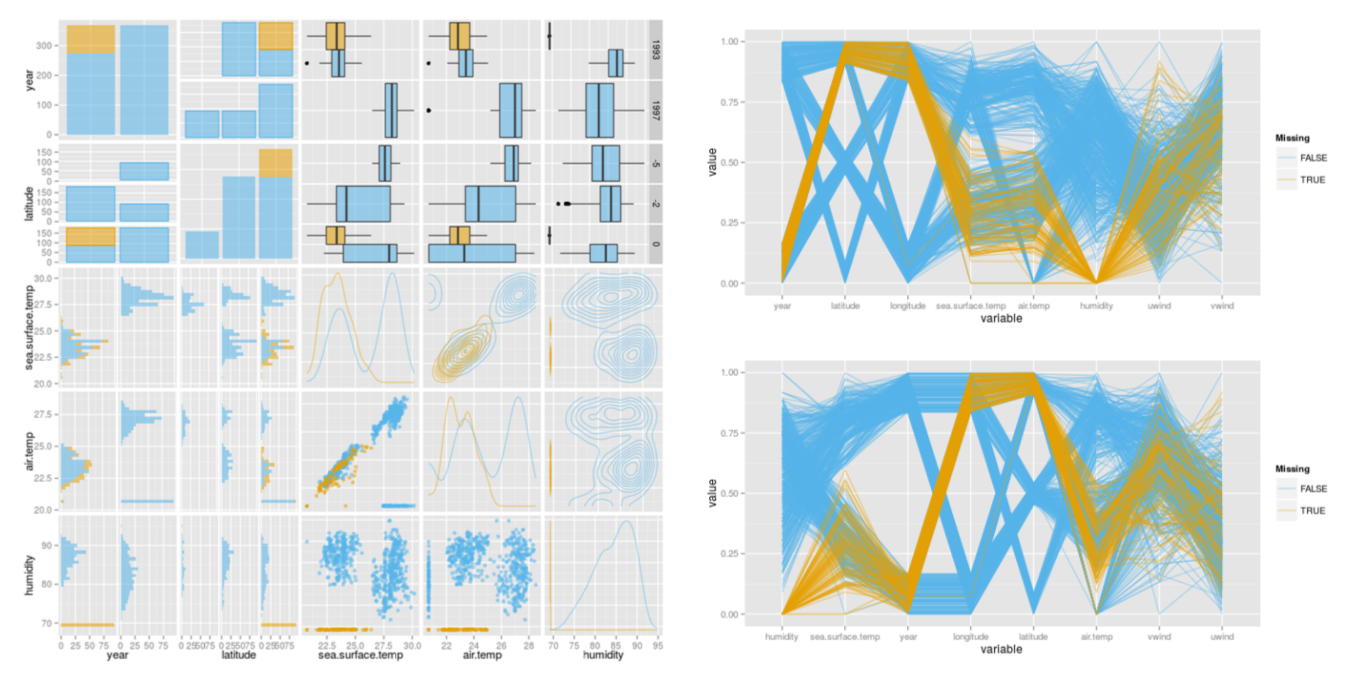}
        \label{Fig:cheng2}
        }
    \caption{Examples of missing data visualizations suggested by Cheng et al.\ \cite{Cheng}.}%
       \label{Fig:ChengSub}%
\end{figure}

Cheng et al.\ \cite{Cheng} presents the development of a missing data graphical user interface (GUI), within an R package, that allows users to visualize and handle missing data. The GUI includes features such as variable selection, imputation method selection, conditional variable checkboxes, and graphics display options. They support  visualization to explore patterns of missing values in data using various types of graphs, including barcharts, histograms, spineplots, spinograms (Figure \ref{Fig:cheng}), missingness map (Figure \ref{Fig:Ch1}), pairwise plots, and parallel coordinates plots (Figure \ref{Fig:cheng2}). These visualizations aid in understanding the patterns of missingness in the data. 

\begin{figure}[h!]
        \centering
       \subfloat[][The missingness map shows an overall overview of missing data based on the groups of variables. Black cells represent missing values.]{
       	\includegraphics[width=7.5 cm]{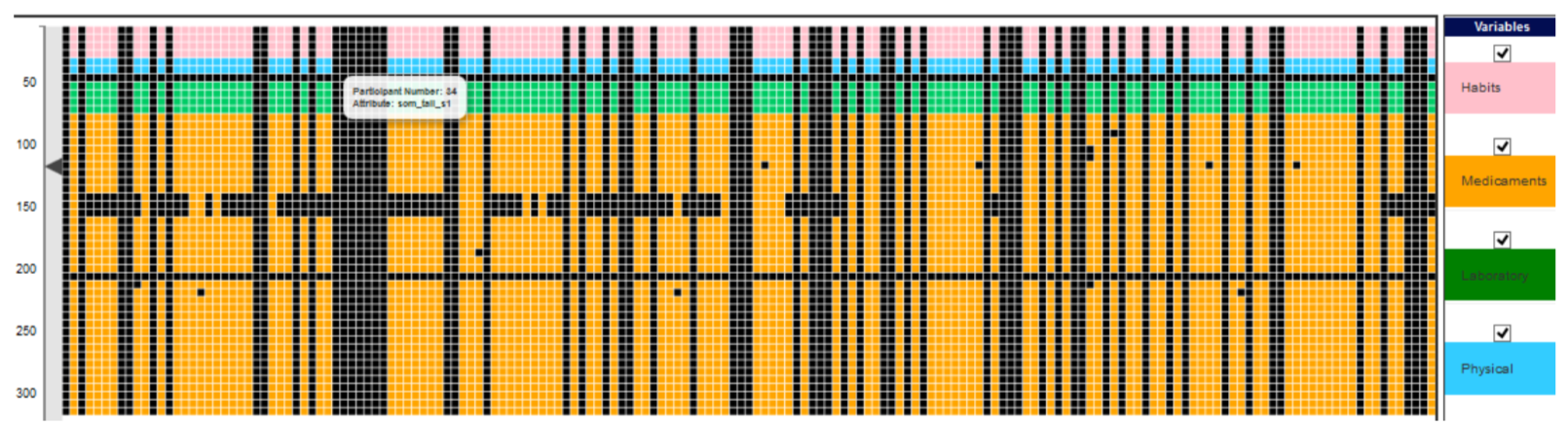}
        \label{Fig:missMap2}
        }%
       \qquad
       \subfloat[][Bean plots are used to compare the distributions of imputed and observed values. The red lines show imputed values and gray lines display observed values. ]{
       	\includegraphics[width=7.5 cm]{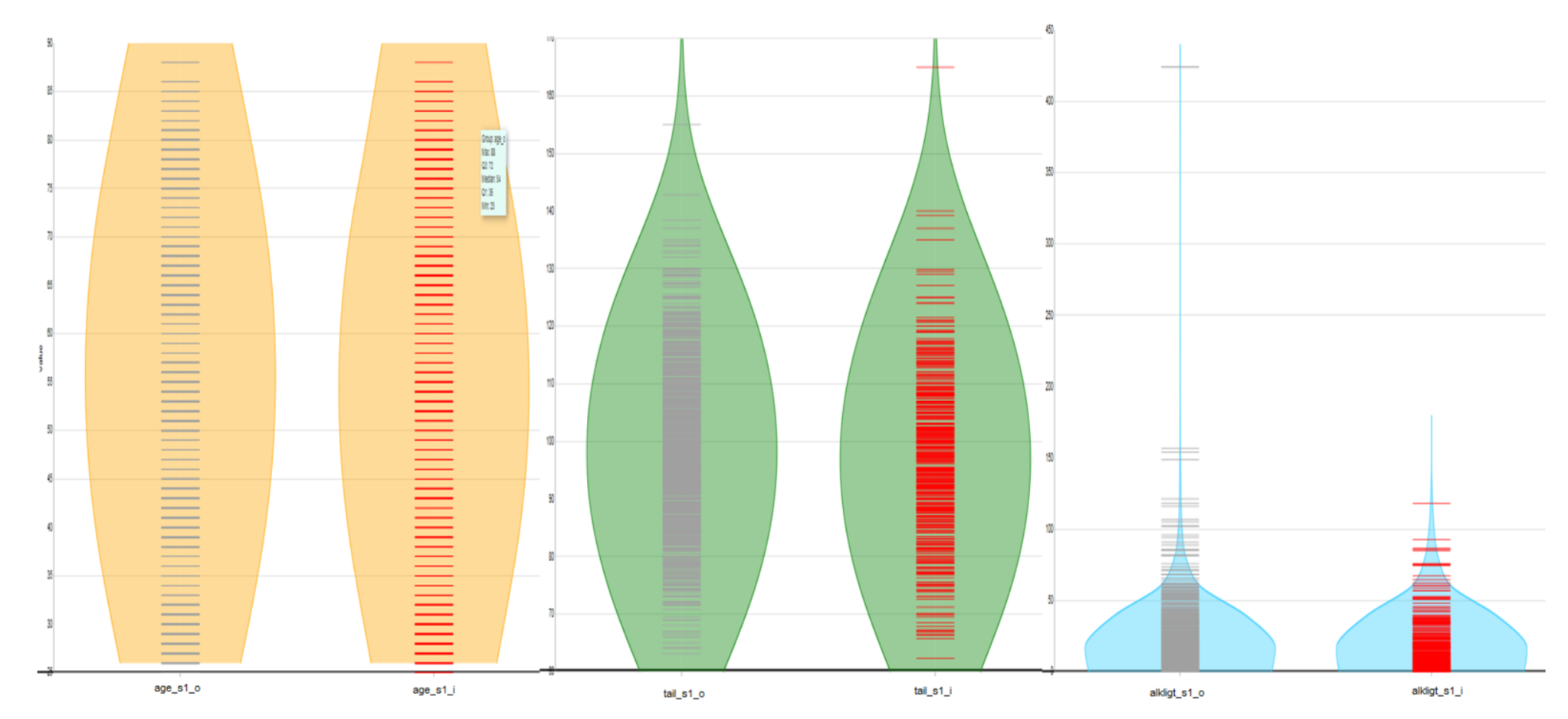}
        \label{Fig:beanPlot}
        }%
       \qquad
         \subfloat[][The strip plot shows the distribution of imputed values over observed values. The gray points represent observed values and the red points represent imputed values.]{
       	\includegraphics[width=7.5 cm]{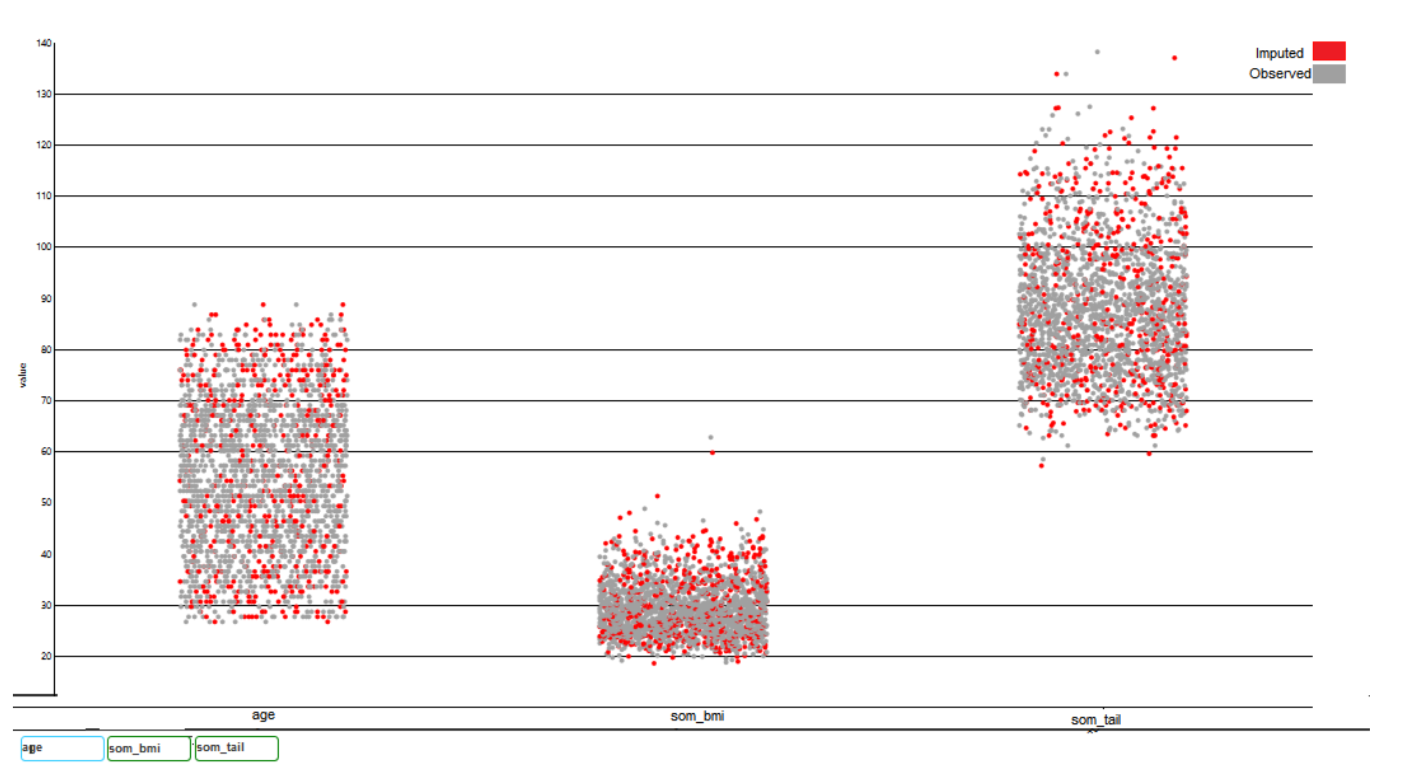}
        \label{Fig:stripPlot}
        }
    \caption{Examples of missing data visualizations suggested by Alemzadeh et al.\ \cite{Alemzadeh2017}.}%
       
       \label{Fig:Alemzadeh}%
\end{figure}

\begin{figure}[H]
       \centering
       	\includegraphics[width=7.5 cm]{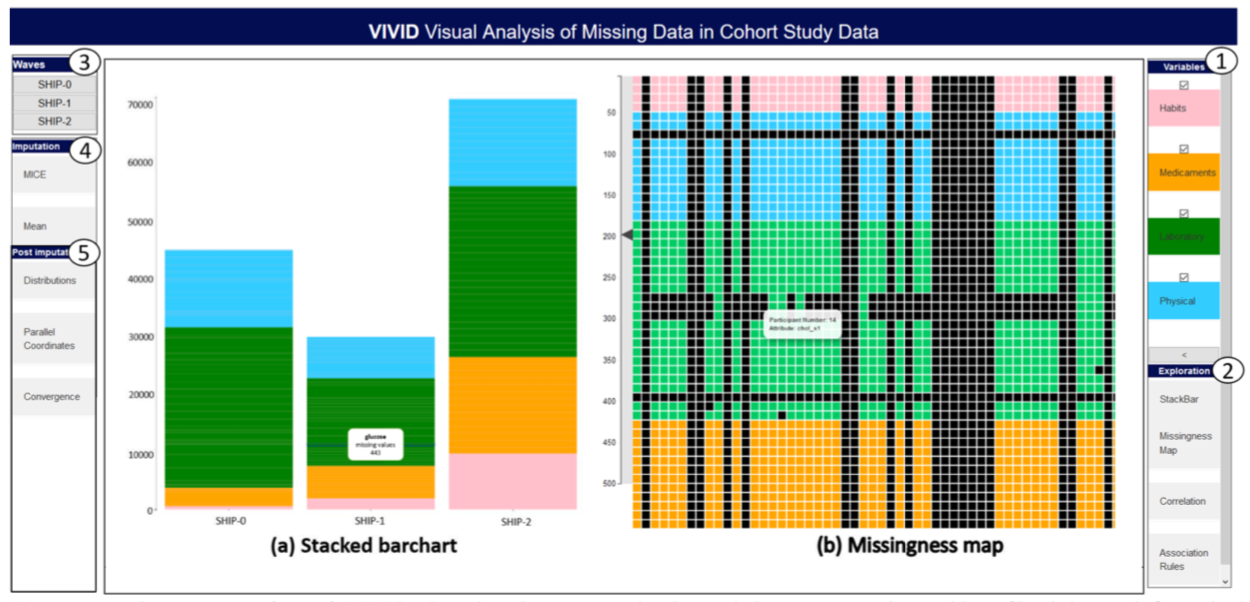}
       \caption{The interface of VIVID including the missingness map \cite{Alemzadeh2020}.}
        \label{Fig:VIVID}
\end{figure}

Alemzadeh et al.\ \cite{Alemzadeh2017} introduces a visual analytics solution for analyzing and managing missing values in epidemiological cohort study data. The approach incorporates visual components to complement existing statistics-based methods used by epidemiologists. It provides a web-based system, with a user interface for exploration, imputation, and quality checking of imputations. It allows the user to visually analyze the patterns of missing data and gain insights into the source of missingness by using missingness maps (Figure \ref{Fig:missMap2}). It also offers a visual representations of missing data among different variables before and after imputation, allowing the analysts to compare different imputation methods and check the quality of imputations, compare the distribution of observed and imputed values, by using bean plots (Figure \ref{Fig:beanPlot}) and strip plot (Figure \ref{Fig:stripPlot}). Building on the work in \cite{Alemzadeh2017} they develop VIVID \cite{Alemzadeh2020}, a web-based system to visually explore and impute missing values in epidemiological studies. It also assets the plausibility of predictions for each imputed dataset after imputation by providing suggestions of the list of the missing values predictors based on the variables correlation. The framework is developed in close collaboration with epidemiologists to ensure the relevance and effectiveness of the visual analysis tools. They developed a missingness map (Figure \ref{Fig:VIVID}) that visualizes the data in a matrix format with colour-coded cells to represent missing and observed values. Each row in the matrix represents a variable, and each column represents a data item. Missing values are depicted as black cells and one completely blacked column represents a drop-out participant. The map provides a compact overview of missing and observed values in the dataset.

\begin{figure}[H]
        \centering
        \subfloat[][Sankey diagram of missing rate in time series data.]{
        \includegraphics[width=4 cm]{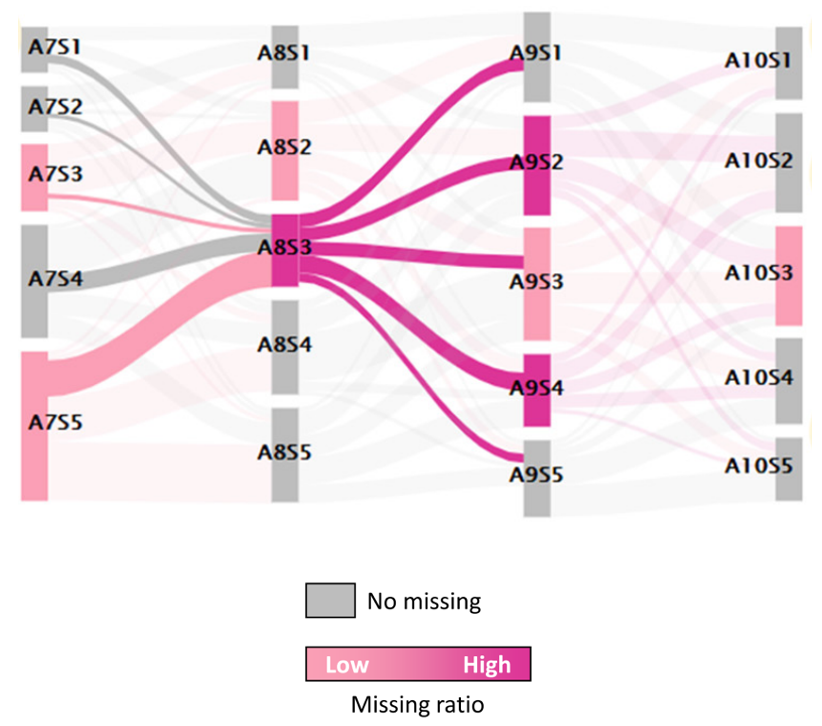}
        \label{Fig:Sankey}
         }%
       \qquad
         \subfloat[][Analysis of missingness patterns using a heatmap.]{
        \includegraphics[width=4 cm]{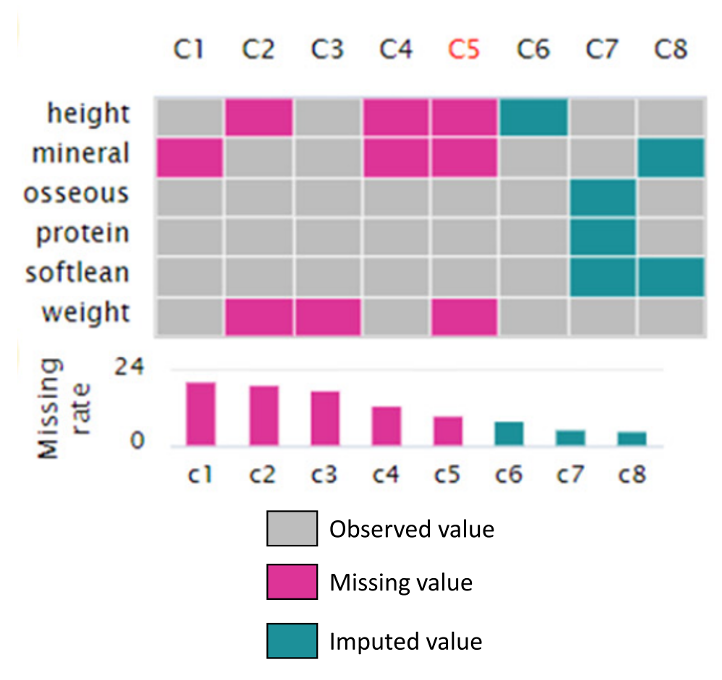}
        \label{Fig:Sankey2}
         }
    \caption{Examples of missing data visualizations for Bayesian networks, as suggested by Yeon et al.\ \cite{Yeon2020}.}%
      
       \label{Fig:Yeon}%
\end{figure}

Yeon et al.\ \cite{Yeon2020} use a Sankey diagram (Figure \ref{Fig:Sankey}) to visualize the Bayesian network with the missing rate of observed time series data. It helps users to understand the transition probabilities between different states in the Bayesian network over time. Each node in the diagram represents a state, and the colour of each state indicates the missing rate in the data or the rate of imputed data, providing a visual representation of missing data patterns. The user can interact with the Sankey diagram to explore missing patterns, identify states with missing data, and analyze the transition probabilities between states. They also suggest using heatmaps to represent missingness patterns \ref{Fig:Sankey2}.

 \section{\textbf{Missing data visualization evaluation}} \label{sec:VisEval}
 This section describes 6 focus papers on Missing data visualization evaluation found in our literature search. 

\begin{figure}[h!]
        \centering
       \subfloat[][\textit{Misleading} display.]{
       	\includegraphics[width=3.5 cm]{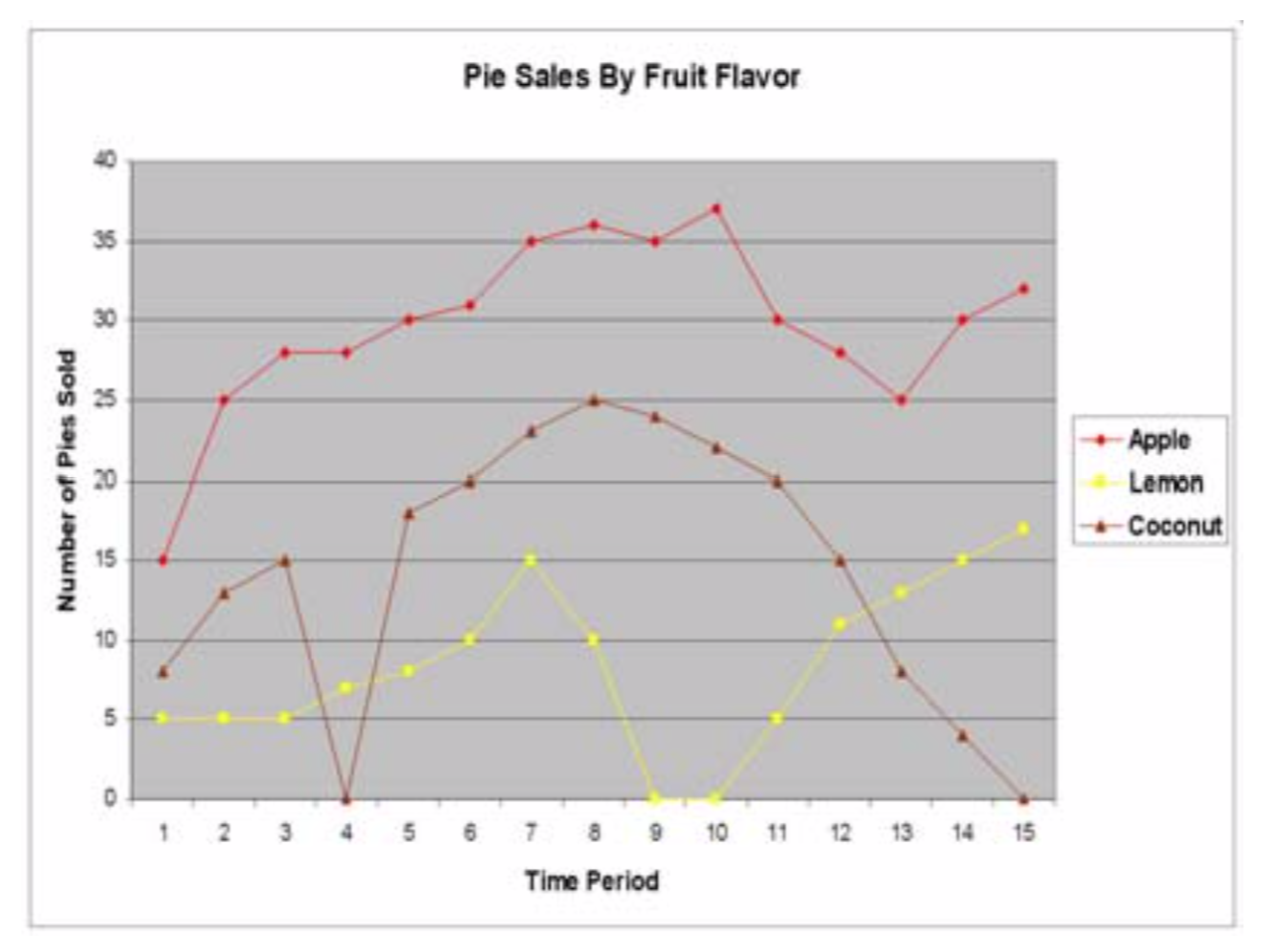}
        \label{Fig:Missleading}
         }%
       \qquad
         \subfloat[][\textit{Absent} display.]{
       	\includegraphics[width=3.5 cm]{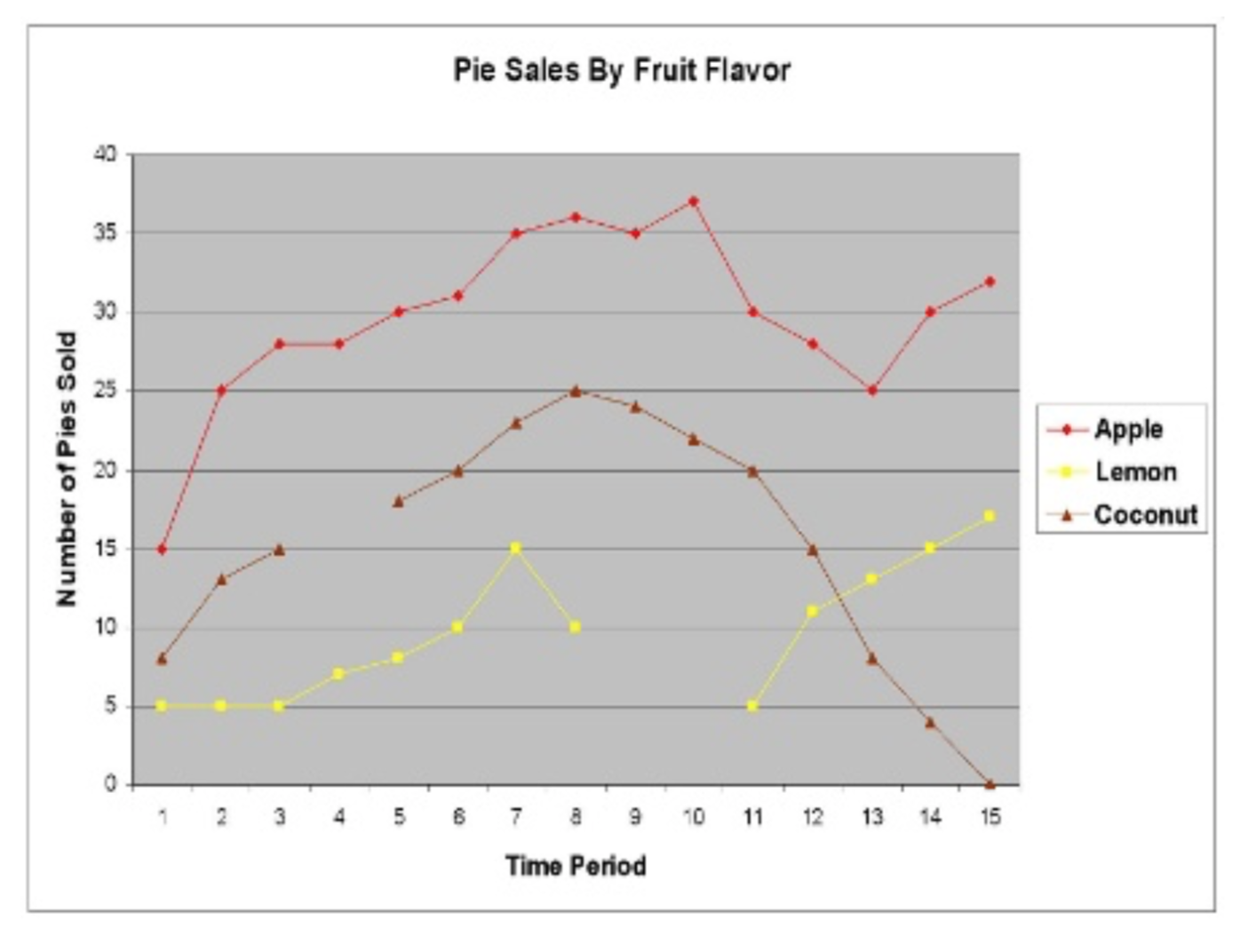}
        \label{Fig:Absent}
        }%
       \qquad
         \subfloat[][\textit{Coded} display.]{
       	\includegraphics[width=3.5 cm]{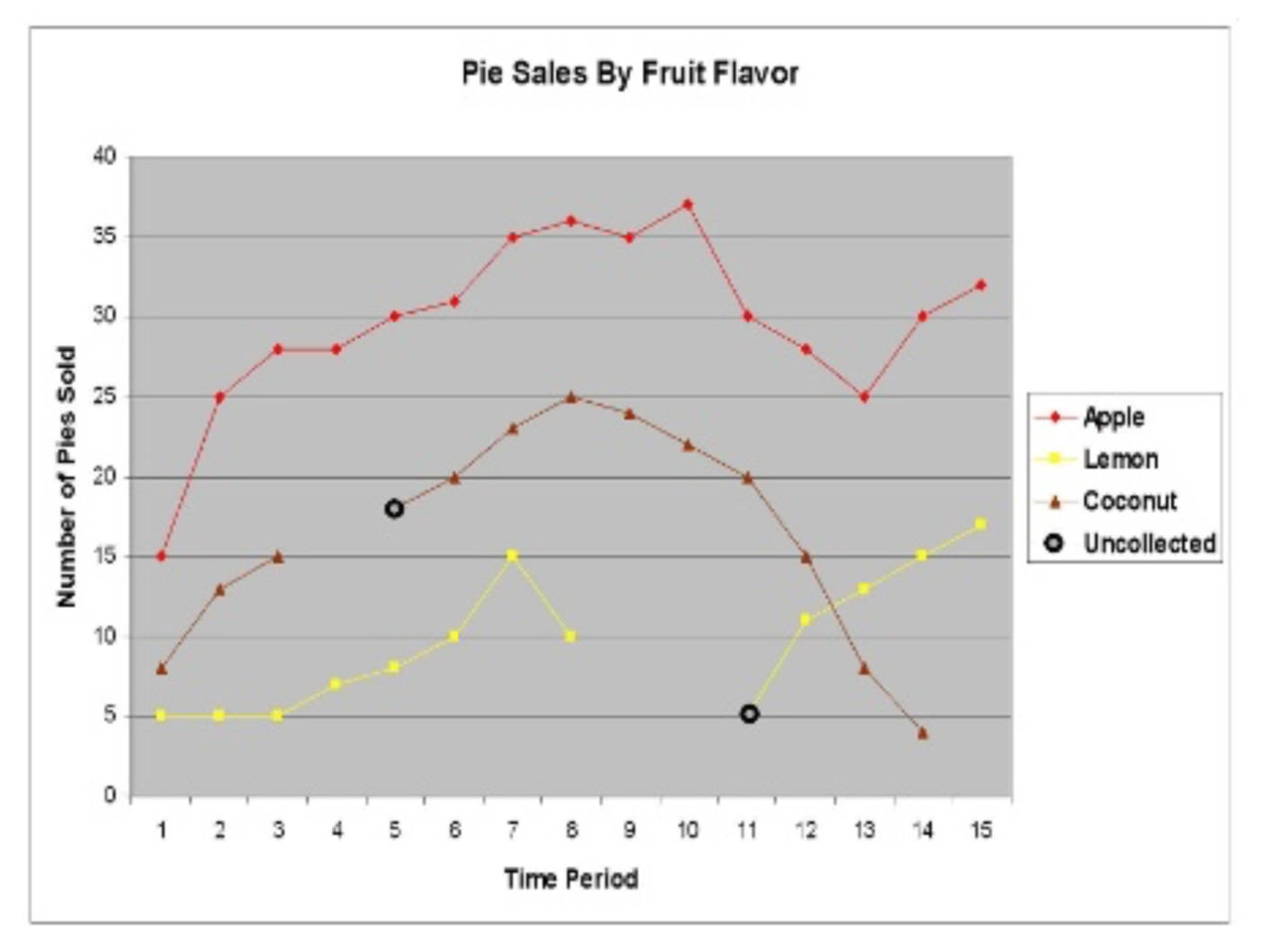}
        \label{Fig:Coded}
         }
    \caption{The three visual techniques for representing missing values that are evaluated by Eaton et al.\ \cite{Eaton2005} }%
       \vspace{-2mm}
       \label{Fig:Eaton}%
\end{figure}

Eaton et al.\ \cite{Eaton2005} provides one of the first user studies related to visualization of missing data. It focuses on three different techniques for missing data: \textit{Misleading}, \textit{Absent}, and \textit{Coded}. The misleading display (Figure \ref{Fig:Missleading}) encodes missing data as 0, while the absent display (Figure \ref{Fig:Absent}) omits missing data and breaks the line graph for missing data, and the coded display (Figure \ref{Fig:Coded}) omits missing data and adds an icon in the next present point which provides the reason why prior points are missing. Their study aims to understand users' ability to interpret graphs including missing data and draw accurate conclusions about trends. The study found that users may not notice that data is missing when it is replaced by a default value. Moreover, users may be compelled to make general conclusions with partial data even if the missing data is realized. Lastly, the coded display was preferred by participants because it provides a reason for missing values.

\begin{figure}[H]
       \centering
       	\includegraphics[width=8 cm]{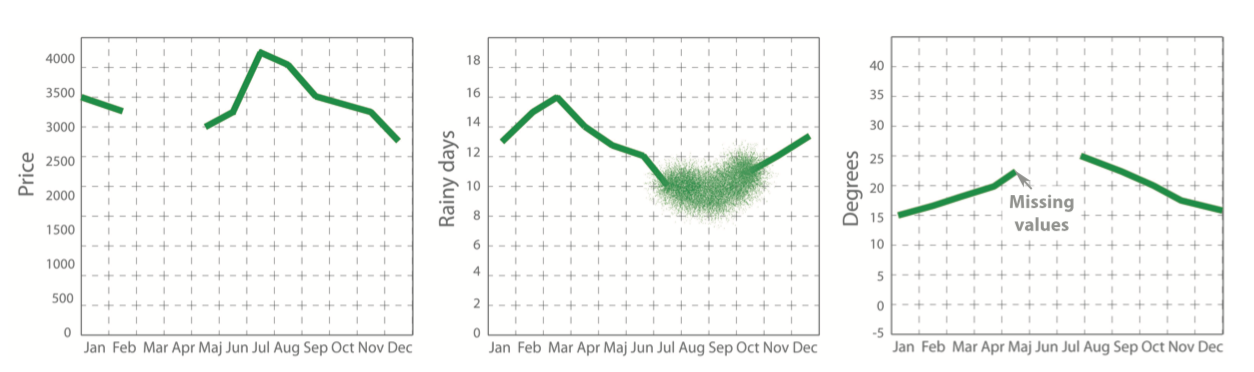}
       \caption{Illustration of the three techniques evaluated by Andreasson and Riveiro \cite{Andreasson2014}: \textit{Emptiness} (left), \textit{Fuzziness} (middle) and \textit{Emptiness plus explanation} (right).}
        \label{Fig:Effect3vis}
\end{figure}

Andreasson and Riveiro \cite{Andreasson2014} evaluates the effects on decision-making of missing value visualization, using three different visualization techniques: \textit{Emptiness}, \textit{Fuzziness} and \textit{Emptiness plus explanation} (Figure \ref{Fig:Effect3vis}). 
Emptiness is using empty space to show missing values, corresponding to \textit{Absent} in \cite{Eaton2005}; \textit{Fuzziness} uses fuzzy depictions to show uncertain information such as missing values; and \textit{Emptiness plus explanation} uses empty space to show missing values and add some explanation to show that values are missing, similar to \textit{Coded} in \cite{Eaton2005}. Their empirical study found that emptiness plus explanation was the most preferred technique with the highest degree of decision confidence.

\begin{figure}[h!]
\centering
\subfloat[][The categories to encode missing values.]{
       	\includegraphics[width=8 cm]{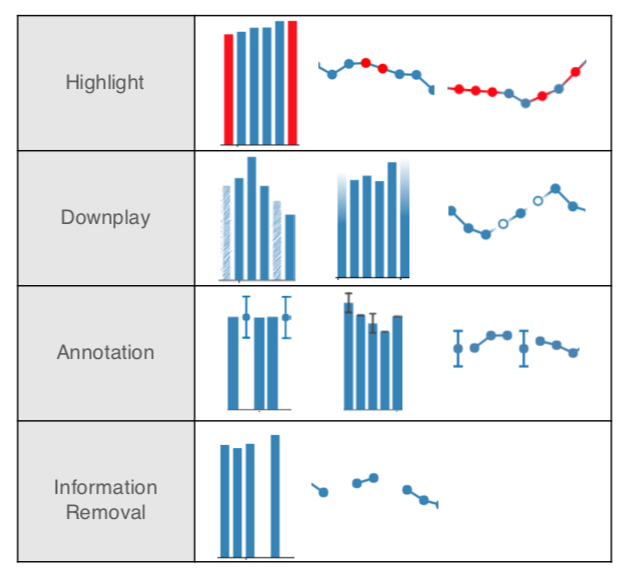}
        \label{Fig:where1}
         }%
       \qquad
         \subfloat[][Three different imputation methods.]{
       	\includegraphics[width=8 cm]{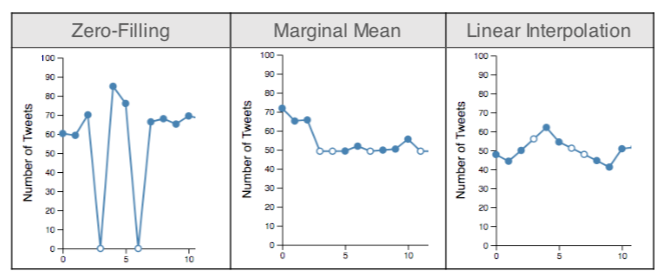}
        \label{Fig:where2}
         }
    \caption{The visual encodings and imputation approaches evaluated by Song and Szafir \cite{Song2019}}%
       
       \label{Fig:Song2019}%
\end{figure}

\begin{figure}[h!]
       \centering
       	\includegraphics[width=8 cm]{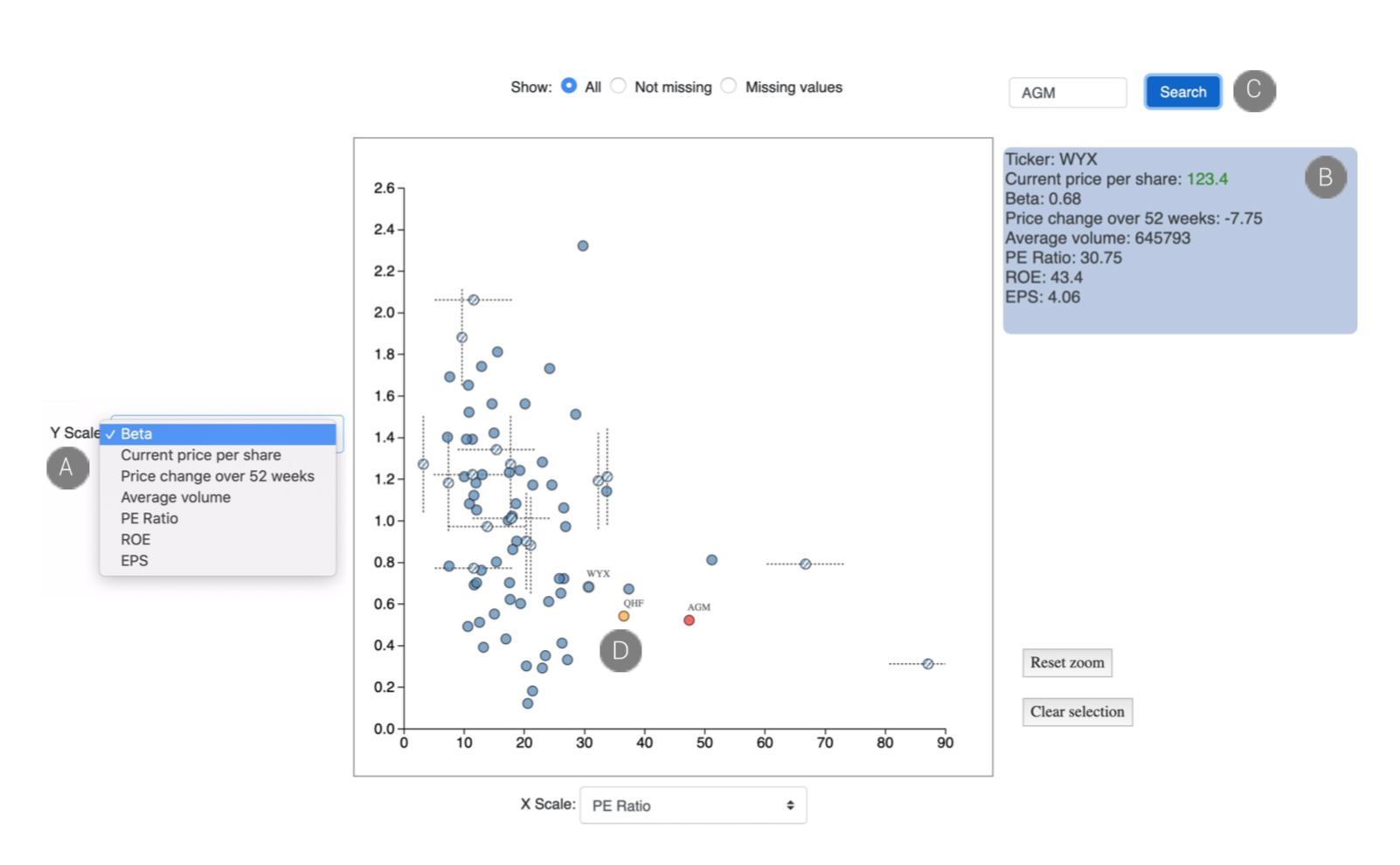}
       \caption{The error bar visualization used by Song et al. \cite{Song2021}.}
        \label{Fig:errorBars}
\end{figure}

Song and Szafir \cite{Song2019} discusses how imputation and visualization  methods of missing values can influence analysts' perceptions of data quality and their confidence in their conclusions. They use three categories to encode imputed values (highlight, downplay, and annotation) and information removal; use three imputation methods (zero filling, local linear interpolation, and marginal means); and two common visualizations (bar charts and line graphs), as displayed in Figure \ref{Fig:Song2019}. They found that there is an impact on the analysts' perceptions by the choice of imputation method and missing data visualization. Their study found that highlighting the missing data and local linear imputation resulted in higher perceived confidence and data quality, while those that break the visual continuity of a graph reduce the perception of quality in data and can bias interpretation. 
In a second study \cite{Song2021}, Song et al.\ examined how the visualization of missing values during the data exploration task affects the decision-making process. They used two versions of interactive scatter plots: \textit{Baseline} and \textbf{Error bars}. Baseline visualization does not visually represent any missing values and only represents the number of missing values, while error bars visualization (Figure \ref{Fig:errorBars}) shows the estimated values and imputed values as points with error bars. Their study indicates that visually representing missing values encourages participants to reason about data quality, resulting in more consistent and regular decision-making processes. In addition, the participants' confidence in their decisions increased when the missing values were represented visually. 

Johansson Fernstad \cite{Fernstad} defined three missingness patterns of relevance for missing data analysis, namely the \textit{Amount Missing} in variables, the \textit{Joint Missingness} across variables, and the \textit{Conditional Missingness} relationship between recorded values of items with missing values in another variable. The usability in context of these missingness patterns was evaluated for three visualization methods from the VIM package \cite{Templ2012}: Marginplot Matrix, which is a scatterplot matrix with missing values represented along the margins of the scatterplots; Matrix Plot, which are heatmaps with missing values represented by colour; and Parallel Coordinates where missing values are represented above the axis. Taking the visual attributes used when representing missing values into account, the study indicated that size based frequency representation, as used in the Matrix Plot, seemed particularly relevant for understanding the Amount Missing and Joint Missingness, while representation of connections between variables as well as between missing and recorded, as in Parallel Coordinates, seemed particularly relevant for identifying Conditional Missingness trends.

\begin{figure}[H]
       \centering      	
       \includegraphics[width=8 cm]{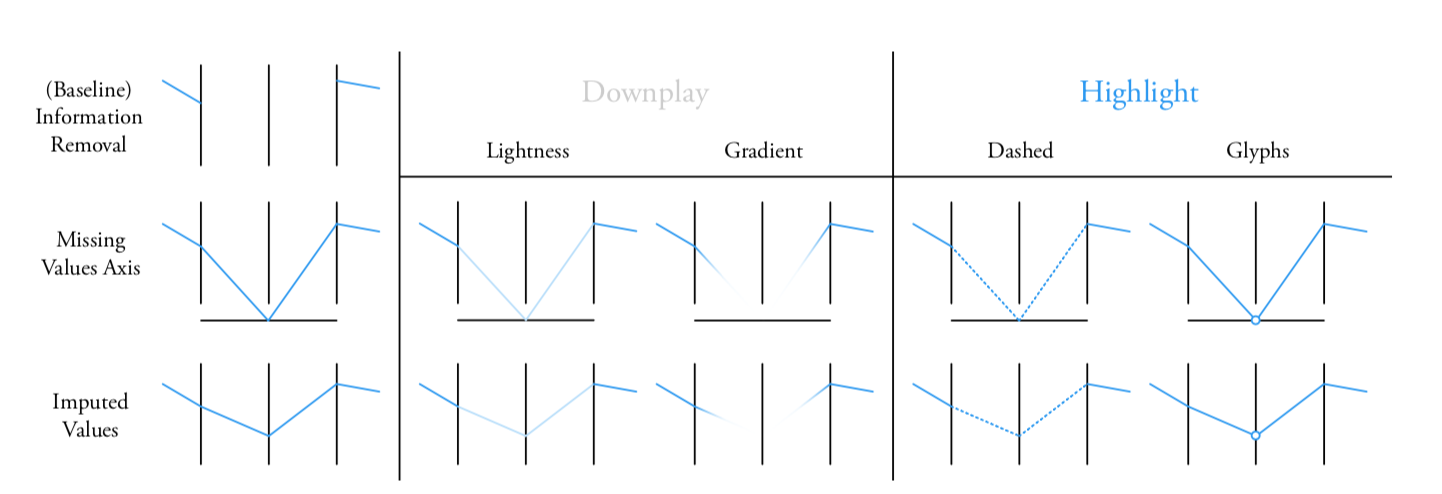}
       \caption{Illustrations of the evaluated concepts (left) and variations (middle, right) for representing missing values in parallel coordinates, studied by B\"{a}uerle et al.\ \cite{Bauerle2022}.}
        \label{Fig:Parallel1}
\end{figure}

B\"{a}uerle et al.\ \cite{Bauerle2022} evaluated three visualization concepts for representing missing values in parallel coordinates: \textit{Information Removal}, \textit{Missing Values Axis}, and \textit{Imputed Values}; with downplay and highlight variations concept for the missing and imputed values (Figure \ref{Fig:Parallel1}). The Information Removal technique means that no line is drawn for missing values. The Missing Values Axis adds an axis separately from coordinates with complete data to show missing values. In the Imputed Values concept, the missing values were imputed and then visualized. The focus of the study was on the trade-off between the ability to perceive missing values and the impact on common tasks. The downplay variation was designed to de-emphasize missing values by reducing their visibility, with two distinct downplay techniques considered by B\"{a}uerle et al.: opacity and gradient. In the the opacity technique, lines representing data points with missing values will be displayed with lower opacity compared to the lines representing complete data points. In the gradient technique, lines are gradually faded away as they approach missing values, making an impression that the coordinate passes behind the axis if the  value is missing. The highlight variation emphasizes the presence of missing values by using dashed lines or glyph to draw attention to the missing values. Their findings indicate that for each parallel coordinates task, there is a preferred visualization concept, for example, the Missing Values Axis is best for missing value estimation, Information Removal is best for outlier detection, and Information Removal and Imputed Values are best for trend estimation. Their study provided the first task based guidelines as to how missing values may be best visualized in parallel coordinates.

\begin{figure}
    \centering
    \subfloat[][The bar chart]{
    \includegraphics[width=0.7\linewidth]{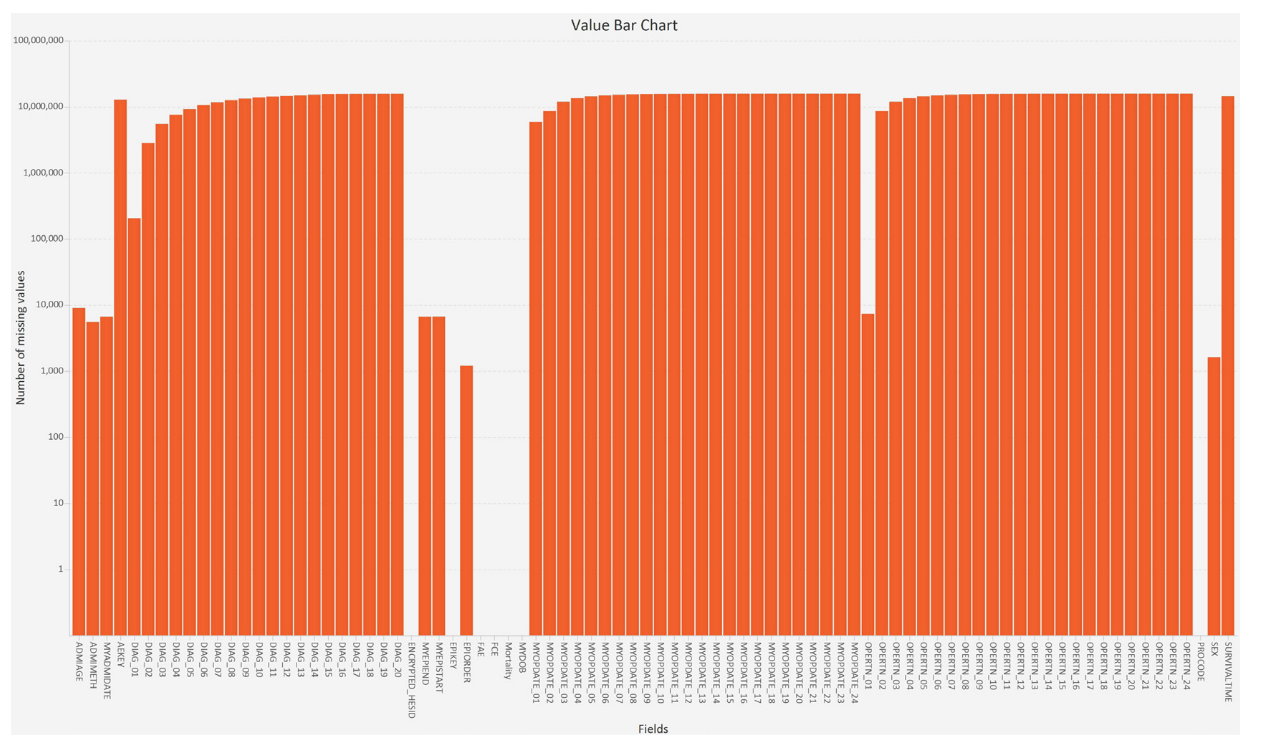}
   \label{RoyBar}}
   \qquad
   \subfloat[][The histogram]{
\includegraphics[width=0.7\linewidth]{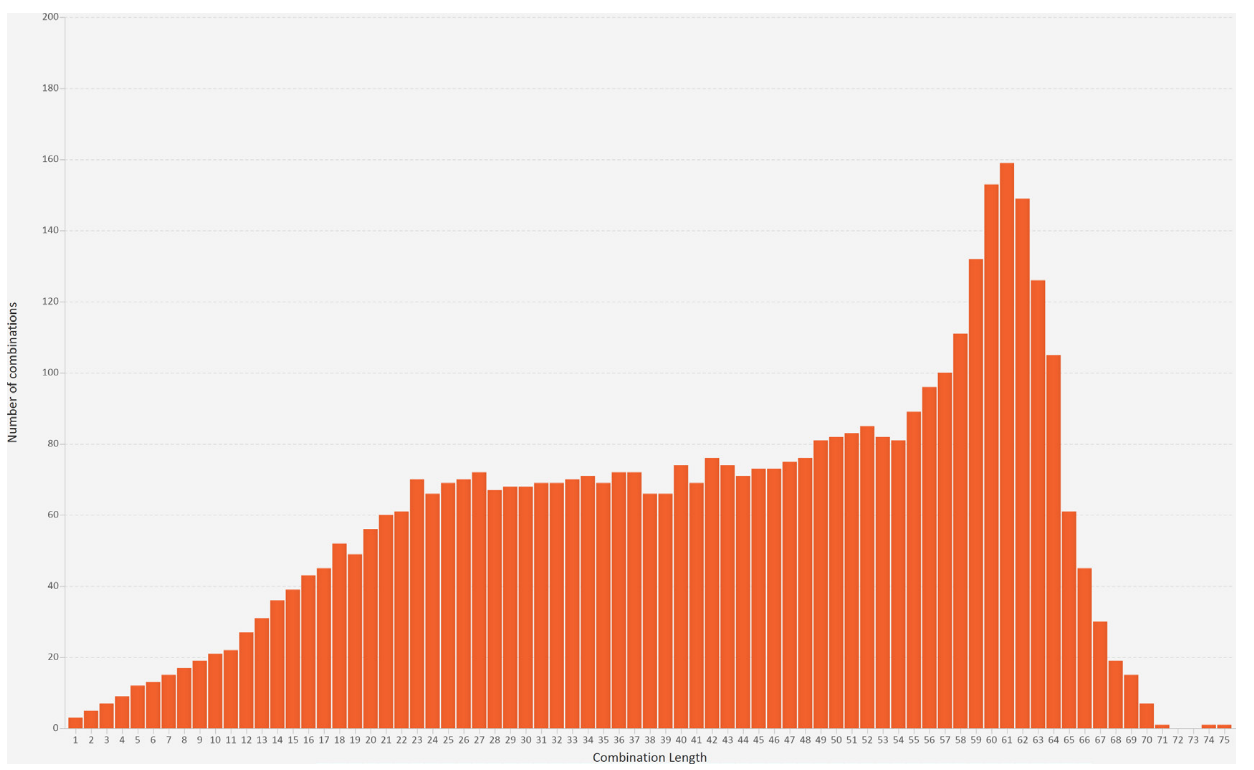}
   \label{RoyHist}}
   \qquad
   \subfloat[][The heatmap]{
   \includegraphics[width=0.7\linewidth]{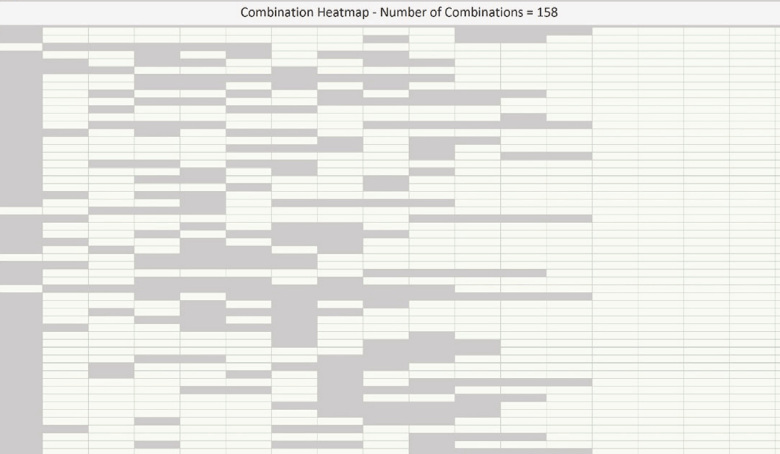}
   \label{RoyHeatmap}
   }
    \caption{ACE techniques: Bar chart, histogram and heatmap.}
    \label{fig:Ruddle}
\end{figure}

Ruddle et al. \ \cite{Ruddle2022} explores a novel approach for analyzing missing data in large datasets, specifically NHS hospital records. The paper introduces an interactive tool, ACE (Analysis of Combinations of Events), that visualizes missing data through bar charts \ref{RoyBar} to display the number of missing values per field, heatmaps \ref{RoyHeatmap} to show intersections of missing values across different combinations of fields, histogram \ref{RoyHist} to represent distributions of missing fields and records, and data mining techniques such as Information Gain Ratio (IGR) and entropy calculations to rank fields and explain patterns. The tool helps uncover complex missing data patterns and provides insights to improve data quality, assisting healthcare organizations in refining data collection methods and enhancing hospital payment systems.

\begin{table*}[]
\caption{The novel Classification Table of the reviewed literature in missing data visualization.} 
\label{table_Classification3}
\scalebox{0.9}{
\begin{tabular}{|c|
>{\columncolor[HTML]{A5A5A5}}c |
>{\columncolor[HTML]{F6DFF5}}c |
>{\columncolor[HTML]{C6E0B4}}c 
>{\columncolor[HTML]{C6E0B4}}c 
>{\columncolor[HTML]{C6E0B4}}c |
>{\columncolor[HTML]{9BC2E6}}c 
>{\columncolor[HTML]{9BC2E6}}c 
>{\columncolor[HTML]{9BC2E6}}c 
>{\columncolor[HTML]{9BC2E6}}c |cc|
>{\columncolor[HTML]{FFFFC7}}c 
>{\columncolor[HTML]{FFFFC7}}c |
>{\columncolor[HTML]{F8CBAD}}l 
>{\columncolor[HTML]{F8CBAD}}l 
>{\columncolor[HTML]{F8CBAD}}l 
>{\columncolor[HTML]{F8CBAD}}l 
>{\columncolor[HTML]{F8CBAD}}l 
>{\columncolor[HTML]{F8CBAD}}l |}
\hline
 &
  \cellcolor[HTML]{A5A5A5} &
  \cellcolor[HTML]{F6DFF5}{\color[HTML]{3F3F3F} } &
  \multicolumn{3}{c|}{\cellcolor[HTML]{C6E0B4}{\color[HTML]{3F3F3F} \textbf{Paper type}}} &
  \multicolumn{4}{c|}{\cellcolor[HTML]{9BC2E6}{\color[HTML]{3F3F3F} \textbf{Data type}}} &
  \multicolumn{2}{c|}{\cellcolor[HTML]{F2F2F2}{\color[HTML]{3F3F3F} \textbf{Interactivity}}} &
  \multicolumn{2}{c|}{\cellcolor[HTML]{FFFFC7}{\color[HTML]{3F3F3F} \textbf{Imputation}}} &
  \multicolumn{6}{c|}{\cellcolor[HTML]{F8CBAD}{\color[HTML]{3F3F3F} \textbf{Tasks}}} \\ \cline{4-20} 
\multirow{-2}{*}{ID} &
  \multirow{-2}{*}{\cellcolor[HTML]{A5A5A5}\textbf{Reference}} &
  \multirow{-2}{*}{\cellcolor[HTML]{F6DFF5}{\color[HTML]{3F3F3F} \textbf{Year}}} &
  \multicolumn{1}{c|}{\cellcolor[HTML]{C6E0B4}{\color[HTML]{3F3F3F} \rotatebox[origin=c]{90}{\textbf{Techniques papers}}}} &
  \multicolumn{1}{c|}{\cellcolor[HTML]{C6E0B4}{\color[HTML]{3F3F3F} \rotatebox[origin=c]{90}{\textbf{Applications and tools}}}} &
  {\color[HTML]{3F3F3F}\rotatebox[origin=c]{90}{ \textbf{Evaluation papers}}} &
  \multicolumn{1}{c|}{\cellcolor[HTML]{9BC2E6}{\color[HTML]{3F3F3F} \rotatebox[origin=c]{90}{\textbf{Numerical}}}} &
  \multicolumn{1}{c|}{\cellcolor[HTML]{9BC2E6}{\color[HTML]{3F3F3F} \rotatebox[origin=c]{90}{\textbf{Categorical}}}} &
  \multicolumn{1}{c|}{\cellcolor[HTML]{9BC2E6}{\color[HTML]{3F3F3F} \rotatebox[origin=c]{90}{\textbf{Temporal}}}} &
  {\color[HTML]{3F3F3F} \rotatebox[origin=c]{90}{\textbf{other}} }&
  \multicolumn{1}{c|}{\cellcolor[HTML]{F2F2F2}{\color[HTML]{3F3F3F} \rotatebox[origin=c]{90}{\textbf{Non-interactivity}}} }&
  \cellcolor[HTML]{F2F2F2}{\color[HTML]{3F3F3F} \rotatebox[origin=c]{90}{\textbf{Interactive}}} &
  \multicolumn{1}{c|}{\cellcolor[HTML]{FFFFC7}{\color[HTML]{3F3F3F} \rotatebox[origin=c]{90}{\textbf{Imputed data}}}} &
  {\color[HTML]{3F3F3F} \rotatebox[origin=c]{90}{ \textbf{Non Imputed data}}} &
  \multicolumn{1}{c|}{\cellcolor[HTML]{F8CBAD}{\color[HTML]{3F3F3F} \rotatebox[origin=c]{90}{\textbf{Missing data visualization}}}} &
  \multicolumn{1}{c|}{\cellcolor[HTML]{F8CBAD}{\color[HTML]{3F3F3F} \rotatebox[origin=c]{90}{\textbf{Missing data exploring}}}} &
  \multicolumn{1}{c|}{\cellcolor[HTML]{F8CBAD}{\color[HTML]{3F3F3F} \rotatebox[origin=c]{90}{\textbf{Missing data structure}}} }&
  \multicolumn{1}{c|}{\cellcolor[HTML]{F8CBAD}{\color[HTML]{3F3F3F} \rotatebox[origin=c]{90}{\textbf{Missing patterns}}}} &
  \multicolumn{1}{c|}{\cellcolor[HTML]{F8CBAD}{\color[HTML]{3F3F3F} \rotatebox[origin=c]{90}{\textbf{Missing mechanism}}}} &
  \multicolumn{1}{c|}{\cellcolor[HTML]{F8CBAD}{\color[HTML]{3F3F3F} \rotatebox[origin=c]{90}{\textbf{Imputation guiding}}}} \\ \hline
1 & \cite{Twiddy1994}
   &
  {\color[HTML]{3F3F3F} 1994} &
  \multicolumn{1}{c|}{\cellcolor[HTML]{C6E0B4}\textbf{•}} &
  \multicolumn{1}{c|}{\cellcolor[HTML]{C6E0B4}\textbf{}} &
  \textbf{} &
  \multicolumn{1}{c|}{\cellcolor[HTML]{9BC2E6}\textbf{}} &
  \multicolumn{1}{c|}{\cellcolor[HTML]{9BC2E6}\textbf{}} &
  \multicolumn{1}{c|}{\cellcolor[HTML]{9BC2E6}\textbf{}} &
  images &
  \multicolumn{1}{c|}{\cellcolor[HTML]{F2F2F2}\textbf{•}} &
  \cellcolor[HTML]{F2F2F2}\textbf{} &
  \multicolumn{1}{c|}{\cellcolor[HTML]{FFFFC7}\textbf{}} &
  \textbf{•} &
  \multicolumn{1}{l|}{\cellcolor[HTML]{F8CBAD}•} &
  \multicolumn{1}{l|}{\cellcolor[HTML]{F8CBAD}} &
  \multicolumn{1}{l|}{\cellcolor[HTML]{F8CBAD}} &
  \multicolumn{1}{l|}{\cellcolor[HTML]{F8CBAD}} &
  \multicolumn{1}{l|}{\cellcolor[HTML]{F8CBAD}} &
   \\ \hline
2 & \cite{Theus1997}
   &
  {\color[HTML]{3F3F3F} 1997} &
  \multicolumn{1}{c|}{\cellcolor[HTML]{C6E0B4}\textbf{}} &
  \multicolumn{1}{c|}{\cellcolor[HTML]{C6E0B4}\textbf{•}} &
  \textbf{} &
  \multicolumn{1}{c|}{\cellcolor[HTML]{9BC2E6}\textbf{•}} &
  \multicolumn{1}{c|}{\cellcolor[HTML]{9BC2E6}\textbf{•}} &
  \multicolumn{1}{c|}{\cellcolor[HTML]{9BC2E6}\textbf{}} &
  \textbf{} &
  \multicolumn{1}{c|}{\cellcolor[HTML]{F2F2F2}\textbf{}} &
  \cellcolor[HTML]{F2F2F2}\textbf{•} &
  \multicolumn{1}{c|}{\cellcolor[HTML]{FFFFC7}\textbf{}} &
  \textbf{•} &
  \multicolumn{1}{l|}{\cellcolor[HTML]{F8CBAD}•} &
  \multicolumn{1}{l|}{\cellcolor[HTML]{F8CBAD}} &
  \multicolumn{1}{l|}{\cellcolor[HTML]{F8CBAD}} &
  \multicolumn{1}{l|}{\cellcolor[HTML]{F8CBAD}} &
  \multicolumn{1}{l|}{\cellcolor[HTML]{F8CBAD}} &
   \\ \hline
3 & \cite{Wang2007}
   &
  {\color[HTML]{3F3F3F} 2007} &
  \multicolumn{1}{c|}{\cellcolor[HTML]{C6E0B4}\textbf{•}} &
  \multicolumn{1}{c|}{\cellcolor[HTML]{C6E0B4}\textbf{}} &
  \textbf{} &
  \multicolumn{1}{c|}{\cellcolor[HTML]{9BC2E6}\textbf{}} &
  \multicolumn{1}{c|}{\cellcolor[HTML]{9BC2E6}\textbf{•}} &
  \multicolumn{1}{c|}{\cellcolor[HTML]{9BC2E6}\textbf{}} &
  \textbf{} &
  \multicolumn{1}{c|}{\cellcolor[HTML]{F2F2F2}\textbf{•}} &
  \cellcolor[HTML]{F2F2F2}\textbf{} &
  \multicolumn{1}{c|}{\cellcolor[HTML]{FFFFC7}\textbf{}} &
  \textbf{•} &
  \multicolumn{1}{l|}{\cellcolor[HTML]{F8CBAD}} &
  \multicolumn{1}{l|}{\cellcolor[HTML]{F8CBAD}•} &
  \multicolumn{1}{l|}{\cellcolor[HTML]{F8CBAD}} &
  \multicolumn{1}{l|}{\cellcolor[HTML]{F8CBAD}} &
  \multicolumn{1}{l|}{\cellcolor[HTML]{F8CBAD}} &
   \\ \hline
4 & \cite{Honaker2011}
   &
  {\color[HTML]{3F3F3F} 2011} &
  \multicolumn{1}{c|}{\cellcolor[HTML]{C6E0B4}\textbf{•}} &
  \multicolumn{1}{c|}{\cellcolor[HTML]{C6E0B4}\textbf{}} &
  \textbf{} &
  \multicolumn{1}{c|}{\cellcolor[HTML]{9BC2E6}\textbf{•}} &
  \multicolumn{1}{c|}{\cellcolor[HTML]{9BC2E6}\textbf{}} &
  \multicolumn{1}{c|}{\cellcolor[HTML]{9BC2E6}\textbf{•}} &
  \textbf{} &
  \multicolumn{1}{c|}{\cellcolor[HTML]{F2F2F2}\textbf{}} &
  \cellcolor[HTML]{F2F2F2}\textbf{•} &
  \multicolumn{1}{c|}{\cellcolor[HTML]{FFFFC7}\textbf{•}} &
  \textbf{} &
  \multicolumn{1}{l|}{\cellcolor[HTML]{F8CBAD}} &
  \multicolumn{1}{l|}{\cellcolor[HTML]{F8CBAD}} &
  \multicolumn{1}{l|}{\cellcolor[HTML]{F8CBAD}} &
  \multicolumn{1}{l|}{\cellcolor[HTML]{F8CBAD}•} &
  \multicolumn{1}{l|}{\cellcolor[HTML]{F8CBAD}} &
  • \\ \hline
5 & \cite{Templ2011}
   &
  {\color[HTML]{3F3F3F} 2011} &
  \multicolumn{1}{c|}{\cellcolor[HTML]{C6E0B4}\textbf{}} &
  \multicolumn{1}{c|}{\cellcolor[HTML]{C6E0B4}\textbf{•}} &
  \textbf{} &
  \multicolumn{1}{c|}{\cellcolor[HTML]{9BC2E6}\textbf{}} &
  \multicolumn{1}{c|}{\cellcolor[HTML]{9BC2E6}\textbf{}} &
  \multicolumn{1}{c|}{\cellcolor[HTML]{9BC2E6}\textbf{•}} &
  \textbf{} &
  \multicolumn{1}{c|}{\cellcolor[HTML]{F2F2F2}\textbf{•}} &
  \cellcolor[HTML]{F2F2F2}\textbf{} &
  \multicolumn{1}{c|}{\cellcolor[HTML]{FFFFC7}\textbf{}} &
  \textbf{•} &
  \multicolumn{1}{l|}{\cellcolor[HTML]{F8CBAD}•} &
  \multicolumn{1}{l|}{\cellcolor[HTML]{F8CBAD}•} &
  \multicolumn{1}{l|}{\cellcolor[HTML]{F8CBAD}•} &
  \multicolumn{1}{l|}{\cellcolor[HTML]{F8CBAD}•} &
  \multicolumn{1}{l|}{\cellcolor[HTML]{F8CBAD}•} &
  • \\ \hline
6 & \cite{Eaton2005}
   &
  2011 &
  \multicolumn{1}{c|}{\cellcolor[HTML]{C6E0B4}} &
  \multicolumn{1}{c|}{\cellcolor[HTML]{C6E0B4}} &
  • &
  \multicolumn{1}{c|}{\cellcolor[HTML]{9BC2E6}} &
  \multicolumn{1}{c|}{\cellcolor[HTML]{9BC2E6}} &
  \multicolumn{1}{c|}{\cellcolor[HTML]{9BC2E6}•} &
   &
  \multicolumn{1}{c|}{•} &
   &
  \multicolumn{1}{c|}{\cellcolor[HTML]{FFFFC7}} &
   &
  \multicolumn{1}{l|}{\cellcolor[HTML]{F8CBAD}•} &
  \multicolumn{1}{l|}{\cellcolor[HTML]{F8CBAD}} &
  \multicolumn{1}{l|}{\cellcolor[HTML]{F8CBAD}} &
  \multicolumn{1}{l|}{\cellcolor[HTML]{F8CBAD}} &
  \multicolumn{1}{l|}{\cellcolor[HTML]{F8CBAD}} &
   \\ \hline
7 & \cite{Lu2012}
   &
  2012 &
  \multicolumn{1}{c|}{\cellcolor[HTML]{C6E0B4}•} &
  \multicolumn{1}{c|}{\cellcolor[HTML]{C6E0B4}} &
   &
  \multicolumn{1}{c|}{\cellcolor[HTML]{9BC2E6}•} &
  \multicolumn{1}{c|}{\cellcolor[HTML]{9BC2E6}•} &
  \multicolumn{1}{c|}{\cellcolor[HTML]{9BC2E6}} &
   &
  \multicolumn{1}{c|}{•} &
   &
  \multicolumn{1}{c|}{\cellcolor[HTML]{FFFFC7}} &
  • &
  \multicolumn{1}{l|}{\cellcolor[HTML]{F8CBAD}•} &
  \multicolumn{1}{l|}{\cellcolor[HTML]{F8CBAD}} &
  \multicolumn{1}{l|}{\cellcolor[HTML]{F8CBAD}} &
  \multicolumn{1}{l|}{\cellcolor[HTML]{F8CBAD}} &
  \multicolumn{1}{l|}{\cellcolor[HTML]{F8CBAD}} &
   \\ \hline
8 & \cite{Templ2012}
   &
  {\color[HTML]{3F3F3F} 2012} &
  \multicolumn{1}{c|}{\cellcolor[HTML]{C6E0B4}\textbf{}} &
  \multicolumn{1}{c|}{\cellcolor[HTML]{C6E0B4}\textbf{•}} &
  \textbf{} &
  \multicolumn{1}{c|}{\cellcolor[HTML]{9BC2E6}\textbf{•}} &
  \multicolumn{1}{c|}{\cellcolor[HTML]{9BC2E6}\textbf{•}} &
  \multicolumn{1}{c|}{\cellcolor[HTML]{9BC2E6}\textbf{}} &
  \textbf{} &
  \multicolumn{1}{c|}{\cellcolor[HTML]{F2F2F2}\textbf{}} &
  \cellcolor[HTML]{F2F2F2}\textbf{•} &
  \multicolumn{1}{c|}{\cellcolor[HTML]{FFFFC7}\textbf{•}} &
  \textbf{} &
  \multicolumn{1}{l|}{\cellcolor[HTML]{F8CBAD}•} &
  \multicolumn{1}{l|}{\cellcolor[HTML]{F8CBAD}•} &
  \multicolumn{1}{l|}{\cellcolor[HTML]{F8CBAD}•} &
  \multicolumn{1}{l|}{\cellcolor[HTML]{F8CBAD}•} &
  \multicolumn{1}{l|}{\cellcolor[HTML]{F8CBAD}•} &
  • \\ \hline
9 & \cite{Andreasson2014}
   &
  {\color[HTML]{3F3F3F} 2014} &
  \multicolumn{1}{c|}{\cellcolor[HTML]{C6E0B4}\textbf{}} &
  \multicolumn{1}{c|}{\cellcolor[HTML]{C6E0B4}\textbf{}} &
  \textbf{•} &
  \multicolumn{1}{c|}{\cellcolor[HTML]{9BC2E6}\textbf{•}} &
  \multicolumn{1}{c|}{\cellcolor[HTML]{9BC2E6}\textbf{•}} &
  \multicolumn{1}{c|}{\cellcolor[HTML]{9BC2E6}\textbf{•}} &
  \textbf{} &
  \multicolumn{1}{c|}{\cellcolor[HTML]{F2F2F2}\textbf{•}} &
  \cellcolor[HTML]{F2F2F2}\textbf{} &
  \multicolumn{1}{c|}{\cellcolor[HTML]{FFFFC7}\textbf{•}} &
  \textbf{} &
  \multicolumn{1}{l|}{\cellcolor[HTML]{F8CBAD}•} &
  \multicolumn{1}{l|}{\cellcolor[HTML]{F8CBAD}} &
  \multicolumn{1}{l|}{\cellcolor[HTML]{F8CBAD}} &
  \multicolumn{1}{l|}{\cellcolor[HTML]{F8CBAD}} &
  \multicolumn{1}{l|}{\cellcolor[HTML]{F8CBAD}} &
   \\ \hline
10 & \cite{Bogl2015}
   &
  2015 &
  \multicolumn{1}{c|}{\cellcolor[HTML]{C6E0B4}•} &
  \multicolumn{1}{c|}{\cellcolor[HTML]{C6E0B4}} &
   &
  \multicolumn{1}{c|}{\cellcolor[HTML]{9BC2E6}} &
  \multicolumn{1}{c|}{\cellcolor[HTML]{9BC2E6}} &
  \multicolumn{1}{c|}{\cellcolor[HTML]{9BC2E6}•} &
   &
  \multicolumn{1}{c|}{} &
  • &
  \multicolumn{1}{c|}{\cellcolor[HTML]{FFFFC7}•} &
   &
  \multicolumn{1}{l|}{\cellcolor[HTML]{F8CBAD}} &
  \multicolumn{1}{l|}{\cellcolor[HTML]{F8CBAD}} &
  \multicolumn{1}{l|}{\cellcolor[HTML]{F8CBAD}} &
  \multicolumn{1}{l|}{\cellcolor[HTML]{F8CBAD}} &
  \multicolumn{1}{l|}{\cellcolor[HTML]{F8CBAD}} &
   \\ \hline
11 & \cite{Cheng}
   &
  {\color[HTML]{3F3F3F} 2015} &
  \multicolumn{1}{c|}{\cellcolor[HTML]{C6E0B4}\textbf{}} &
  \multicolumn{1}{c|}{\cellcolor[HTML]{C6E0B4}\textbf{•}} &
  \textbf{} &
  \multicolumn{1}{c|}{\cellcolor[HTML]{9BC2E6}\textbf{}} &
  \multicolumn{1}{c|}{\cellcolor[HTML]{9BC2E6}\textbf{}} &
  \multicolumn{1}{c|}{\cellcolor[HTML]{9BC2E6}\textbf{•}} &
  \textbf{} &
  \multicolumn{1}{c|}{\cellcolor[HTML]{F2F2F2}\textbf{}} &
  \cellcolor[HTML]{F2F2F2}\textbf{•} &
  \multicolumn{1}{c|}{\cellcolor[HTML]{FFFFC7}\textbf{•}} &
  \textbf{} &
  \multicolumn{1}{l|}{\cellcolor[HTML]{F8CBAD}} &
  \multicolumn{1}{l|}{\cellcolor[HTML]{F8CBAD}} &
  \multicolumn{1}{l|}{\cellcolor[HTML]{F8CBAD}} &
  \multicolumn{1}{l|}{\cellcolor[HTML]{F8CBAD}•} &
  \multicolumn{1}{l|}{\cellcolor[HTML]{F8CBAD}} &
   \\ \hline
12 & \cite{Alemzadeh2017}
   &
  2017 &
  \multicolumn{1}{c|}{\cellcolor[HTML]{C6E0B4}} &
  \multicolumn{1}{c|}{\cellcolor[HTML]{C6E0B4}•} &
   &
  \multicolumn{1}{c|}{\cellcolor[HTML]{9BC2E6}} &
  \multicolumn{1}{c|}{\cellcolor[HTML]{9BC2E6}} &
  \multicolumn{1}{c|}{\cellcolor[HTML]{9BC2E6}} &
  Epidemiological data &
  \multicolumn{1}{c|}{} &
  • &
  \multicolumn{1}{c|}{\cellcolor[HTML]{FFFFC7}•} &
   &
  \multicolumn{1}{l|}{\cellcolor[HTML]{F8CBAD}•} &
  \multicolumn{1}{l|}{\cellcolor[HTML]{F8CBAD}•} &
  \multicolumn{1}{l|}{\cellcolor[HTML]{F8CBAD}} &
  \multicolumn{1}{l|}{\cellcolor[HTML]{F8CBAD}•} &
  \multicolumn{1}{l|}{\cellcolor[HTML]{F8CBAD}} &
  • \\ \hline
13 & \cite{Sjobergh2017}
   &
  2017 &
  \multicolumn{1}{c|}{\cellcolor[HTML]{C6E0B4}•} &
  \multicolumn{1}{c|}{\cellcolor[HTML]{C6E0B4}} &
   &
  \multicolumn{1}{c|}{\cellcolor[HTML]{9BC2E6}•} &
  \multicolumn{1}{c|}{\cellcolor[HTML]{9BC2E6}•} &
  \multicolumn{1}{c|}{\cellcolor[HTML]{9BC2E6}•} &
   &
  \multicolumn{1}{c|}{} &
  • &
  \multicolumn{1}{c|}{\cellcolor[HTML]{FFFFC7}} &
  • &
  \multicolumn{1}{l|}{\cellcolor[HTML]{F8CBAD}•} &
  \multicolumn{1}{l|}{\cellcolor[HTML]{F8CBAD}} &
  \multicolumn{1}{l|}{\cellcolor[HTML]{F8CBAD}} &
  \multicolumn{1}{l|}{\cellcolor[HTML]{F8CBAD}} &
  \multicolumn{1}{l|}{\cellcolor[HTML]{F8CBAD}} &
   \\ \hline
14 & \cite{Song2019}
   &
  2019 &
  \multicolumn{1}{c|}{\cellcolor[HTML]{C6E0B4}} &
  \multicolumn{1}{c|}{\cellcolor[HTML]{C6E0B4}} &
  • &
  \multicolumn{1}{c|}{\cellcolor[HTML]{9BC2E6}} &
  \multicolumn{1}{c|}{\cellcolor[HTML]{9BC2E6}} &
  \multicolumn{1}{c|}{\cellcolor[HTML]{9BC2E6}•} &
   &
  \multicolumn{1}{c|}{•} &
   &
  \multicolumn{1}{c|}{\cellcolor[HTML]{FFFFC7}} &
  • &
  \multicolumn{1}{l|}{\cellcolor[HTML]{F8CBAD}•} &
  \multicolumn{1}{l|}{\cellcolor[HTML]{F8CBAD}} &
  \multicolumn{1}{l|}{\cellcolor[HTML]{F8CBAD}} &
  \multicolumn{1}{l|}{\cellcolor[HTML]{F8CBAD}} &
  \multicolumn{1}{l|}{\cellcolor[HTML]{F8CBAD}} &
   \\ \hline
15 & \cite{Fernstad}
   &
  2019 &
  \multicolumn{1}{c|}{\cellcolor[HTML]{C6E0B4}} &
  \multicolumn{1}{c|}{\cellcolor[HTML]{C6E0B4}} &
  • &
  \multicolumn{1}{c|}{\cellcolor[HTML]{9BC2E6}•} &
  \multicolumn{1}{c|}{\cellcolor[HTML]{9BC2E6}•} &
  \multicolumn{1}{c|}{\cellcolor[HTML]{9BC2E6}} &
   &
  \multicolumn{1}{c|}{•} &
   &
  \multicolumn{1}{c|}{\cellcolor[HTML]{FFFFC7}} &
  • &
  \multicolumn{1}{l|}{\cellcolor[HTML]{F8CBAD}} &
  \multicolumn{1}{l|}{\cellcolor[HTML]{F8CBAD}} &
  \multicolumn{1}{l|}{\cellcolor[HTML]{F8CBAD}} &
  \multicolumn{1}{l|}{\cellcolor[HTML]{F8CBAD}•} &
  \multicolumn{1}{l|}{\cellcolor[HTML]{F8CBAD}} &
   \\ \hline
16 & \cite{Valero-Mora2019}
   &
  2019 &
  \multicolumn{1}{c|}{\cellcolor[HTML]{C6E0B4}•} &
  \multicolumn{1}{c|}{\cellcolor[HTML]{C6E0B4}} &
   &
  \multicolumn{1}{c|}{\cellcolor[HTML]{9BC2E6}•} &
  \multicolumn{1}{c|}{\cellcolor[HTML]{9BC2E6}} &
  \multicolumn{1}{c|}{\cellcolor[HTML]{9BC2E6}} &
   &
  \multicolumn{1}{c|}{•} &
   &
  \multicolumn{1}{c|}{\cellcolor[HTML]{FFFFC7}•} &
   &
  \multicolumn{1}{l|}{\cellcolor[HTML]{F8CBAD}} &
  \multicolumn{1}{l|}{\cellcolor[HTML]{F8CBAD}} &
  \multicolumn{1}{l|}{\cellcolor[HTML]{F8CBAD}} &
  \multicolumn{1}{l|}{\cellcolor[HTML]{F8CBAD}•} &
  \multicolumn{1}{l|}{\cellcolor[HTML]{F8CBAD}} &
   \\ \hline
17 & \cite{Alemzadeh2020}
   &
  2020 &
  \multicolumn{1}{c|}{\cellcolor[HTML]{C6E0B4}} &
  \multicolumn{1}{c|}{\cellcolor[HTML]{C6E0B4}•} &
   &
  \multicolumn{1}{c|}{\cellcolor[HTML]{9BC2E6}} &
  \multicolumn{1}{c|}{\cellcolor[HTML]{9BC2E6}} &
  \multicolumn{1}{c|}{\cellcolor[HTML]{9BC2E6}} &
  Longitudinal data &
  \multicolumn{1}{c|}{} &
  • &
  \multicolumn{1}{c|}{\cellcolor[HTML]{FFFFC7}•} &
   &
  \multicolumn{1}{l|}{\cellcolor[HTML]{F8CBAD}} &
  \multicolumn{1}{l|}{\cellcolor[HTML]{F8CBAD}•} &
  \multicolumn{1}{l|}{\cellcolor[HTML]{F8CBAD}} &
  \multicolumn{1}{l|}{\cellcolor[HTML]{F8CBAD}} &
  \multicolumn{1}{l|}{\cellcolor[HTML]{F8CBAD}} &
  • \\ \hline
18  &\cite{Yeon2020}
   &
  {\color[HTML]{3F3F3F} 2020} &
  \multicolumn{1}{c|}{\cellcolor[HTML]{C6E0B4}\textbf{}} &
  \multicolumn{1}{c|}{\cellcolor[HTML]{C6E0B4}\textbf{•}} &
  \textbf{} &
  \multicolumn{1}{c|}{\cellcolor[HTML]{9BC2E6}\textbf{}} &
  \multicolumn{1}{c|}{\cellcolor[HTML]{9BC2E6}\textbf{}} &
  \multicolumn{1}{c|}{\cellcolor[HTML]{9BC2E6}\textbf{•}} &
  \textbf{} &
  \multicolumn{1}{c|}{\cellcolor[HTML]{F2F2F2}\textbf{}} &
  \cellcolor[HTML]{F2F2F2}\textbf{•} &
  \multicolumn{1}{c|}{\cellcolor[HTML]{FFFFC7}\textbf{}} &
  \textbf{} &
  \multicolumn{1}{l|}{\cellcolor[HTML]{F8CBAD}} &
  \multicolumn{1}{l|}{\cellcolor[HTML]{F8CBAD}•} &
  \multicolumn{1}{l|}{\cellcolor[HTML]{F8CBAD}} &
  \multicolumn{1}{l|}{\cellcolor[HTML]{F8CBAD}•} &
  \multicolumn{1}{l|}{\cellcolor[HTML]{F8CBAD}} &
   \\ \hline
19 & \cite{Song2021}
   &
  {\color[HTML]{3F3F3F} 2021} &
  \multicolumn{1}{c|}{\cellcolor[HTML]{C6E0B4}\textbf{}} &
  \multicolumn{1}{c|}{\cellcolor[HTML]{C6E0B4}\textbf{}} &
  \textbf{•} &
  \multicolumn{1}{c|}{\cellcolor[HTML]{9BC2E6}\textbf{•}} &
  \multicolumn{1}{c|}{\cellcolor[HTML]{9BC2E6}\textbf{•}} &
  \multicolumn{1}{c|}{\cellcolor[HTML]{9BC2E6}\textbf{}} &
  \textbf{} &
  \multicolumn{1}{c|}{\cellcolor[HTML]{F2F2F2}\textbf{}} &
  \cellcolor[HTML]{F2F2F2}\textbf{•} &
  \multicolumn{1}{c|}{\cellcolor[HTML]{FFFFC7}\textbf{•}} &
  \textbf{} &
  \multicolumn{1}{l|}{\cellcolor[HTML]{F8CBAD}•} &
  \multicolumn{1}{l|}{\cellcolor[HTML]{F8CBAD}} &
  \multicolumn{1}{l|}{\cellcolor[HTML]{F8CBAD}} &
  \multicolumn{1}{l|}{\cellcolor[HTML]{F8CBAD}} &
  \multicolumn{1}{l|}{\cellcolor[HTML]{F8CBAD}} &
   \\ \hline
20 & \cite{Bauerle2022}
   &
  {\color[HTML]{3F3F3F} 2022} &
  \multicolumn{1}{c|}{\cellcolor[HTML]{C6E0B4}\textbf{}} &
  \multicolumn{1}{c|}{\cellcolor[HTML]{C6E0B4}\textbf{}} &
  \textbf{•} &
  \multicolumn{1}{c|}{\cellcolor[HTML]{9BC2E6}\textbf{•}} &
  \multicolumn{1}{c|}{\cellcolor[HTML]{9BC2E6}\textbf{•}} &
  \multicolumn{1}{c|}{\cellcolor[HTML]{9BC2E6}\textbf{}} &
  \textbf{} &
  \multicolumn{1}{c|}{\cellcolor[HTML]{F2F2F2}\textbf{}} &
  \cellcolor[HTML]{F2F2F2}\textbf{•} &
  \multicolumn{1}{c|}{\cellcolor[HTML]{FFFFC7}\textbf{•}} &
  \textbf{} &
  \multicolumn{1}{l|}{\cellcolor[HTML]{F8CBAD}•} &
  \multicolumn{1}{l|}{\cellcolor[HTML]{F8CBAD}} &
  \multicolumn{1}{l|}{\cellcolor[HTML]{F8CBAD}} &
  \multicolumn{1}{l|}{\cellcolor[HTML]{F8CBAD}} &
  \multicolumn{1}{l|}{\cellcolor[HTML]{F8CBAD}} &
   \\ \hline
21 & \cite{Jimenez2022}
   &
  2022 &
  \multicolumn{1}{c|}{\cellcolor[HTML]{C6E0B4}•} &
  \multicolumn{1}{c|}{\cellcolor[HTML]{C6E0B4}} &
   &
  \multicolumn{1}{c|}{\cellcolor[HTML]{9BC2E6}} &
  \multicolumn{1}{c|}{\cellcolor[HTML]{9BC2E6}} &
  \multicolumn{1}{c|}{\cellcolor[HTML]{9BC2E6}•} &
   &
  \multicolumn{1}{c|}{•} &
   &
  \multicolumn{1}{c|}{\cellcolor[HTML]{FFFFC7}} &
  • &
  \multicolumn{1}{l|}{\cellcolor[HTML]{F8CBAD}•} &
  \multicolumn{1}{l|}{\cellcolor[HTML]{F8CBAD}} &
  \multicolumn{1}{l|}{\cellcolor[HTML]{F8CBAD}} &
  \multicolumn{1}{l|}{\cellcolor[HTML]{F8CBAD}•} &
  \multicolumn{1}{l|}{\cellcolor[HTML]{F8CBAD}} &
   \\ \hline
22 & \cite{Fernstad2022}
   &
  2022 &
  \multicolumn{1}{c|}{\cellcolor[HTML]{C6E0B4}•} &
  \multicolumn{1}{c|}{\cellcolor[HTML]{C6E0B4}} &
   &
  \multicolumn{1}{c|}{\cellcolor[HTML]{9BC2E6}•} &
  \multicolumn{1}{c|}{\cellcolor[HTML]{9BC2E6}•} &
  \multicolumn{1}{c|}{\cellcolor[HTML]{9BC2E6}} &
   &
  \multicolumn{1}{c|}{} &
  • &
  \multicolumn{1}{c|}{\cellcolor[HTML]{FFFFC7}} &
  • &
  \multicolumn{1}{l|}{\cellcolor[HTML]{F8CBAD}} &
  \multicolumn{1}{l|}{\cellcolor[HTML]{F8CBAD}} &
  \multicolumn{1}{l|}{\cellcolor[HTML]{F8CBAD}•} &
  \multicolumn{1}{l|}{\cellcolor[HTML]{F8CBAD}•} &
  \multicolumn{1}{l|}{\cellcolor[HTML]{F8CBAD}} &
   \\ \hline
   
23 & \cite{Ruddle2022}
   &
  2022 &
  \multicolumn{1}{c|}{\cellcolor[HTML]{C6E0B4}} &
  \multicolumn{1}{c|}{\cellcolor[HTML]{C6E0B4}•} &
   &
  \multicolumn{1}{c|}{\cellcolor[HTML]{9BC2E6}•} &
  \multicolumn{1}{c|}{\cellcolor[HTML]{9BC2E6}•} &
  \multicolumn{1}{c|}{\cellcolor[HTML]{9BC2E6}} &
   &
  \multicolumn{1}{c|}{} &
  • &
  \multicolumn{1}{c|}{\cellcolor[HTML]{FFFFC7}} &
  • &
  \multicolumn{1}{l|}{\cellcolor[HTML]{F8CBAD}•} &
  \multicolumn{1}{l|}{\cellcolor[HTML]{F8CBAD}•} &
  \multicolumn{1}{l|}{\cellcolor[HTML]{F8CBAD}} &
  \multicolumn{1}{l|}{\cellcolor[HTML]{F8CBAD}•} &
  \multicolumn{1}{l|}{\cellcolor[HTML]{F8CBAD}} &
   \\ \hline
24 & \cite{Tierney2023}
   &
  2023 &
  \multicolumn{1}{c|}{\cellcolor[HTML]{C6E0B4}•} &
  \multicolumn{1}{c|}{\cellcolor[HTML]{C6E0B4}} &
   &
  \multicolumn{1}{c|}{\cellcolor[HTML]{9BC2E6}•} &
  \multicolumn{1}{c|}{\cellcolor[HTML]{9BC2E6}•} &
  \multicolumn{1}{c|}{\cellcolor[HTML]{9BC2E6}•} &
  \makecell{spatial, networks,\\ and longitudinal data} &
  \multicolumn{1}{c|}{} &
  • &
  \multicolumn{1}{c|}{\cellcolor[HTML]{FFFFC7}} &
  • &
  \multicolumn{1}{l|}{\cellcolor[HTML]{F8CBAD}•} &
  \multicolumn{1}{l|}{\cellcolor[HTML]{F8CBAD}•} &
  \multicolumn{1}{l|}{\cellcolor[HTML]{F8CBAD}•} &
  \multicolumn{1}{l|}{\cellcolor[HTML]{F8CBAD}•} &
  \multicolumn{1}{l|}{\cellcolor[HTML]{F8CBAD}} &
   \\ \hline
25 & \cite{Alsufyani2024}
   &
  2024 &
  \multicolumn{1}{c|}{\cellcolor[HTML]{C6E0B4}•} &
  \multicolumn{1}{c|}{\cellcolor[HTML]{C6E0B4}} &
   &
  \multicolumn{1}{c|}{\cellcolor[HTML]{9BC2E6}•} &
  \multicolumn{1}{c|}{\cellcolor[HTML]{9BC2E6}•} &
  \multicolumn{1}{c|}{\cellcolor[HTML]{9BC2E6}} &
  &
  \multicolumn{1}{c|}{} &
  • &
  \multicolumn{1}{c|}{\cellcolor[HTML]{FFFFC7}} &
  • &
  \multicolumn{1}{l|}{\cellcolor[HTML]{F8CBAD}•} &
  \multicolumn{1}{l|}{\cellcolor[HTML]{F8CBAD}•} &
  \multicolumn{1}{l|}{\cellcolor[HTML]{F8CBAD}•} &
  \multicolumn{1}{l|}{\cellcolor[HTML]{F8CBAD}•} &
  \multicolumn{1}{l|}{\cellcolor[HTML]{F8CBAD}} &
   \\ \hline
\end{tabular}
}
\end{table*}

\section{\textbf{Future Research challenges}} \label{sec:Future}
Many of the publications included in this work focus on other aspects in addition to missing data visualization. Nonetheless, the literature provides insights into potential future research directions. Some of the main challenges facing the designers of missing data visualisation include: 
\begin{itemize}
   \item \textbf{Conducting evaluations with increase generalizability and realism}: The generalizability of studies to date are limited by participant numbers, dataset size and a lack of heterogeneity in data and visualization type. Most evaluations focus on visualization of missing values in numerical data, and is conducted as controlled experiments. Evaluations in more realistic settings with a wider range of data is crucial to fully understand the usefulness and utility of missing data visualization.
    \item \textbf{Supporting a variety of data types:} There is a lack of missing data visualization covering a wider variety of data types, with most of the reviewed visualization methods supporting only specific data types. Current work mainly addresses numerical, categorical or time series data, while less work is done on missing values in, e.g., networks, geometric or heterogeneous data. 
    \item \textbf{Applying the techniques and approaches to real-world datasets:} This was the most common limitation in the reviewed research, with a large number of papers relying on synthetic data. Some papers generated the missing values in the dataset, which, on the one hand enable control, but may also produce inaccurate results and reduce the generalizability and applicability of findings. Since the correlation between different variables and the distribution are considered the main factors in missing data visualization, using real-world data may often be preferable.   
    \item \textbf{Providing high-quality visualization with no loss of useful information and no clutter:} While the majority of current visualization methods tend to lose valuable information and become cluttered when dealing with large data sizes, some reviewed papers have addressed this challenge by focusing on subsets of data and incorporating interactive features for zooming in. However, designing high-quality visualizations that appropriately represent missing data while retaining all useful information, encompass the entire dataset, and avoid clutter remains a significant challenge.
    \item \textbf{Integrating interactive features and facilitating seamless integration with other tools:} Using interactive visualizations can be highly effective and beneficial, as they allow users to explore more details, generate new insights, and enhance decision-making. Integrating visualizations with other tools can significantly enhance data-driven decision-making and reporting, and increase the take-up and impact of visualization. With regards to missing data, the integration of novel visualization methods with established imputation packages and data profiling tools is a core concern. 
\end{itemize}

\section{\textbf{Conclusion }} \label{sec:Conclusion}
This paper presented the state of the art in missing data visualization. This is a growing and underdeveloped field with a comparably small number of significant publications. Based on the collected set of publications comprising all these techniques and tools, we developed a novel taxonomy consisting of five main categories: paper type, data type, interactivity, imputation,  and tasks \ref{table_Classification3}. We also classified the visualization type (common visualisation \ref{table_Classification2} and specialized visualization \ref{table_Classification}) based on the paper type (Techniques, applications and tools, and evaluation papers). The taxonomy presents what has been accomplished in this field and indicates possible gaps in the research literature. This survey makes approaches and techniques more comparable and highlights the similarities and differences, through supporting
visualization designers and data analysts in exploring and comparing new tools and techniques for visualizing missing data. The work presented in this paper indicates that the visualization of missing data is still a comparably rare topic, it highlights the lack of visualization approaches for missing data and highlights gaps in current visualization research that are relevant to address.

\bibliographystyle{abbrv}
\bibliography{mainTemplate}

\end{document}